\documentclass{aa}  

\usepackage{amsmath, amssymb, mathrsfs}
\usepackage{hyperref} 
\usepackage{mathrsfs}
\usepackage{dsfont}

\usepackage{pdflscape}
\usepackage{afterpage}
\usepackage{capt-of}
\usepackage{tikz}
\usepackage{graphicx}
\usepackage{nicefrac}
\usepackage{subcaption}
\usepackage{txfonts}
\begin{document}

   \title{Charting nearby dust clouds using Gaia data only}

   \author{R. H. Leike\inst{1}\inst{2}
          \and
	  T. A. En\ss lin \inst{1}\inst{2}
          }

   \institute{Max Planck Institute for Astrophysics, Karl-Schwarzschildstra\ss e 1, 85748 Garching, Germany
         \and
	 Ludwig-Maximilians-Universit\"at, Geschwister-Scholl Platz 1, 80539 Munich, Germany
             }

   \date{Received XXXX, accepted XXXX}

 
  \abstract
  {}
   {Highly resolved maps of the local Galactic dust are an important ingredient for sky emission models.
   In nearly the whole electromagnetic spectrum one can see imprints of dust, many of which originate from dust clouds within 300pc.
   Having a detailed 3D reconstruction of these local dust clouds enables detailed studies, helps to quantify the impact on other observables and is a milestone necessary to enable larger reconstructions, as every sightline for more distant objects will pass through the local dust.}
   {To infer the dust density we use parallax and extinction estimates published by the Gaia collaboration in their second data release.
   We model the dust as a log-normal process using a hierarchical Bayesian model.
   We also infer non-parametrically the kernel of the log-normal process, which corresponds to the physical spatial correlation power spectrum of the log-density.}
   {Using only Gaia data of the second Gaia data release, we reconstruct the 3D dust density and its
   spatial correlation spectrum in a 600pc cube centered on the Sun. 
   We report a spectral index of the logarithmic dust density of $3.1$ on Fourier scales with wavelengths between $2$pc and $125$pc.
   The resulting 3D dust map as well as the power spectrum 
   and posterior samples are publicly available for download.}
   {}

   \keywords{ISM: dust, extinction --
                Galaxy: local interstellar matter --
		methods: data analysis
               }

   \maketitle
%

\section{Introduction}
\label{sec:introduction}

Emission and extinction by Galactic dust is a prominent astronomical foreground at many wavelengths.
Therefore, knowing its distribution on the 2D sky and in 3D is essential for many astronomical observations.
However, dust is also interesting to be studied on its own, as it provides information
about the physical conditions in the interstellar medium and informs us about star forming regions.
Dust has been mapped out by surveys for a long time, the first notable contribution being \cite{burstein1978hi}.
Their dust reconstruction, as most reconstructions of the dust distribution so far, was focused on 
mapping the dust in 2D on the sky.
This can be done by looking at the sky in infra-red wavelengths,
where it is dominated by dust emission.
However, when mapping out dust using infra-red emission one is biased by the radiation field, as dust emission is not only proportional to the dust density but also to the amount of starlight absorbed by the dust.
Furthermore, dust maps that were produced by mapping infrared light might contain extended infrared sources that are not within our galaxy, as was shown by \cite{chiang2018extragalactic}.
On the other hand, a hypothetical cold dust cloud cannot be seen in infra-red,
leading to systematic errors in the analysis of distant targets, for example quasars or the Cosmic Microwave Background (CMB).

For accurate analyses of objects in our galactic vicinity,
it is vital to have a 3D dust map as a foreground model, which informs us about regions that cannot be observed, or only be observed with less fidelity, due to dust obscuration.
The first non-parametric reconstruction of galactic dust in 3D published is 
\cite{arenou1992tridimensional}.
Since then there have been many attempts to chart the dust density in 3D in increasing resolution, accuracy and for an ever greater part of our galaxy \citep{vergely1997extinction,vergely2001nai, chen2018three, kh2017can, kh2018three, kh2018detection, gontcharov20123d, gontcharov20173d, sale20143d, sale2018large}.
A large driving force for 3D dust reconstruction and astronomy in general are large surveys like 2MASS \citep{skrutskie2006two}, Pan-STARRS \citep{kaiser2002pan} and SDSS/APOGEE \citep{albareti201713th} and WISE \citep{wright2010wide}. 
These surveys provide photometric measurements and some of them spectra for thousands of stars, from which the calculation of photometric distances is possible.
There are two 3D dust reconstructions based on these data sets that are closely related to the approach taken in this paper.

In \cite{lallement20183d} reddening data from 71\,000 sources has been used to perform a 3D reconstruction using Gaussian process regression.
The resulting dust map covers a $4$kpc square of the galactic plane and $600$pc in perpendicular direction with a voxel size of $(5\,\text{pc})^3$.

In \cite{green2018galactic} a 3D dust map is produced by combining the star data of Pan-STARRS and 2MASS, binning it
in angular and distance bins, and performing independent 
Bayesian reconstructions per angular bin.
The result is a dust map that covers three quarters of the sky to a distance up to 2kpc.
This reconstruction shows artificial radial structures
called the "fingers of God effect" in analogy to the well known phenomenon in cosmology \citep{jackson1972critique, tully1978nearby}.
One way to mitigate this effect is to use more accurate parallax information.

A prominent new survey is performed by the Gaia collaboration \citep{brown2016gaia}.
In its second data release (DR2, \cite{brown2018gaia}) accurate parallaxes for roughly 2 billion stars are published.
The provided catalogue also contains estimates of extinction coefficients for a subset of about 88 million stars, using spectral information of the Gaia satellite's three energy bands.
Due to the limited spectral information, the accuracy of the extinction coefficients estimated for individual sightlines is quite low. 
For this reason it is recommended by \cite{andrae2018gaia} to not use the information of individual sightlines but only the joint information of several sightlines.
Even though the data quality of individual sightlines is rather low, the sheer amount of data points and the accuracy of the parallaxes outweigh this limitation as our work shows.

So far, 3D dust reconstructions have never been performed using solely Gaia data, instead Gaia data has been used for its accurate parallax measurements only and the more accurate spectral information of other surveys was used \citep[for example]{lallement20183d}.

In this paper we present a 3D dust reconstruction using Gaia DR2 data only.
The results of the reconstruction are provided online on \url{https://wwwmpa.mpa-garching.mpg.de/~ensslin/research/data/dust.html} or by its doi:10.5281/zenodo.2577337, and can be used under the terms of the ODC-By 1.0 license.
The inference of the unknown dust density from the extinction data is performed by a critical filter, a method for Gaussian process regression with non-parametric kernel learning, first published in \cite{ensslin2011reconstruction}. 
While the statistical model used here is up to minor details equivalent to the model introduced in that paper, the algorithm to arrive at an approximate posterior summary statistics is quite different.
The relevant numerical method used in this paper is outlined in \cite{knollmuller2019metric}, which describes a general method to derive posterior summary statistics for high dimensional Bayesian inference problems.
For a theoretical discussion of the underlying inference framework of information field theory we refer to \cite{ensslin2018information}. 
The algorithm was implemented using the Python package NIFTy5, which is the newest version of the software package NIFTy \citep{SeligMaxEnt2013,steininger2017nifty, arras2019nifty5}\footnote{The version of NIFTy used for this reconstruction is available on \url{https://gitlab.mpcdf.mpg.de/ift/NIFTy}\ .}.
Even though mathematical theory, statistical motivations, and numerical details are distributed over the aforementioned papers, this paper is entirely self-contained by describing the whole method.

In Sec.\,\ref{sec:data} we discuss which part of the Gaia data we used.
We introduce our statistical model of the interstellar dust density as well as of the measurement in Sec.\,\ref{sec:model}.
In Sec.\,\ref{sec:simulated-data-test} we present a test application of the algorithm using synthetic data, verifying the predictive power of the algorithm.
The main results of the dust density reconstruction using Gaia data are presented in Sec.\,\ref{sec:results}. 
This section also contains a brief recommendation on how to use our results.
Our dust reconstruction is compared to other 3D dust density reconstructions in Sec.\,\ref{sec:discussion}.
In Sec.\,\ref{sec:conclusions} we summarize the findings of this paper.

\section{Data}
\label{sec:data}

\begin{figure}[ht!]
	\centering
	\begin{subfigure}[t]{.4\textwidth}
	\includegraphics[trim={1.3cm 1.3cm .5cm 2.5cm}, clip, width=\textwidth]{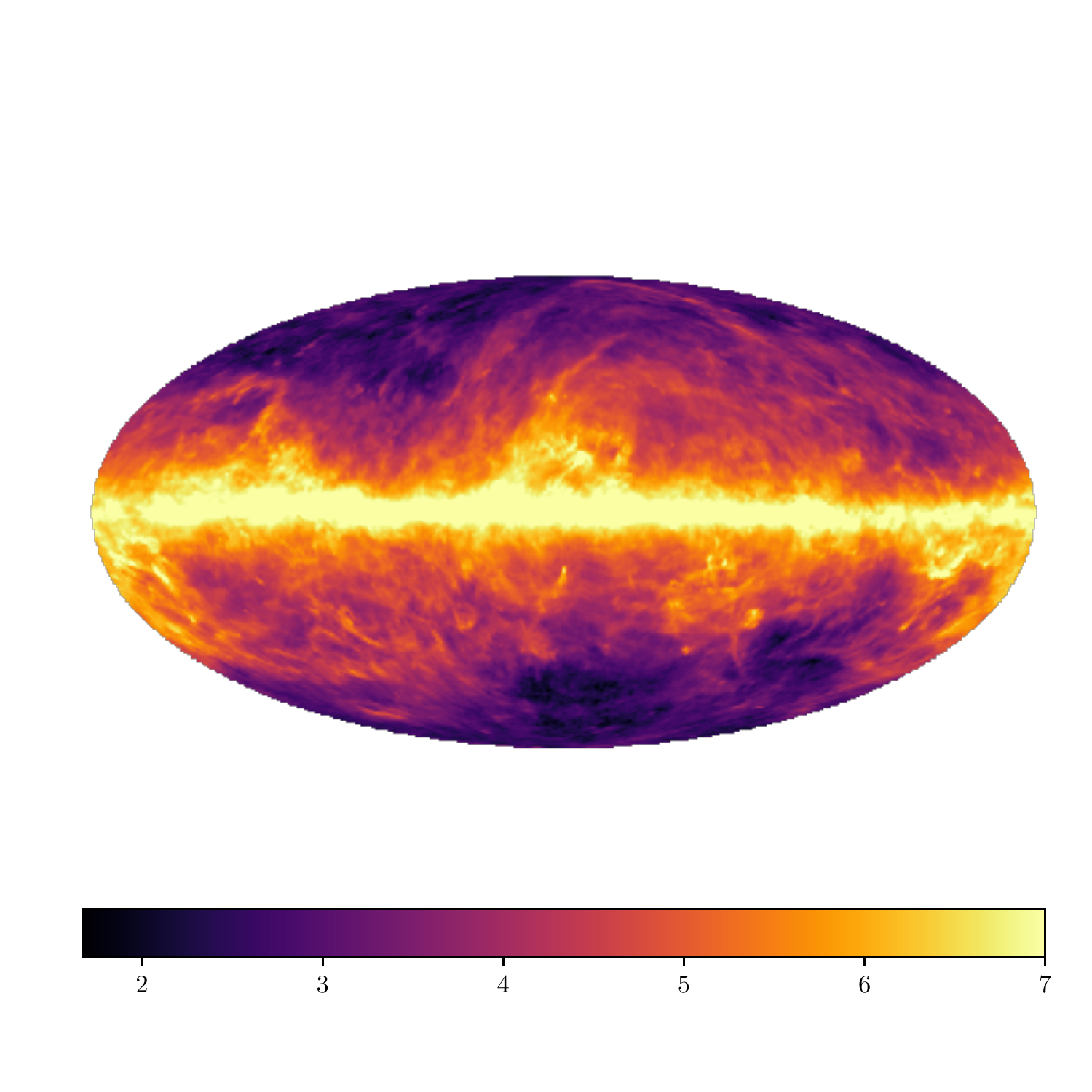}
	\end{subfigure}
	\\
	\begin{subfigure}[t]{.4\textwidth}
	\includegraphics[trim={1.3cm 1.3cm .5cm 2.5cm}, clip, width=\textwidth]{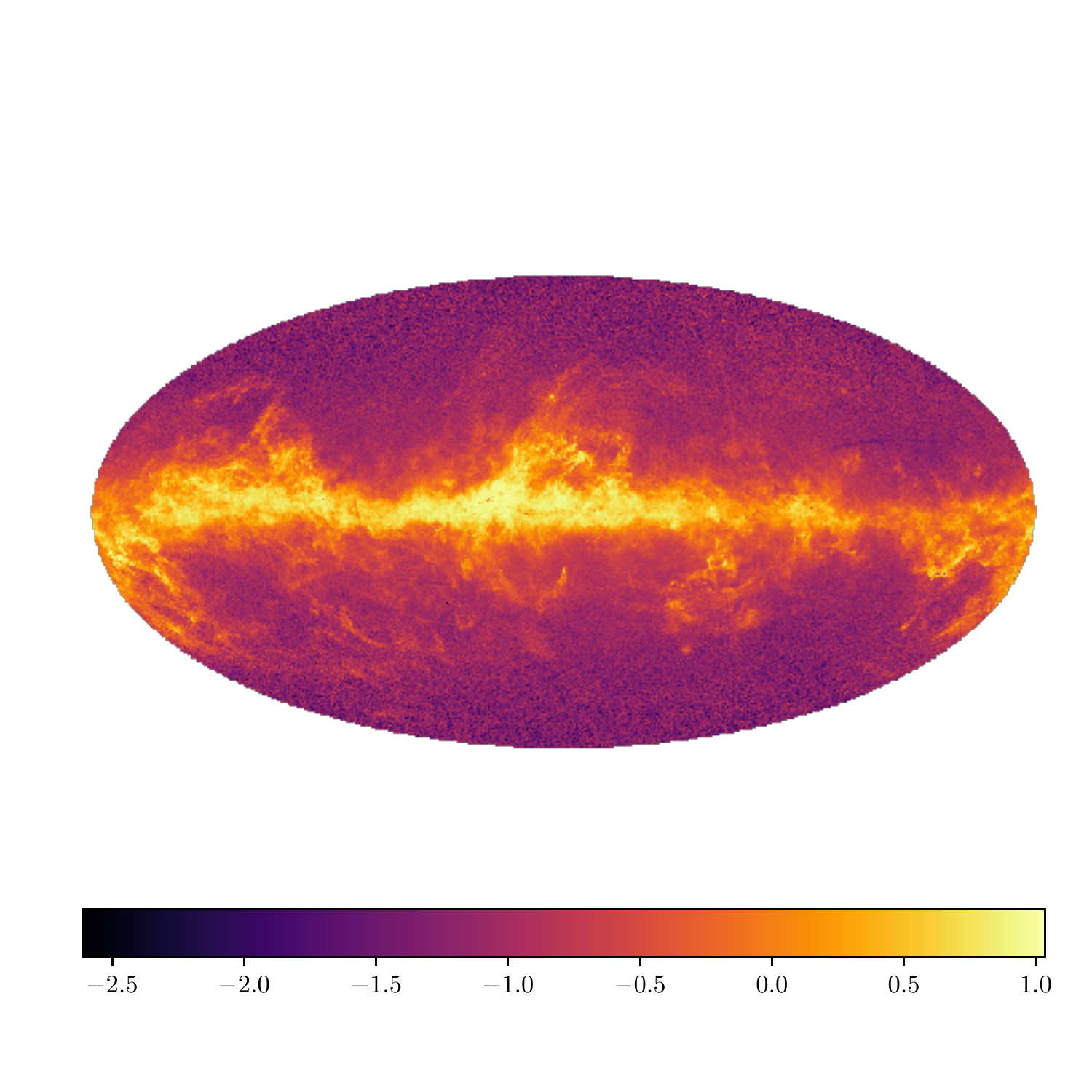}
	\end{subfigure}
	\\
	\begin{subfigure}[t]{.4\textwidth}
	\includegraphics[trim={1.3cm 1.3cm 0.5cm 2.5cm}, clip, width=\textwidth]{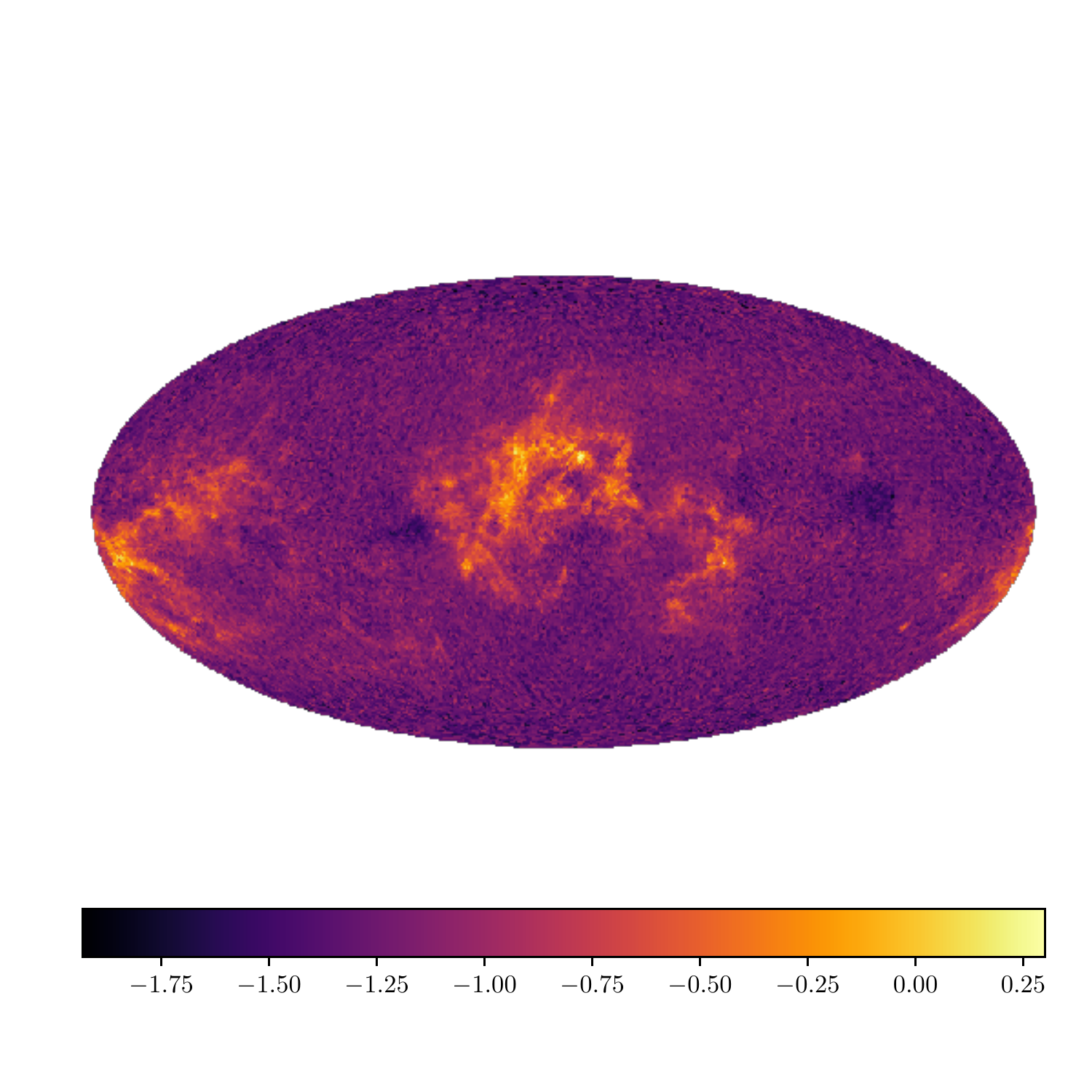}
	\end{subfigure}
	\caption{\label{fig:Planck-and-data}
		The top panel shows the natural logarithm of the thermal dust emission map produced by the Planck collaboration \citep{akrami2018planck}.
	The color-scheme was saturated to visually match that of the Gaia dust extinction data shown in the middle panel.
	The bottom panel shows the subset of Gaia extinction data within a 600pc cube centered on the Sun; the data used in this paper.
	The scale is natural logarithm of the extinction data in magnitudes.
	Data points in the same direction were averaged.
	}
\end{figure}

We used the data from the Gaia DR2 catalogue by \cite{brown2018gaia},
to reconstruct the galactic dust in the nearby interstellar medium.
From the Gaia data archive we extract the parallaxes, the G-band
extinction, the latitude and longitude as well as their respective uncertainties. A plot of the full Gaia extinction data set can be seen in Fig.\,\ref{fig:Planck-and-data}.

We select sources according to the following criteria:
\begin{enumerate}
	\item{the above mentioned data are available for the source}
	\item{the parallax $\widetilde{\omega}$ is inside a 600pc cube around the Sun}
	\item{the relative parallax error is sufficiently low, $\widetilde{\omega}/\sigma_{\widetilde{\omega}} > 5$}
	\item{Priam flag 0100001 or 0100002}
\end{enumerate}
The last two criteria are suggested by \cite{andrae2018gaia}.
There are about 3.7 million stars selected by these criteria.
Fig.\,\ref{fig:Planck-and-data} shows a sky average of the data points used in the reconstruction. 
In this data plot one can observe structures present also in other dust maps, for example the Planck dust map (Fig\,\ref{fig:Planck-and-data}).

\section{Model}
\label{sec:model}
\subsection{Algorithm}
\label{sec:algorithm}

The algorithm is derived from Bayesian reasoning.
In Bayesian reasoning information some data $d$ provides about a quantity of interest $s$
is calculated according to Bayes theorem:
\begin{align}
	P(s|d) = \frac{P(d|s)P(s)}{P(d)} \label{eq:Bayes}
\end{align}

Note that the quantity of interest can be a (possibly high-dimensional) vector, in our case it is the dust density for every point in space ($256^3$ degrees of freedom after discretization).
There are three main ingredients necessary for the inference of the quantity of interest $s$:
\begin{enumerate}
	\item The likelihood $P(d|s)$ of the data $d$ given a realization of the quantity of interest $s$. We describe our likelihood in Subsec.\,\ref{sec:likelihood}.
	\item The prior $P(s)$ describing the best available knowledge about the quantity of interest $s$ in absence of data. We describe our prior in Subsec.\,\ref{sec:prior}. 	
	\item An inference algorithm that yields a statistical summary of $P(s|d)$ given the joint distribution $P(d,s) = P(d|s)P(s)$ of $d$ and $s$. We use the inference algorithm described in \cite{knollmuller2019metric}.
\end{enumerate}

The main quantity of interest $s$ is the logarithmic G-band dust extinction cross-section density $s=\text{ln}(\alpha \rho/\text{pc})$, henceforth called the logarithmic dust density.
Hereby $\rho$ denotes the actual dust mass density and $\alpha$ the average G-band dust cross section per mass. 
The value of $\alpha$ is uncertain, which is why we report extinction densities, also called dust pseudo-densities, instead.

We approximate the posterior with a Gaussian 
\begin{align}
	Q(s) &=\mathscr{G}(s-m, D) \\
	&= \frac{\text{exp}\left(-\frac{1}{2}(s-m)^\dagger D^{-1}(s-m)\right)}{\left|2\pi D\right|^{\frac{1}{2}}}\ ,\label{eq:approximant}
\end{align}
by adopting a suitable mean $m$ and uncertainty dispersion $D$.
The approximation is obtained by minimizing the Kullback-Leibler divergence \citep{Kullback1951}
\begin{align}
	\text{KL}(Q,P) = \int \text{d}Q\,\text{ln}\frac{Q}{P}\label{eq:KL}
\end{align}
with respect to the parameters of $Q$.
This approach is known as variational Bayes \citep{nasrabadi2007pattern} or Gibbs free energy approach \citep{ensslin2010inference}.
The approach we take in finding the unknown approximate posterior mean $m$ and covariance $D$ of Eq.\,\ref{eq:approximant} is described in detail in \cite{knollmuller2019metric}.
It can be summarized as follows:
\begin{enumerate}
	\item{Calculate the negative log-probability $-\text{log}(P(s,d))$ for the problem, disregarding normalization terms like the evidence $P(d)$.}
	\item{Perform coordinate transformations of the unknown quantities until those are a-priori Gaussian distributed with unit covariance\label{enum:whitening} \citep{kucukelbir2017automatic, knollmuller2018encoding}.}
	\item{Choose the class of approximating distributions to be Gaussian with variable mean $m$ and covariance $D = (\mathds{1}+M_m)^{-1}$, where $M_m$ is the Fisher information metric at the current $m$.
		Here $\mathds{1}$ is the contribution of the prior which was transformed in step \ref{enum:whitening} to have unit covariance. This uncertainty dispersion is a lower bound to the uncertainty \citep{cramer1946mathematical, rao1947minimum} and has been shown to be an efficient technique to take cross-correlations between all degrees of freedom into account without having to parameterize them explicitly \citep{knollmuller2019metric}}
	\item{Minimize Eq.\,(\ref{eq:KL}) with respect to $m$ using Newton Conjugate gradient as second order scheme with the covariance $D$ of $Q$ as curvature. 
		The expectation value with respect to $Q$ is hereby approximated through a set of samples drawn from the approximating distribution $Q$. 
		Second order minimization by preconditioning with the inverse Fisher metric is also called natural gradient descent \citep{amari1997neural}  in the literature.}
\end{enumerate}
A description of the used likelihood and the prior follows.

\subsection{Likelihood}
\label{sec:likelihood}

The likelihood $P(d|s)$ can be split into two parts,
one part states how the true extinction depends on the dust density and one part that states how the actual data is distributed given the true extinction on that line of sight.
The first part, which we call the response $R$, states how the unknown dust extinction density $\rho$ imprints itself on the data.
The extinction of light on the $i$-th line of sight $L_i$ is given by the line integral
\begin{align}
	(A_G)_i = \left[R(\rho)\right]_i = \int_{L_i}\text{d}l\,\alpha\,\rho(l)\ .
\end{align}
Here $\alpha$ is the average dust cross section per unit of mass and the line of sight $L_i = L_i(\omega)$ is dependent on the true parallax $\omega$. 
As noted in subsection \ref{sec:algorithm}, the value of $\alpha$ is uncertain and we reconstruct the dust extinction density $s=\text{ln}\left(\alpha \rho/\text{pc}\right)$ instead.

The extinction is additive because the extinction data are given in the magnitudes scale, which is logarithmic. 
The true parallax $\omega$ of the star is uncertain. 
We assume the true parallax $\omega$ to be Gaussian distributed around the published parallax $\widetilde{\omega}$ with a standard deviation equal to the published parallax error $\sigma_{\omega}$:
\begin{align}
	P(\omega|\widetilde{\omega}, \sigma_\omega) = \mathscr{G}(\omega-\widetilde{\omega}, \sigma_\omega^2)
\end{align}
The parallaxes of Gaia DR2 were shown to be Gaussian distributed with incredible reliability by \cite{luri2018gaia}.
However, it was also noted by the same authors that there can be outliers.
By restricting ourselves to close-by sources for which G-band extinction values are published, we
expect to have cut out most of the outliers.

We do not reconstruct the actual positions of the stars in our reconstruction, thus we have to marginalize them out to obtain the response:
\begin{align}
	&P((A_G)_i|\widetilde{\omega}_i, \sigma_{\omega_1}, \rho) = \nonumber\\
	&=\int \text{d}\omega_i\,P((A_G)_i, \omega_i|\widetilde{\omega}_i, \sigma_{\omega_i}, \rho) \nonumber\\
	&= \int \text{d}\omega_i\,P((A_G)_i|\omega_i, \widetilde{\omega}_i, \sigma_{\omega_i}, \rho)P(\omega_i|\widetilde{\omega_i}, \sigma_{\omega_i},\rho) \nonumber\\
	&= \int \text{d}\omega_i\,P((A_G)_i|\omega_i, \rho)P(\omega_i|\widetilde{\omega}_i, \sigma_{\omega_i}) \nonumber\\
	&= \int \text{d}\omega_i\,\delta\left((A_G)_i - \int_{L_i(\omega_i)}\text{d}l_i\,\alpha\rho(l_i)\right)P(\omega_i|\widetilde{\omega}_i, \sigma_{\omega_i}) \nonumber\\
	&\,\approx\,\delta\left((A_G)_i - \int \text{d}\omega_i\,\int_{L_i(\omega_i)}\text{d}l_i\,\alpha\rho(l_i)\,P(\omega_i|\widetilde{\omega}_i, \sigma_{\omega_i})\right) \nonumber\\
	&=\delta\left((A_G)_i - \int \text{d}\omega_i\,\int_{L_i(0)}\text{d}l_i\,\alpha\mathds{1}_{[0,\frac{1}{\omega}]}(l_i)\,\rho(l_i)\,\mathscr{G}(\omega_i-\widetilde{\omega}_i, \sigma_{\omega_i}^2)\right) \nonumber\\
	&= \delta\left((A_G)_i - \int_{L_i(0)}\text{d}l_i\,\alpha\rho(l_i)\text{ sf}_{\mathscr{G}}\left(\frac{\frac{1}{l_i} - \widetilde{\omega_i}}{\sigma_{\omega_i}}\right)\right)\label{eq:los-response}\ .
\end{align}
Here 
\begin{align}
	\text{sf}_\mathscr{G}(x) = 1-\int_{-\infty}^x \text{d}t\,\frac{1}{\sqrt{2\pi}}\text{exp}\left(-\frac{1}{2} t^2\right)
\end{align}
denotes the survival function of a standard normal distribution 
and 
\begin{align}
	\mathds{1}_{[a,b]}(x) = 
	\begin{cases}
		1 &\text{ for } x\in [a,b]\\
		0 &\text{ otherwise}
	\end{cases}
\end{align}
denotes the indicator function for the closed interval $[a,b]$.
Note that we did an approximation where we replace the true extinction by the expected extinction.
As a consequence the lines of sight are smeared out by the parallax uncertainty in our approximation.
This smoothing can be regarded as a first order correction for the uncertainty of the parallax and was already used by \cite{vergely2001nai}.
A fully Bayesian analysis would treat the true parallax as unknown and infer these along the other unknowns, but this is beyond the scope of this work.

For the algorithm, the integral in Eq.\,(\ref{eq:los-response}) is discretized into a weighted sum, such that each voxel contributes to the line integral over $L_i$ exactly equal to the length of the line segment of $L_i$ within that voxel while being discounted by the probability $P(l|\widetilde{\omega}, \sigma_\omega)$ of that voxel being on the line of sight.
Applying the response $R$ thus takes $\mathscr{O}(N_\text{data}N_\text{side})$ operations, where $N_\text{data}=3\,661\,286$ is the number of data points used in the reconstruction and $N_\text{side}=256$ is the number of voxels per axis.
Due to the large number of data points, evaluating the response on a computer turns out to be numerically expensive.
The inference algorithm (see Sec.\,\ref{sec:algorithm}) is a minimization for which this response has to be evaluated many times. 
To make the dust inference feasible in a reasonable amount of time we restricted our reconstruction to a 600pc cube centered on the Sun.

The second part of the likelihood states how the published data $\widetilde{A_G}$ is distributed given the true extinctions $(A_G)_i$. 
We use the data likelihood recommended by the Gaia collaboration in \cite{andrae2018gaia}.
This likelihood assumes the data $\widetilde{A_G}$ to be distributed according to a truncated Gaussian with a global variance $N=\left(0.46\,\text{mag}\right)^2\mathds{1}$.
This leads to the likelihood
\begin{align}
	P(\widetilde{A_G}|\rho) &=
		\prod_i \frac{\mathscr{G}(\widetilde{A_G}_i-R(\rho)_i, N_{ii})}{\text{cdf}_{\mathscr{G}(R(\rho)_i, N_{ii})}(A_G^\text{max})-\text{cdf}_{\mathscr{G}(R(\rho)_i, N_{ii})}(A_G^\text{min})} \\
		\text{for } d&\in [A_G^\text{min}, A_G^\text{max}]\ .
\end{align}
Here $\text{cdf}_{\mathscr{G}(R(\rho)_i, N_{ii})}$ denotes the cumulative density function of a normal distribution with mean $R(s)_i$ and variance $N_{ii}$.
We took the boundaries of the truncated Gaussian to be $A_G^\text{min}=0$ and $A_G^\text{max}=3.609\,\text{mag}$ as recommended in \cite{andrae2018gaia}.

\subsection{Prior}
\label{sec:prior}

We assume the dust density to be a positive quantity that can vary over orders of magnitude.
The dust is assumed to be spatially correlated and statistically homogeneous and isotropic.
The statistical model is constructed to be as general as possible with these two properties in mind.

To reflect the positivity and to allow variations of the dust density by orders of magnitude we assume the dust density $\rho$ to be a-priori log-normal distributed with 
\begin{align}
	\alpha\rho &= \rho_0\,\text{exp}(s)\label{eq:exp}\ ,\\
	\text{where }	s &\curvearrowleft \mathscr{G}(s, S)\label{eq:Gauss}
\end{align}
is assumed to be Gaussian distributed with Gaussian process kernel $S$.
Here $\rho_0 = \nicefrac{1}{1000\,\text{pc}}$ is a constant introduced to give $\rho$ the correct unit and to bring it to roughly the right order of magnitude.
By using an exponentiated Gaussian process we allow the dust density to vary by orders of magnitude
while simultaneously ensuring that it is a positive quantity.
In Eq.\,(\ref{eq:Gauss}) $S$ is the prior covariance. 
If we assume no point or direction to be special a-priori, then according to the Wiener-Khinchin theorem $S$ can be fully characterized by its spatial power spectrum $S_{kk^\prime} = 2\pi\delta(k-k^\prime)P_s(k)$.
We non-parametrically infer this power spectrum $P_s(k)$ as well. 
There are two main motivations to reconstruct the power spectrum.
From a physical perspective the power spectrum provides valuable insights into the underlying processes.
From a signal processing point of view, many linear filters can be identified with a Bayesian filter that assumes a certain prior power spectrum.
The optimal linear filter is obtained when the power spectrum used for the filter is exactly equal to the power spectrum of the unknown quantity \citep{ensslin2011reconstruction}.
However, the power spectrum of the unknown quantity is usually also unknown, thus one has to reconstruct it as well.
While this argumentation holds for linear filters, certainly many aspects of it carry over to nonlinear filters such as the reconstruction performed in this paper.

\begin{figure}[t]
	\begin{centering}
	\begin{tikzpicture}
		\node[draw, thick] (L1) at (7.5,0) {$\widetilde{A_G}|A_G$};
		\node[draw, thick] (L2) at (5,0) {$A_G|\rho$};
		\node[draw, thick] (s) at (2.5,0) {$\text{log}(\rho)|S_{kk}$};
		\node[draw, thick] (tau) at (0,0) {$S_{kk}$};
		\draw [<-] (L1) edge (L2) (L2) edge (s) (s) edge (tau);
	\end{tikzpicture}
	\end{centering}
	\caption{Graphical representation of the data model for our reconstruction.  \label{fig:model1}
	The logarithmic dust density $\text{log}(\rho)$ is a Gaussian process with a smooth Gaussian process power spectrum $S_{kk^\prime} = 2\pi\delta(k-k^\prime)P_s(k)$. 
	The true extinctions $A_G$ are directly dependent on the dust density $\rho$ on each line of sight.
	The measured extinctions $\widetilde{A_G}$ are assumed to be distributed around the true extinctions $A_G$ following a truncated Gaussian distribution as described in section \ref{sec:likelihood}.
	}
\end{figure}
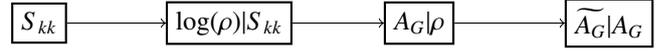

Fig.\,\ref{fig:model1} depicts the hierarchical Bayesian model for the extinction data $\widetilde{A_G}$ resulting from the logarithmic dust density $\text{ln}(\rho)$, which itself is shaped by the power spectrum $P_s(k)$.

Our statistical model for the power spectrum $P_s(k)$ is a falling power law with Gaussian distributed slope and offset modified by differentiable non-parametric deviations. It is up to minor details\footnote{The amplitude model given by Eq.\,\ref{eq:amplitude-model} is not exactly equivalent to an integrated Wiener process, but shown by \citep{arras2019unified} to be equivalent to it in a certain limit while still allowing a numerically stable transformation of the prior to a white Gaussian.} an integrated Wiener process \citep{doob1953stochastic} on log-log-scale.

This is realized by the following formula:
\begin{align}
	&\sqrt{P_s(k)} = \nonumber\\
	&\text{exp}\left((\phi_m\sigma_m+\bar{m})\text{log}(k) + \phi_y\sigma_y + \bar{y} + \mathbb{F}^{-1}_{\text{log}(k)t}\left(\frac{a}{1+\nicefrac{t^2}{t_0^2}}\tau(t)\right)\right)\label{eq:amplitude-model}
\end{align}
Here $\phi_m$, $\phi_y$ and $\tau(t)$ are the parameters to be reconstructed,
$\sigma_m=1, \bar{m}=-4, \sigma_y=2., \bar{y} = -16, \sigma_y = 3., a=11, t_0=0.2$ are fixed hyperparameters, $\mathbb{F}^{-1}_{\text{log}(k)t}$ denotes the inverse Fourier transform on log-scale, and $V=(600\,\text{pc})^3$ is the total volume of the reconstruction.
These hyperparameters settings were determined by trial and error such that data measured from a prior sample has roughly the same order of magnitude as the actual data and such that the dust density varies by more than one order of magnitude in prior samples.

\begin{figure}[ht]
	\includegraphics[width=0.5\textwidth]{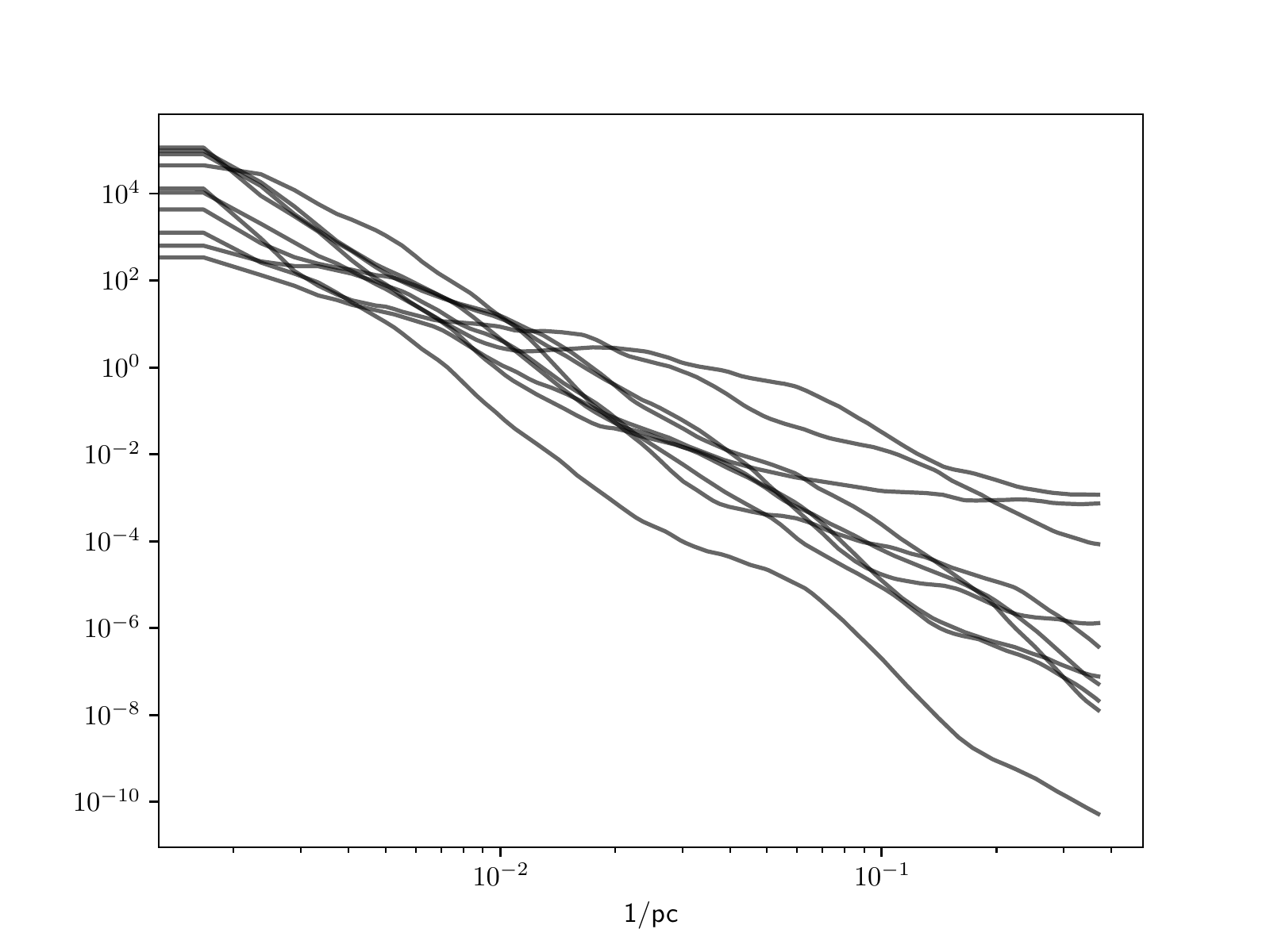}
	\caption{
		Several prior samples of the logarithmic spatial correlation power spectrum in units of $\text{pc}^3$.
		\label{fig:prior-power-sample}
		}
\end{figure}

\begin{figure}[ht]
	\includegraphics[width=0.5\textwidth]{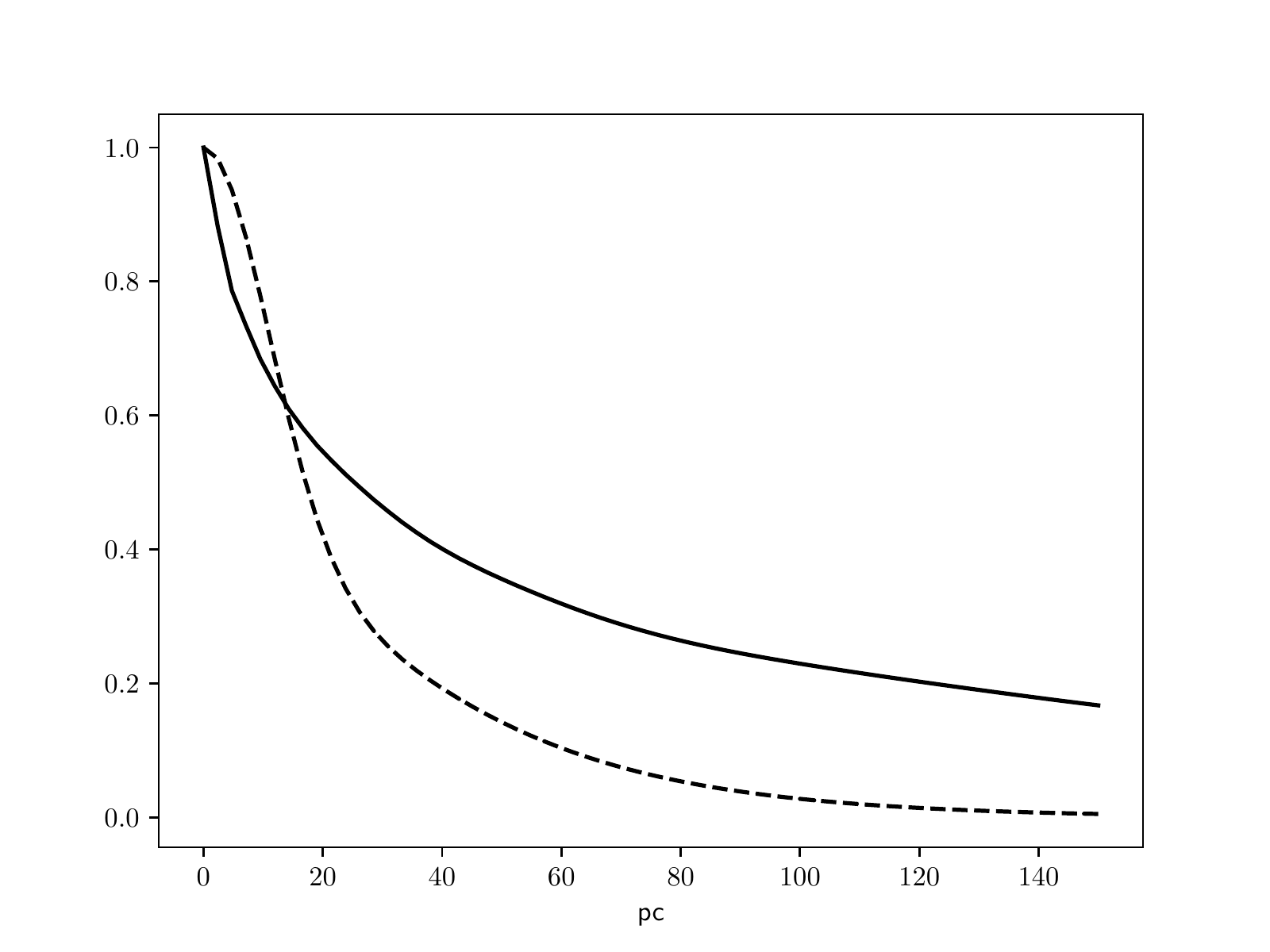}
	\caption{The log-normal process normalized 2-point correlation reconstructed by our method (solid line) and imposed in the reconstruction by \cite{lallement20183d} (dashed line).
	\label{fig:kernels}
	One can see that the dust is assumed to be strongly correlated at a distance scale of up to about 30pc.
	This plot shows normalized one dimensional cuts through the three dimensional Fourier transform of the log-normal spatial correlation power spectrum shown in Fig.\,\ref{fig:prior-spectra}.
	}
\end{figure}

In our reconstruction the parameter $\tau$ for the smooth deviations of the log-log power spectrum was discretized using 128 pixels.
The mathematical motivation to take Eq.\, (\ref{eq:amplitude-model}) as a generative prior for power spectra is discussed in \cite{arras2019unified}.
As a rule of thumb, $k$-modes for which the data constrains the power spectrum very well will be recovered in great detail due to the non-parametric nature of the model.
For $k$-modes on which the data provide little information, the power spectrum will be complemented by the prior which forces it into a falling power law whenever the data is not informative.
If the actual physical process deviates strongly from a falling power law for the unobserved $k$-modes, the prior might artificially suppress or amplify the posterior uncertainty of the result, possibly biasing the uncertainty quantification.
Fig.\,\ref{fig:prior-power-sample} shows a few examples of prior samples of power spectra using our choice of hyperparameters.
While the individual samples might not look too different qualitatively, it should be noted on the one hand that any kind of power spectrum is representable with our model given enough data and on the other hand that the figure depicts the power spectrum of the log-density on log-log scale. A small deviation in this figure can have a huge impact on the actual statistics.
Reconstructing the power spectrum is equivalent to reconstructing the correlation kernel.
We show our reconstructed normalized kernel as well as the one assumed by \cite{lallement20183d} in Fig.\,\ref{fig:kernels}.
Certain biases can appear when using a fixed kernel,
for example introducing a characteristic length scale of the order of the FWHM of the kernel.

Putting together likelihood and prior, the overall joint information Hamiltonian for our parameters $\xi$, $\tau$, and $\phi$ is
\begin{align}
	P(d, \xi, \tau, \phi) &= \text{TG}_{A_G^\text{min},\ A_G^\text{min},\ 0.46,\ d}(R(\alpha\rho))\nonumber\\
	&\ \mathscr{G}((\xi,\phi,\tau)^T,\mathds{1})\label{eq:full-hamiltonian}\\
	\text{where } \left[R(\alpha\rho)\right]_i &= \int_{L_i}\text{d}l\,\alpha\,\rho(l)\\
	\alpha\rho &= \frac{1}{1000}\text{exp}\left(\mathbb{F}^{-1}_{xk}\sqrt{P_s(k)V}(\phi, \tau)\,\xi_k\right)
\end{align}
\begin{align}
	\sqrt{P_s(k)}(\phi, \tau) &= \nonumber\\
	&\hspace{-2.7cm}\text{exp}\left((\phi_m\sigma_m+\bar{m})\text{log}(k) + \phi_y\sigma_y + \bar{y} + \mathbb{F}^{-1}_{\text{log}(k)t}\left(\frac{a}{1+\nicefrac{t^2}{t_0^2}}\tau_t\right)\right)\\
	\text{TG}_{x_\text{min},\ x_\text{max},\ \sigma,\ \bar{x}}(x) &= 
		\prod_i \frac{\mathscr{G}(\bar{x}_i-x_i,\sigma^2)}{\text{cdf}_{\mathscr{G}(x_i, \sigma^2)}(x_\text{max})-\text{cdf}_{\mathscr{G}(x_i, \sigma^2)}(x_\text{min})} \\
		\text{for } x&\in [x_\text{min}, x_\text{max}]\nonumber, 
\end{align}
is a truncated Gaussian, and $V=(600\,\text{pc})^3$.
The application, calculation of the gradient, and the application of the Fisher metric of Eq.\,(\ref{eq:full-hamiltonian}) scales almost linearly with the number of voxels $N_\text{side}^3$, more specifically it takes $\mathscr{O}(N_\text{data}N_\text{side} + N_\text{side}^3\text{log}{N_\text{side}})$ operations to evaluate Eq.\,\ref{eq:full-hamiltonian}, where $N_\text{data}=3\,661\,286$ is the number of data points used in the reconstruction and $N_\text{side}=256$ is the number of voxels per axis.

\section{Simulated Data Test}
\label{sec:simulated-data-test}

\subsection{Data generation}

In this section, a test on simulated data is presented. 
This test enables comparing the results of the reconstruction to a known ground truth.
As ground truth dust density public data from the SILCC collaboration \citep{walch2015silcc} was used, more specifically from the magneto-hydrodynamic simulation of the interstellar medium \emph{B6-1}pc at 50 Myr published by \cite{girichidis2018silcc}.
This simulation result spans a cube with size $(512\,\text{pc})^3$. 
We computed our synthetic ground truth differential absorption $\rho_\text{mock}$ from the gas density of the simulation $\rho_\text{sim}$ via
\begin{align}
	\rho_\text{mock}\left(x-(150,150,0)^T\right) = \sqrt[3]{\rho_\text{sim}\left(\frac{512 x}{600}\right)10^{17}\frac{\text{cm}^3}{\text{g}}} \frac{1}{\text{pc}}\label{eq:mock-sky} \ .
\end{align}
Thus we stretch the $512\,\text{pc}$ simulated cube to the $600\,\text{pc}$ of our reconstruction, scale it with a constant factor, and shift it by $150\,\text{pc}$.
The shift is performed in order to have an underdense region at the center.
We also take the third root of the gas density in Eq.(\ref{eq:mock-sky}).
There are two reasons for this.

A practical motivation for taking the third root is that it reduces the dynamic range. 
If one does not do this, the sky will be dominated by one very small, but very strongly absorbing blob.

A more physical motivation is that very dense regions lead to star formation. 
These forming stars again reduce the density by blowing the material out of these regions.
This feedback mechanism was not included into the simulation by \cite{girichidis2018silcc} but was shown to have a strong impact on the gas density in a followup simulation by \cite{haid2018silcc}.
The third root can be seen as a very crude way of reducing the density in these overdense regions.

To obtain the synthetic data from the ground truth differential extinction cube $\rho_\text{mock}$, the following operations were performed:
\begin{enumerate}
	\item Sampling ground truth parallaxes $\omega_i\curvearrowleft \mathscr{G}(\omega_i-\widetilde{\omega}_i, \sigma_i^2)$ according to the parallax likelihood published by the Gaia collaboration.
	\item Integrating the dust density from the center of the cube to the location of the sampled star location $\nicefrac{1}{\omega}$ using the full resolution of $512^3$ voxels\footnote{Note that the simulation of which the data is used was performed on an adaptive grid. The full resolution of $512^3$ is only realized in the high density regions.}.
	\item Sampling an observed extinction magnitude according to the truncated Gaussian likelihood described in section \ref{sec:likelihood}.
\end{enumerate}

\subsection{Results}
\label{sec:mock-results}

\begin{figure*}[p]
	\centering
	\begin{subfigure}[t]{.46\textwidth}
		\includegraphics[trim={1.5cm .5cm 2cm 1cm}, clip, width=.95\textwidth]{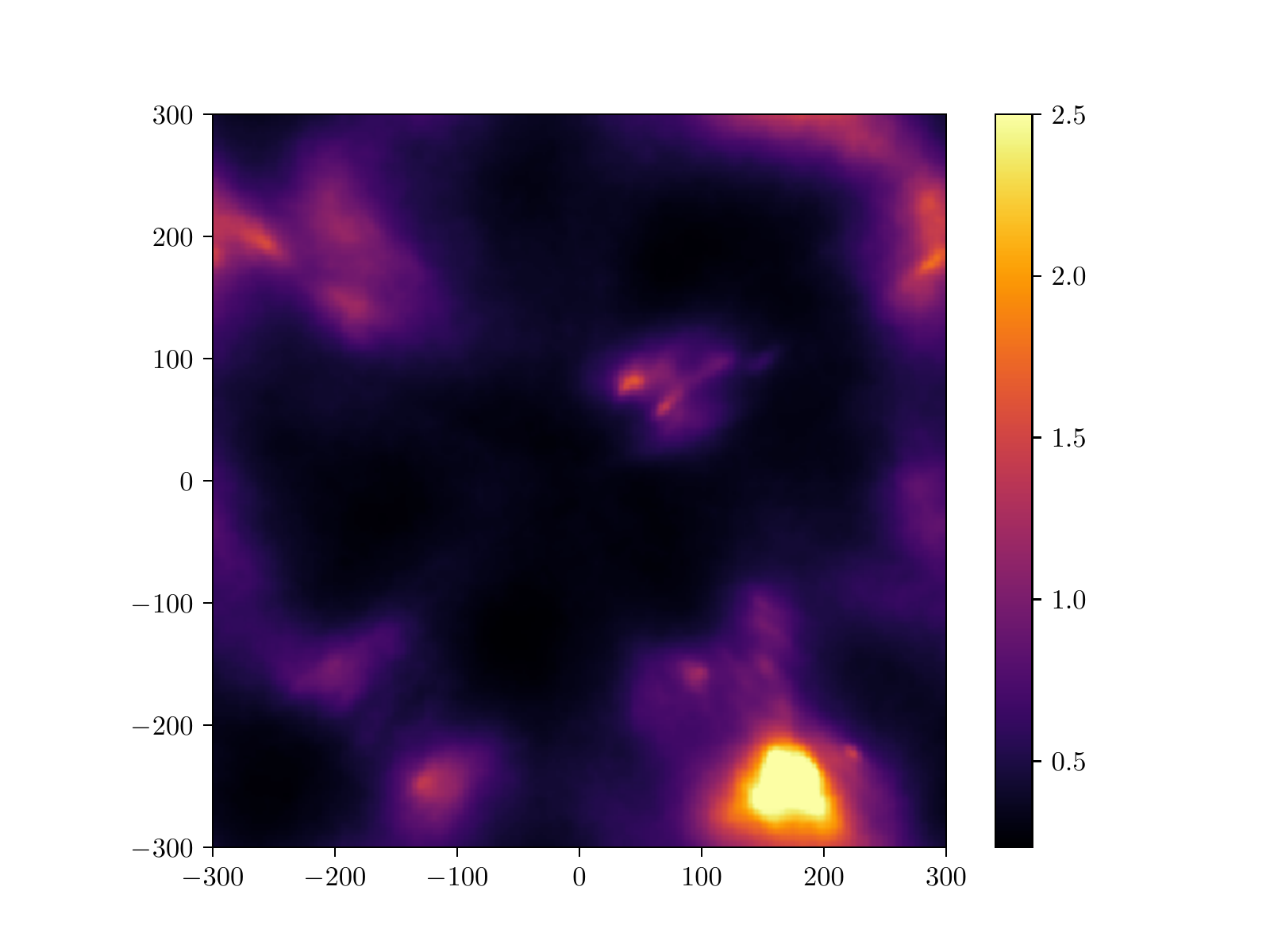}
	\caption{
	\label{fig:test-plane-projection}
	}
	\end{subfigure}
	~
	\begin{subfigure}[t]{.46\textwidth}
		\includegraphics[trim={1.5cm .5cm 2cm 1cm}, clip, width=.95\textwidth]{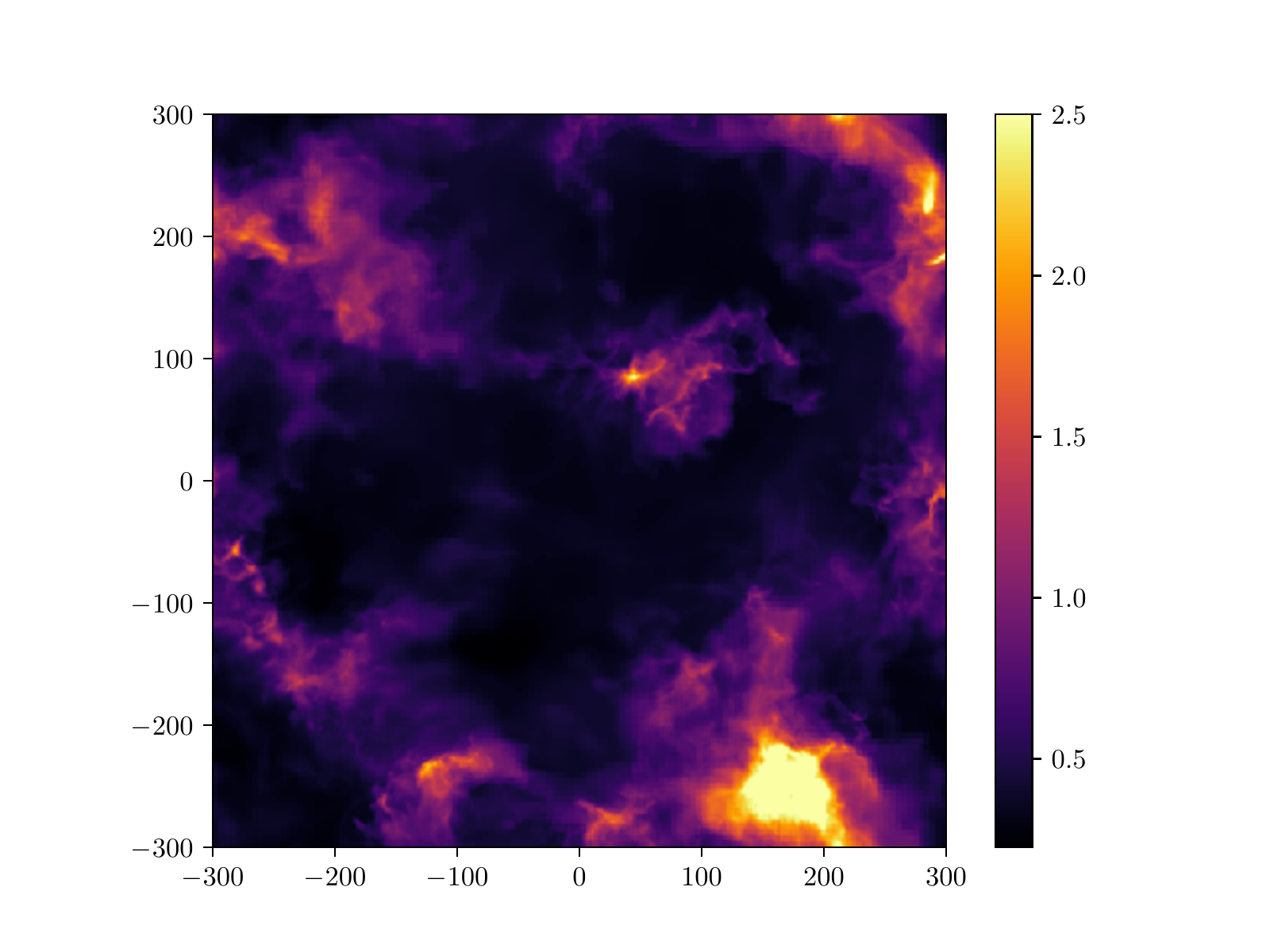}
		\caption{
		\label{fig:mock-plane-projection}}
	\end{subfigure}
	\\
	\begin{subfigure}[t]{.46\textwidth}
		\includegraphics[trim={1.5cm .5cm 2cm 1cm}, clip, width=.95\textwidth]{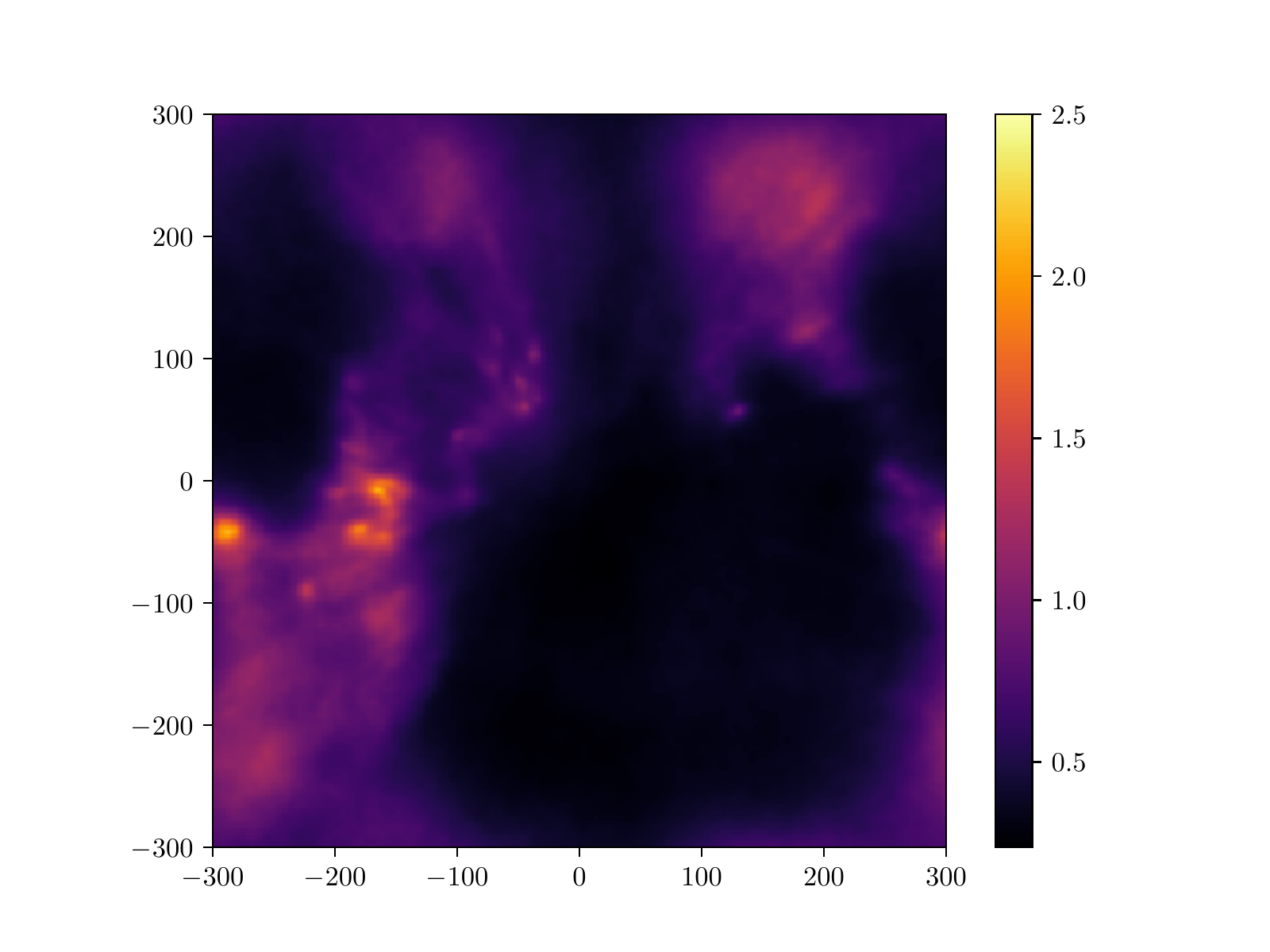}
	\caption{
	\label{fig:test-x-projection}
	}
	\end{subfigure}
	~
	\begin{subfigure}[t]{.46\textwidth}
	\includegraphics[trim= {1.5cm .5cm 2cm 1cm} , clip, width=.95\textwidth]{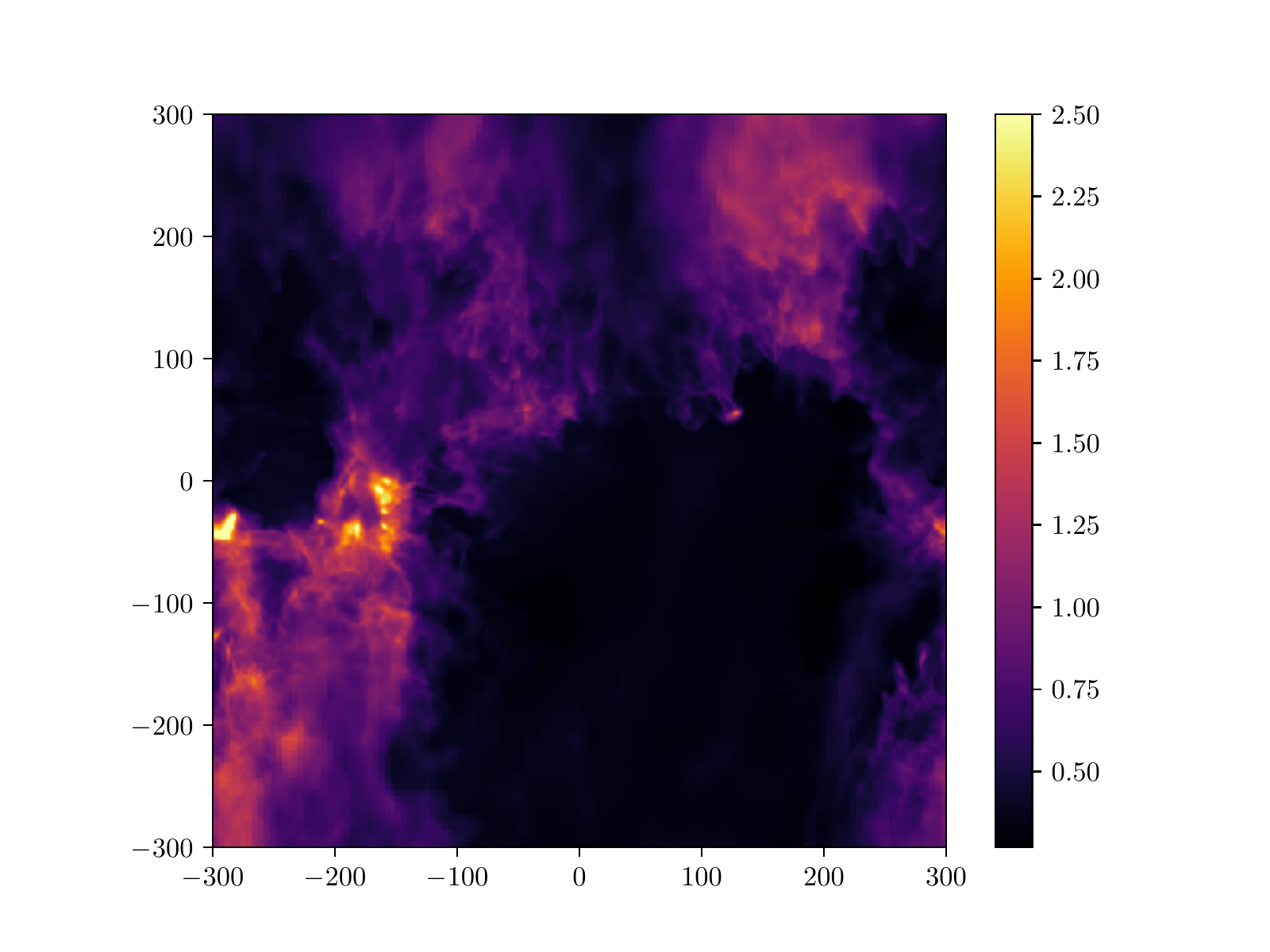}
	\caption{
	\label{fig:mock-x-projection}
		}
	\end{subfigure}
	\\
	\begin{subfigure}[t]{.46\textwidth}
		\includegraphics[trim={1.5cm .5cm 2cm 1cm}, clip, width=.95\textwidth]{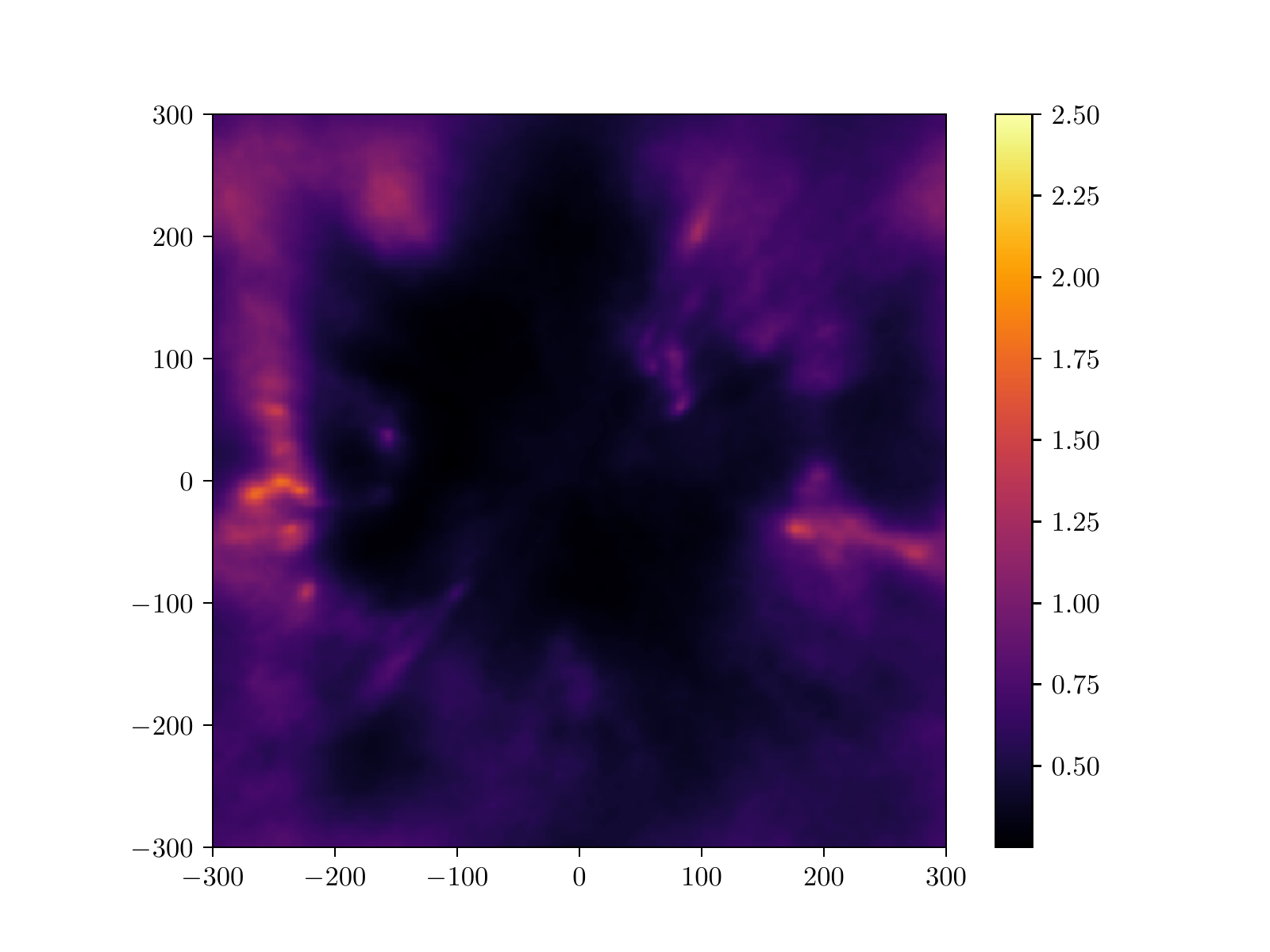}
	\caption{
	\label{fig:test-y-projection}
	}
	\end{subfigure}
	~
	\begin{subfigure}[t]{.46\textwidth}
	\includegraphics[trim= {1.5cm .5cm 2cm 1cm} , clip, width=.95\textwidth]{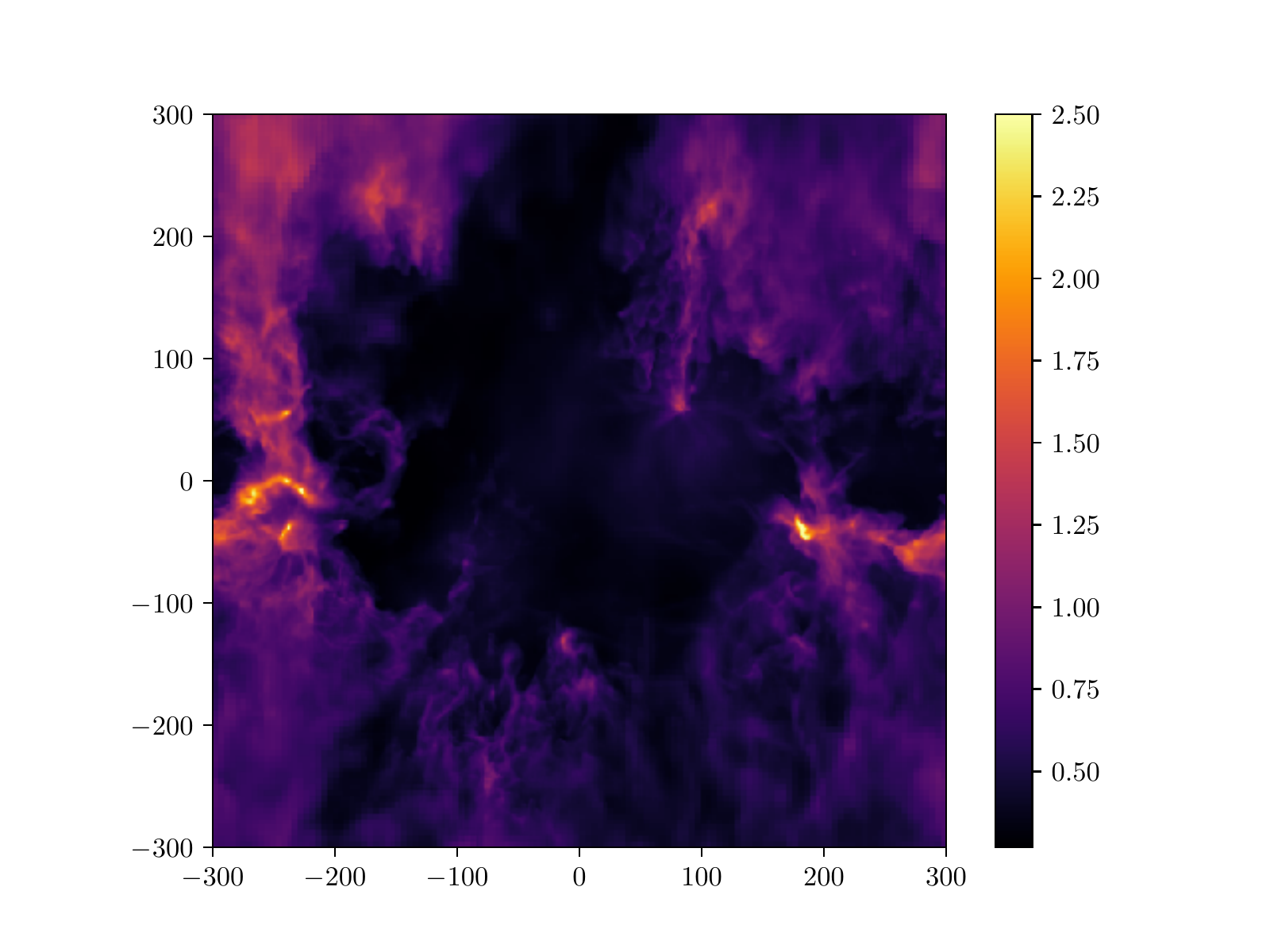}
	\caption{
	\label{fig:mock-y-projection}
	}
	\end{subfigure}
	\caption{
		\label{fig:mock-and-reconstruction}
		Results of our test using simulated data.
	The rows show integrated dust extinction for sightlines parallel to the $z$- $x$- and $y$- axis respectively.
	The first column corresponds to the test reconstruction, the second column is the ground truth synthetic extinction.
	}
\end{figure*}

We were able to recover a slightly smeared out version of the original synthetic extinction cube.
In Fig.\,\ref{fig:mock-and-reconstruction} integrations with respect to the $x$-, $y$-, and $z$-axis of the synthetic extinction cube and the reconstructed extinction cube are shown.
This visually confirms the reliability of the reconstruction.

\begin{figure}[ht]
	\includegraphics[width=0.5\textwidth]{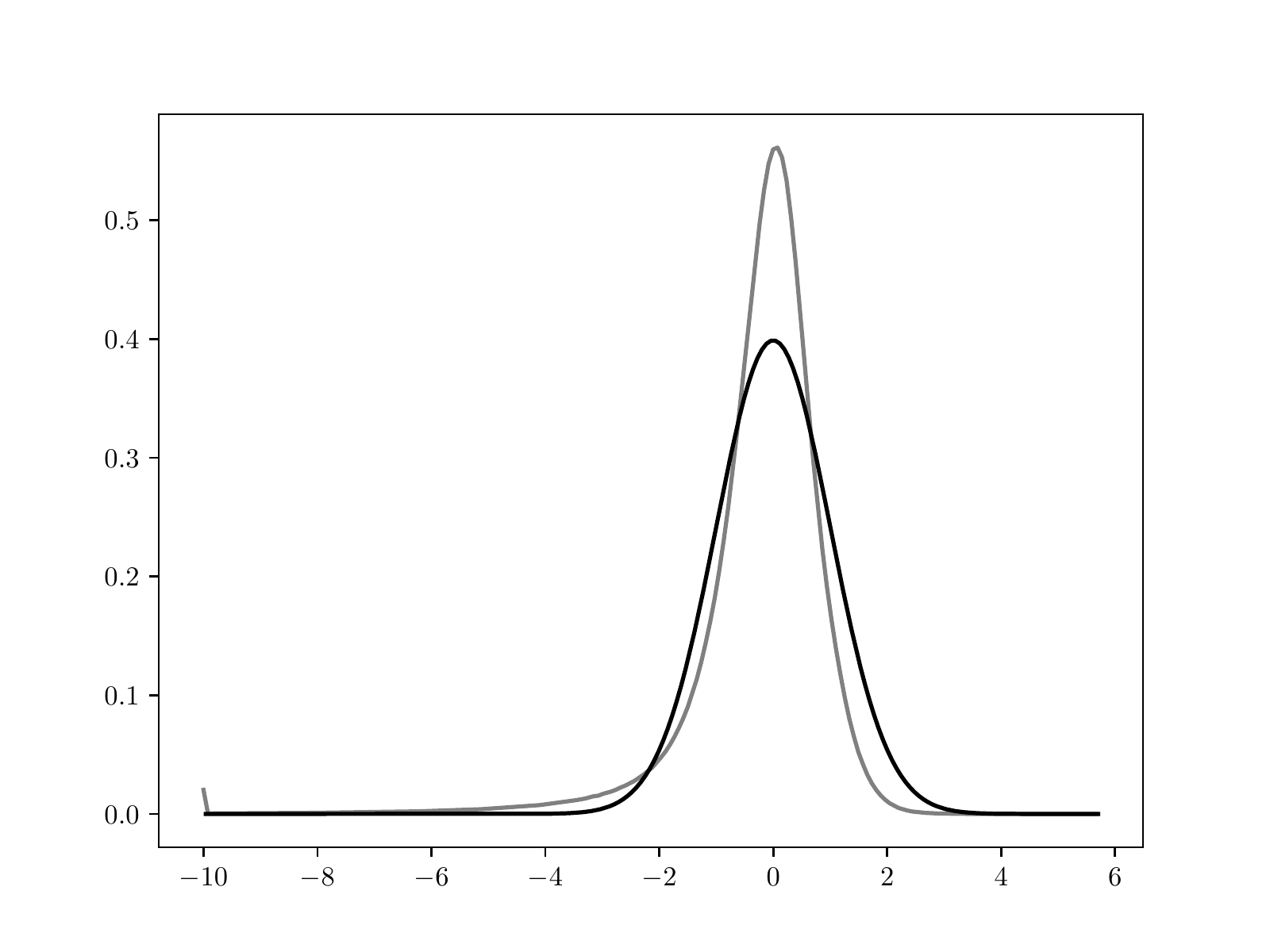}
	\caption{
		The gray curve shows a normalized histogram of the deviation of the reconstruction from the true solution, in sigmas.
		The black curve is the probability density function of a standard normal distribution, which is plotted as a reference.
		Note that the values were clipped to the range from $-10$ to $10$,
		i.e. the bump in the gray curve at $-10$ corresponds to outliers that can be up to $250$ sigmas. These outliers correspond to about $0.15\%$ of all voxels.
	\label{fig:mock-histogram}
	}
\end{figure}

For a more quantitative analysis, we compared the reconstructed differential extinctions with the ground truth voxel-wise. 
More specifically, we computed the uncertainty weighted residual $r$
\begin{align}
	r &= \frac{\rho_\text{reconstruction}-\rho_\text{ground truth}}{\sigma_\rho},
\end{align}
where $\rho_\text{reconstruction}$ and $\sigma_\rho$ are the posterior mean and standard deviation computed from the approximate posterior samples.
The ground truth differential extinction was recovered well within the recovered approximate posterior uncertainty, apart from outliers which make up about $0.15\%$ of the voxels. See Fig.\,\ref{fig:mock-histogram} for a histogram of the uncertainty weighted residual.

Overall, the reconstruction seems very reliable on a qualitative and quantitatively level within the uncertainty for most of the voxels.

\section{Results from Gaia Data}
\label{sec:results}

\begin{figure}[ht]
	\includegraphics[width=0.5\textwidth]{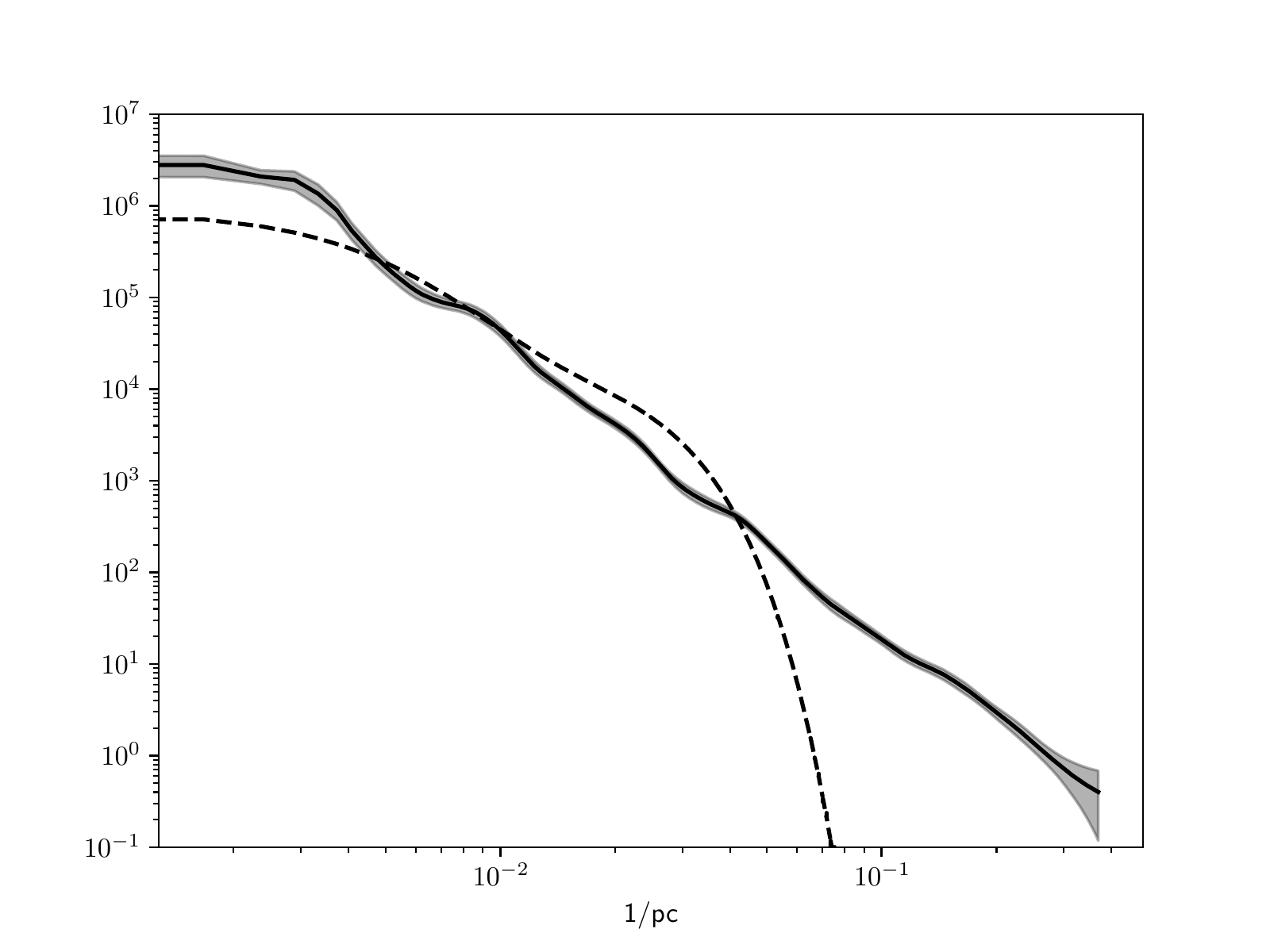}
	\caption{The log-normal process spatial correlation power spectrum inferred in our reconstruction (solid line) as well as the imposed power spectrum of \cite{lallement20183d} (dashed line).
	The shaded area around the solid line indicates $1\sigma$ error bounds.
	The unit of the $y$-axis is $\text{pc}^3$.
	The functions can be interpreted as the a-priori expected value
	of $\left|\mathds{F}\text{ln}(\rho)\right|^2/V$, where $V$ is the volume the density $\rho$ is defined on and $\mathds{F}$ is the Fourier transform.
	The region between $0.0008$/pc and $0.426$/pc is almost power-law like with a slope of $3.1$, the spectral index of the power law.
	\label{fig:prior-spectra}
	}
\end{figure}

We reconstruct the dust density in a 600pc cube using $256^3$ voxels, resulting in a resolution of $(2.34\,\text{pc})^3$ per voxel. 
For our reconstructed volume we also infer the spatial correlation power spectrum  of the log-density, see Fig.\,\ref{fig:prior-spectra}. 

\begin{figure*}[hp]
	\centering
	\begin{subfigure}[t]{.46\textwidth}
		\includegraphics[trim={1.5cm .5cm 2cm 1cm}, clip, width=.95\textwidth]{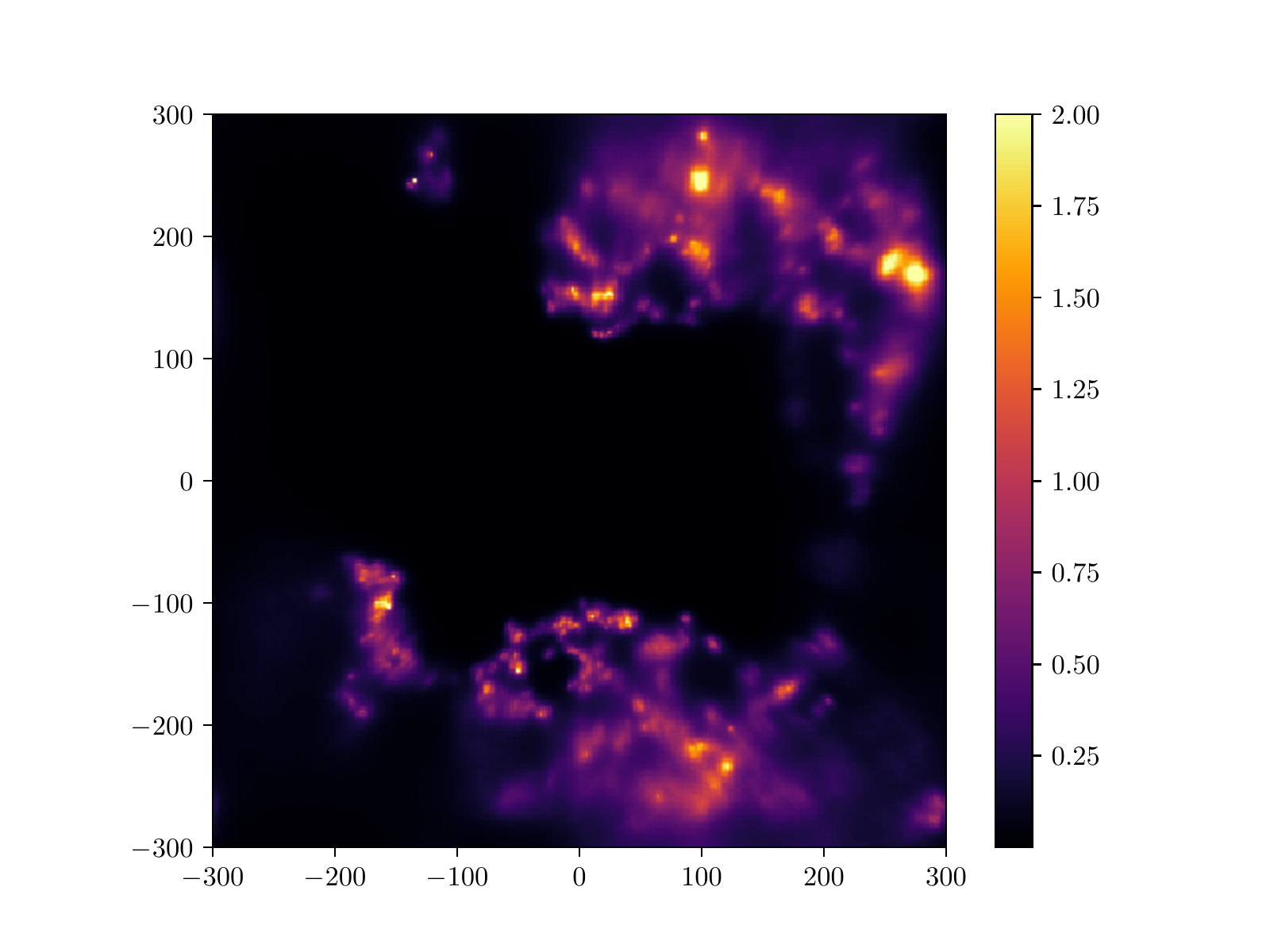}
	\caption{
	\label{fig:plane-projection}
	}
	\end{subfigure}
	~
	\begin{subfigure}[t]{.46\textwidth}
		\includegraphics[trim= {1.1cm 1cm .6cm 3cm} , clip, width=.95\textwidth]{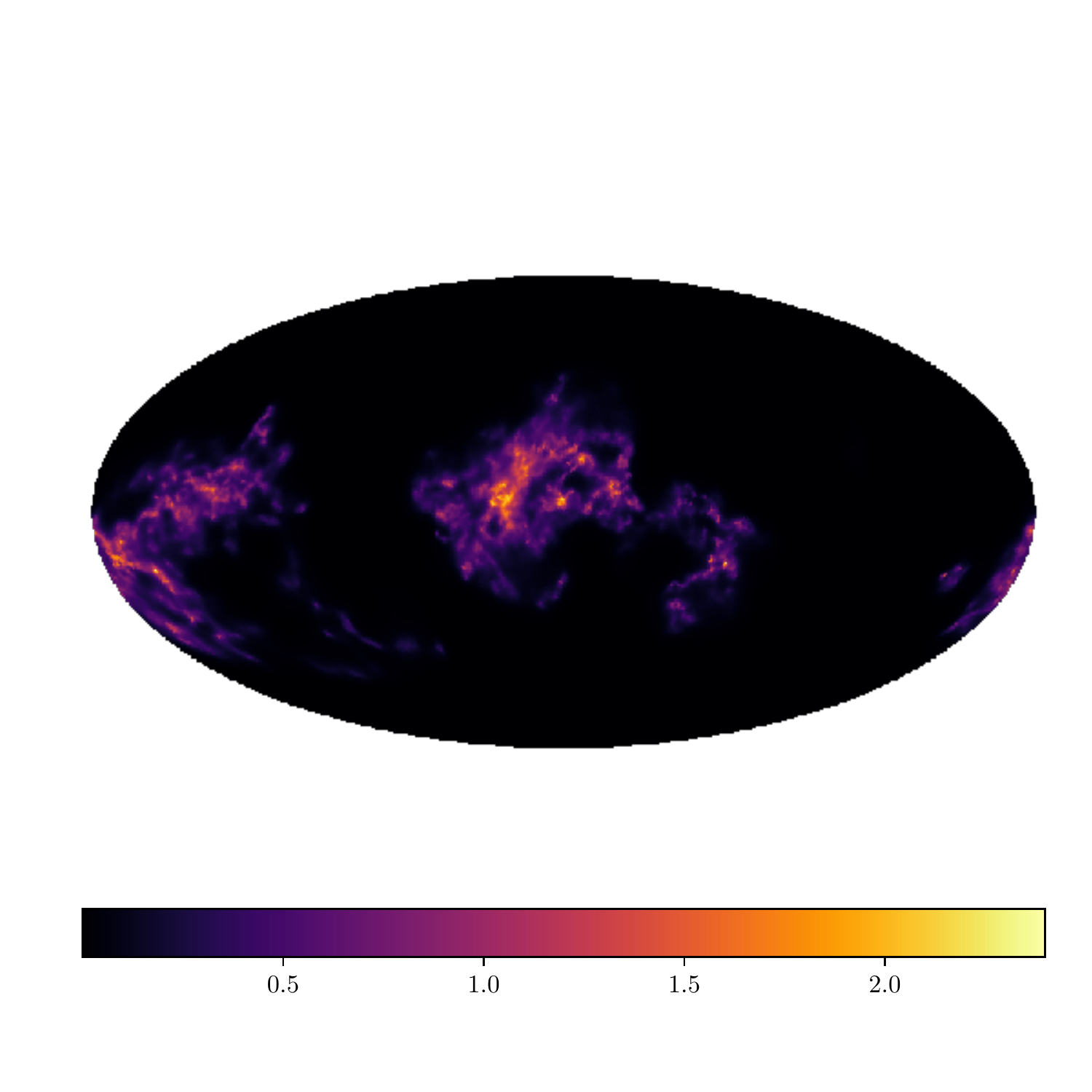}
		\caption{
		\label{fig:hp-projection}}
	\end{subfigure}
	\\
	\begin{subfigure}[t]{.46\textwidth}
		\includegraphics[trim={1.5cm .5cm 2cm 1cm}, clip, width=.95\textwidth]{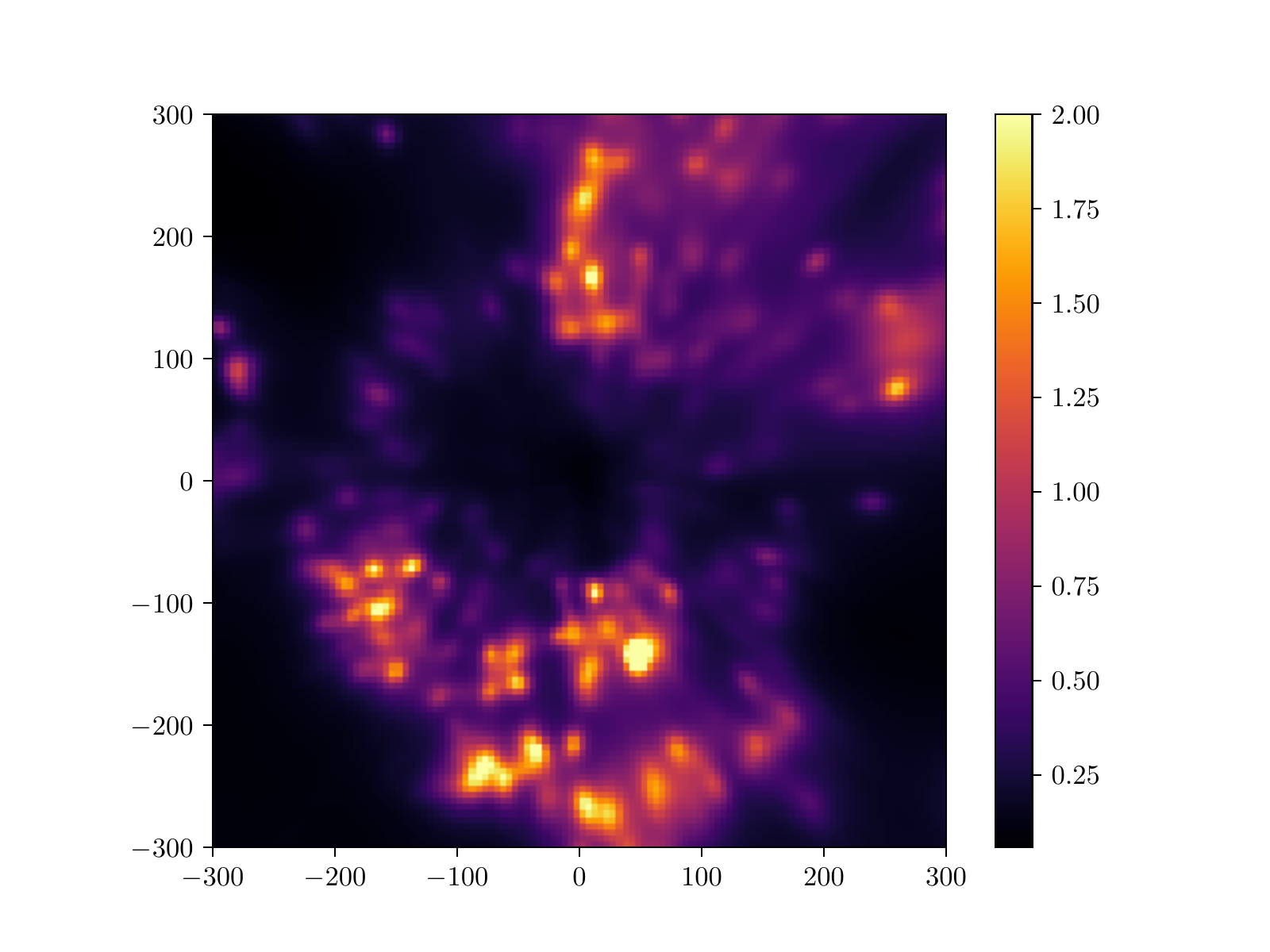}
	\caption{
	\label{fig:plane-projection-lallement}
	}
	\end{subfigure}
	~
	\begin{subfigure}[t]{.46\textwidth}
	\includegraphics[trim= {1.1cm 1cm 0.6cm 3cm} , clip, width=.95\textwidth]{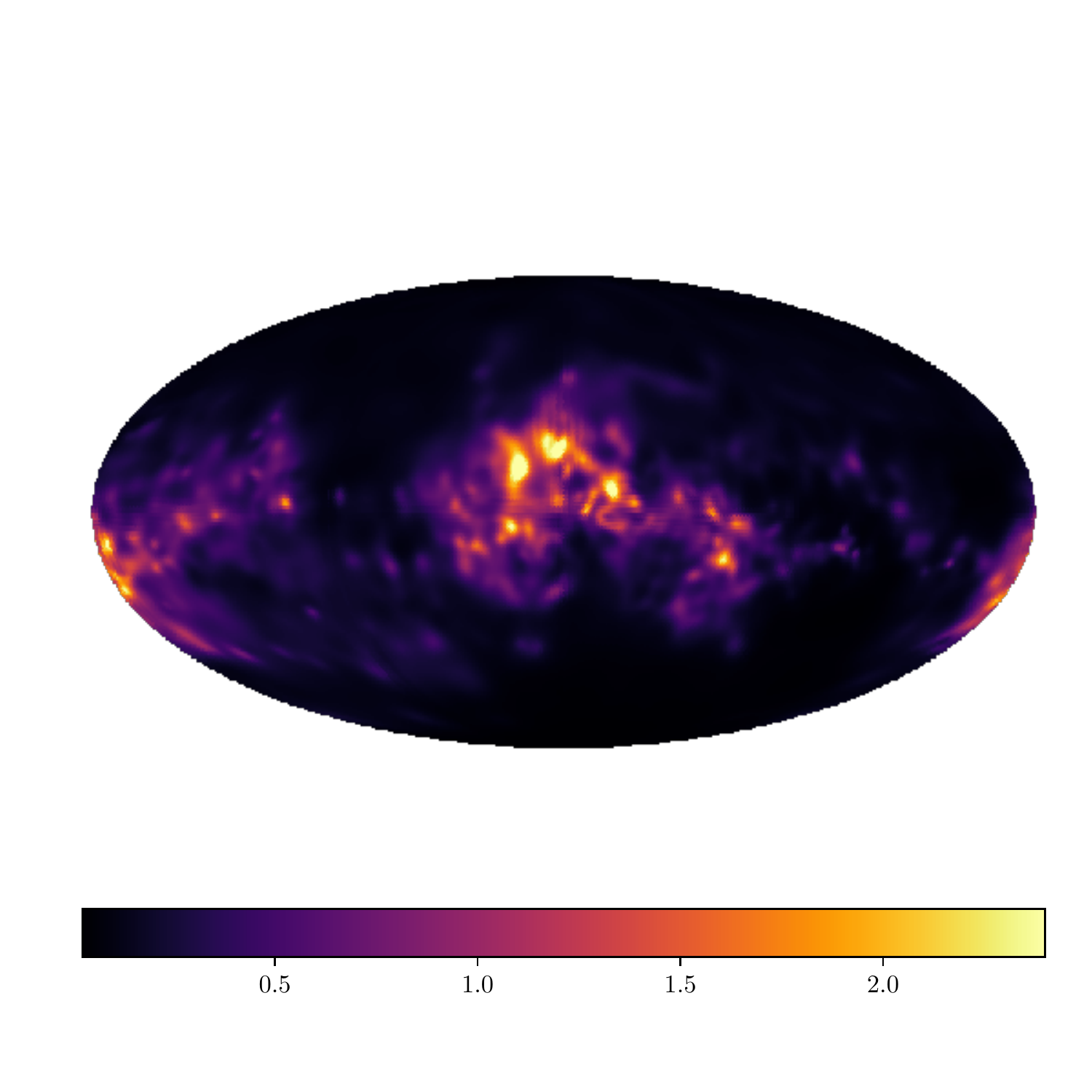}
	\caption{
	\label{fig:hp-projection-lallement}
		}
	\end{subfigure}
	\\
	\begin{subfigure}[t]{.46\textwidth}
		\includegraphics[trim={1.5cm .5cm 2cm 1cm}, clip, width=.95\textwidth]{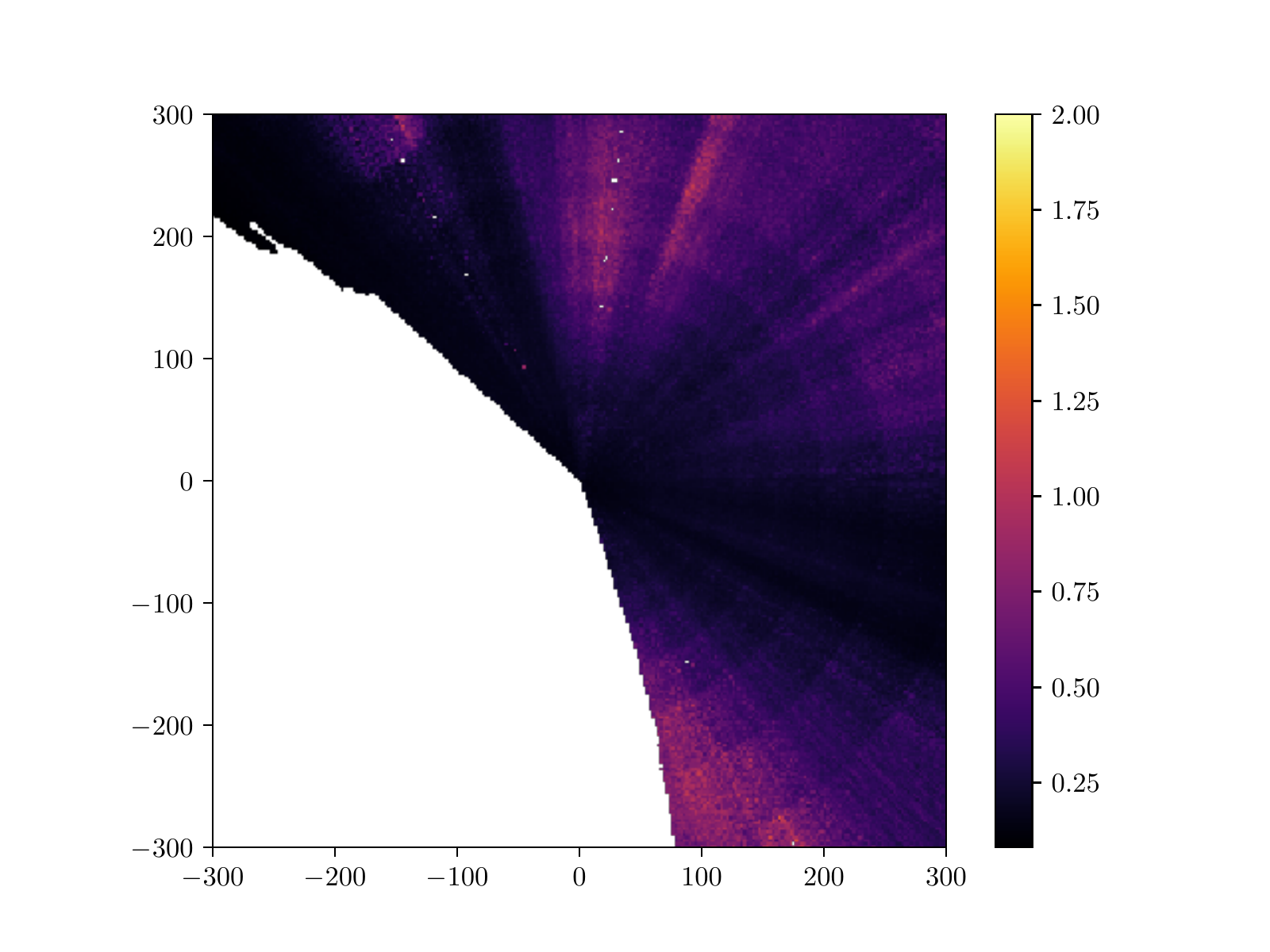}
	\caption{
	\label{fig:plane-projection-finkbeiner}
	}
	\end{subfigure}
	~
	\begin{subfigure}[t]{.46\textwidth}
	\includegraphics[trim= {1.1cm 1cm 0.6cm 3cm} , clip, width=.95\textwidth]{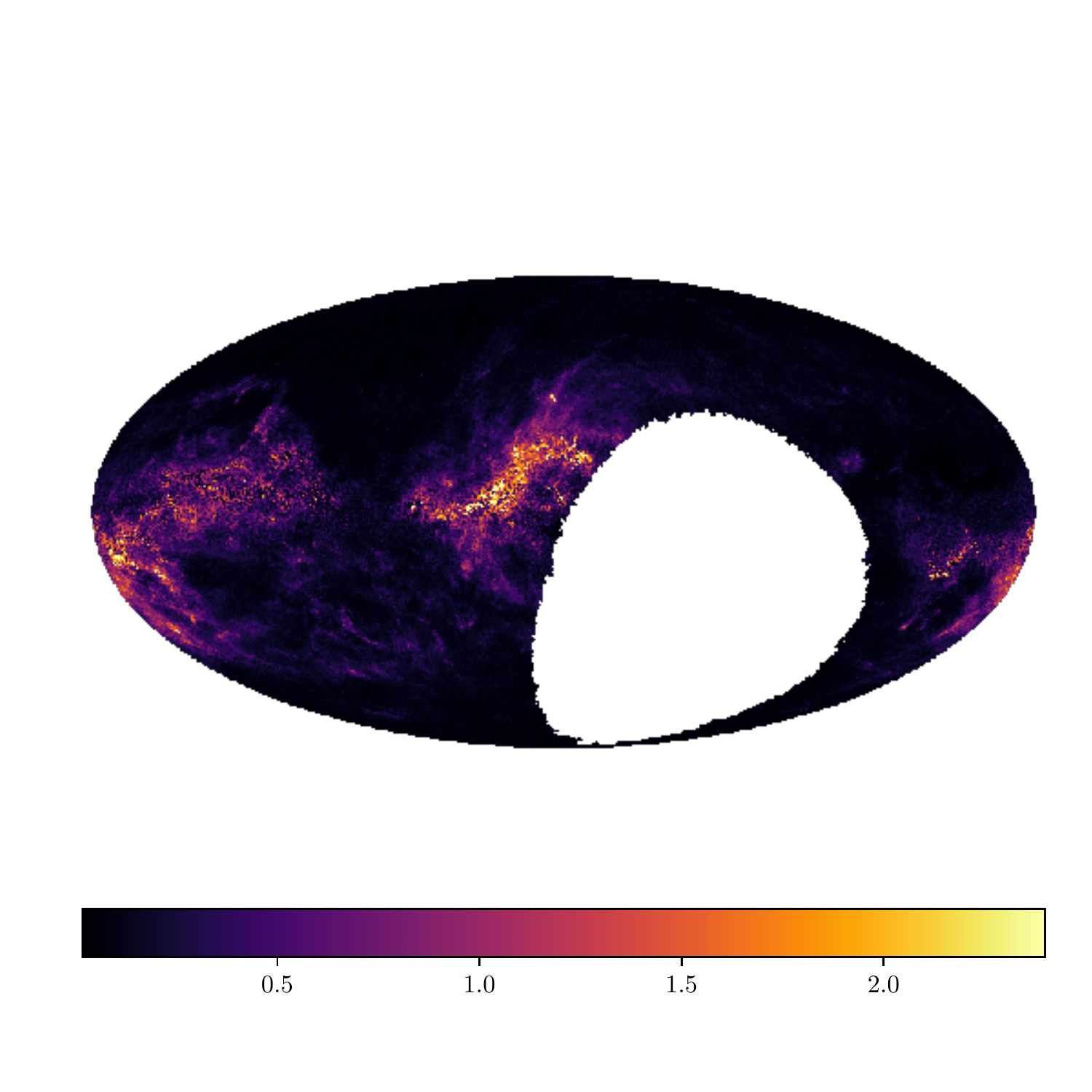}
	\caption{
	\label{fig:hp-projection-finkbeiner}
	}
	\end{subfigure}
	\caption{
		\label{fig:comparison-plot}
	The left column shows integrated dust extinction from $-300\,\text{pc}$ to $300\,\text{pc}$ for sightlines perpendicular to the galactic plane. The image covers a 600pc cube centered around the Sun.
	The units are $e$-folds of extinction.
	The Sun is at coordinate $(0,0)$, the galactic center is located towards the bottom of the plot, and the galactic West is on the left.
		The right column shows all-sky integrated dust extinction maps of the same region, but for sightlines towards the location of the Sun.
	The first row is the result of the reconstruction discussed in this paper, the second row is the reconstruction performed by \cite{lallement20183d},
	the last row shows the reconstruction by \cite{green2018galactic}.
	}
\end{figure*}

\begin{figure*}[hp]
	\centering
	\begin{subfigure}[t]{.45\textwidth}
		\includegraphics[trim={1.5cm .5cm 2cm 1cm}, clip, width=\textwidth]{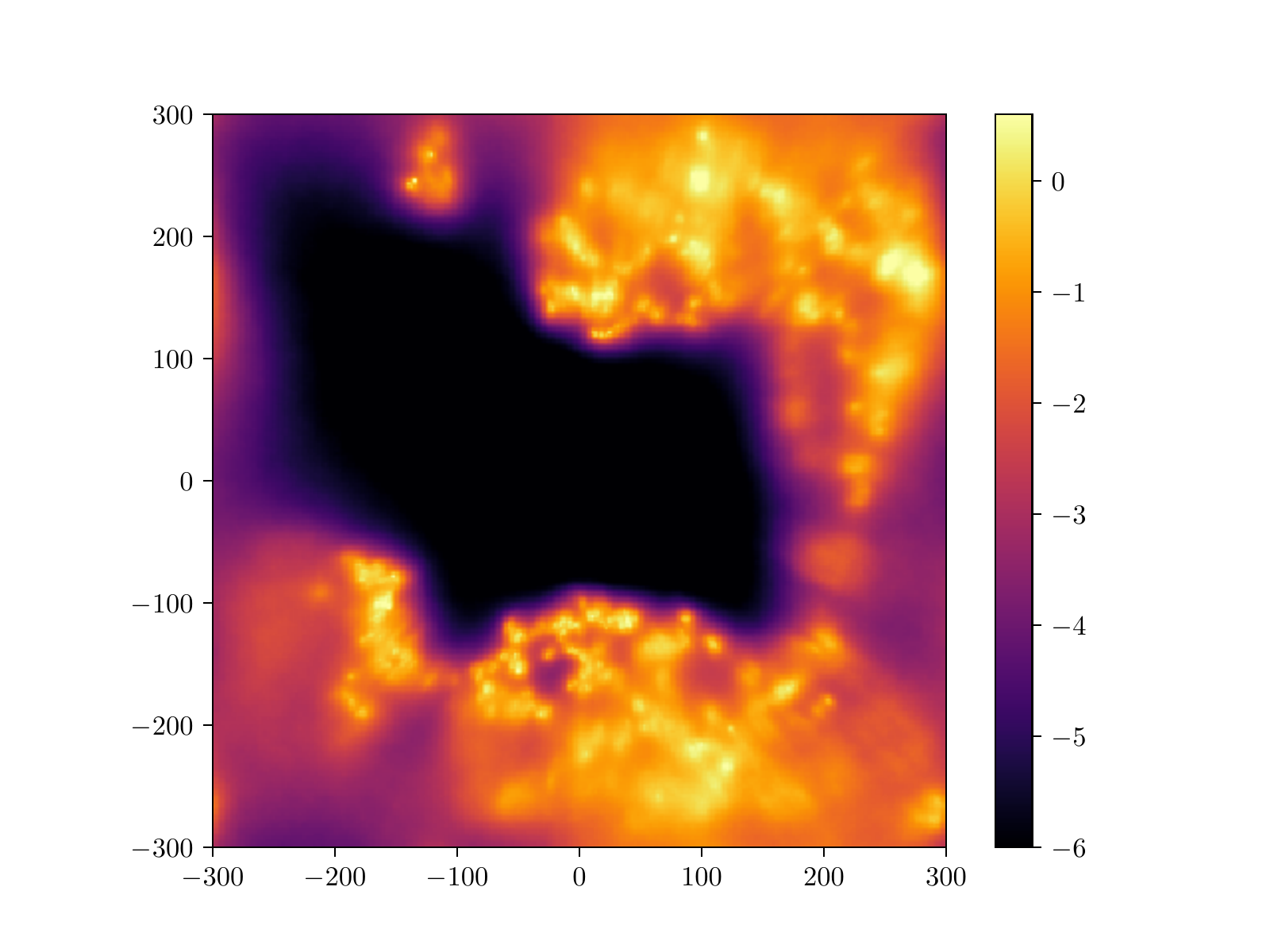}
	\caption{
	\label{fig:log-plane-projection}
	}
	\end{subfigure}
	~
	\begin{subfigure}[t]{.45\textwidth}
		\includegraphics[trim= {1.1cm 1cm 0.6cm 3cm} , clip, width=\textwidth]{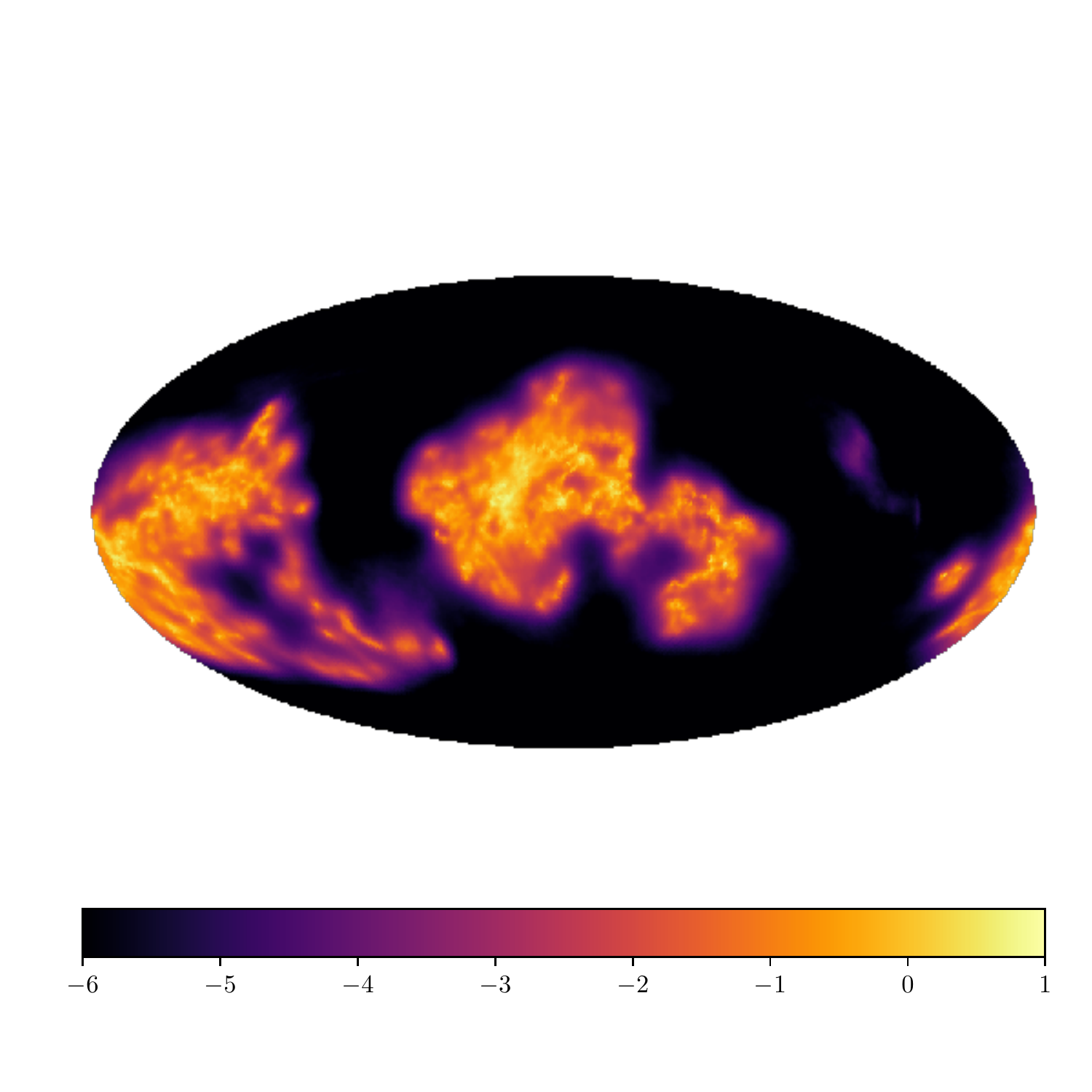}
		\caption{
		\label{fig:log-hp-projection}}
	\end{subfigure}
	\\
	\begin{subfigure}[t]{.45\textwidth}
		\includegraphics[trim={1.5cm .5cm 2cm 1cm}, clip, width=\textwidth]{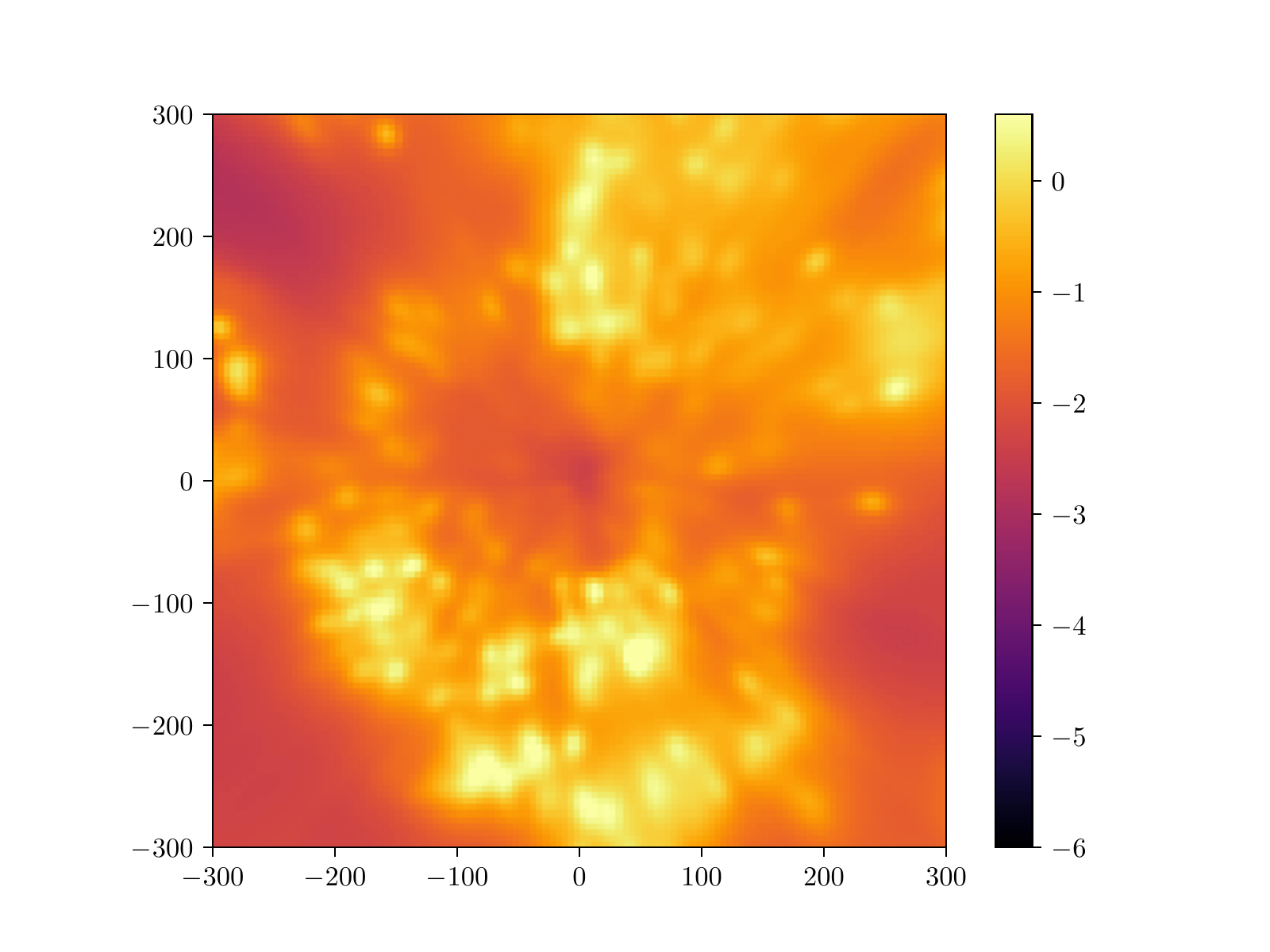}
	\caption{
	\label{fig:log-plane-projection-lallement}
	}
	\end{subfigure}
	~
	\begin{subfigure}[t]{.45\textwidth}
	\includegraphics[trim= {1.1cm 1cm 0.6cm 3cm} , clip, width=\textwidth]{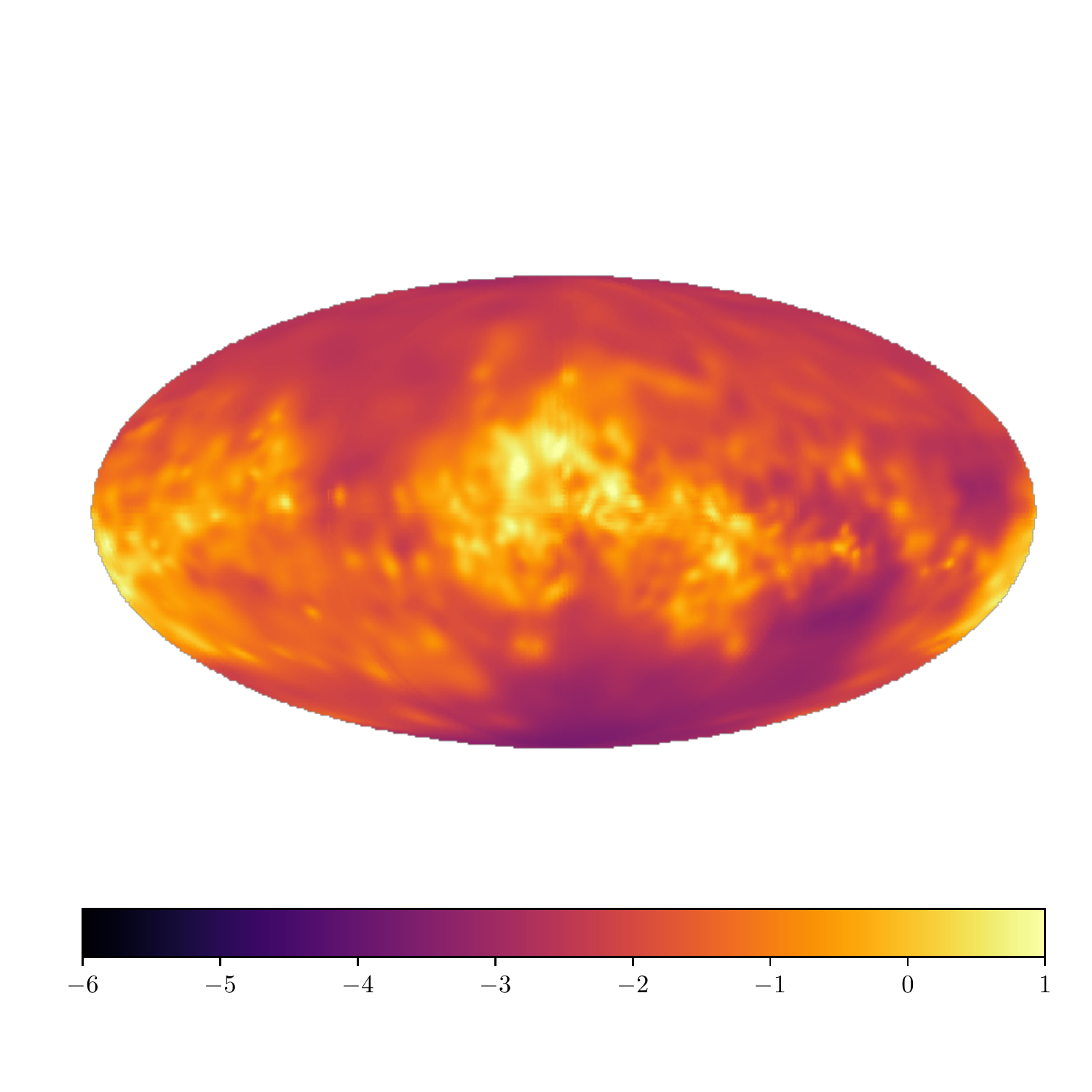}
	\caption{
	\label{fig:log-hp-projection-lallement}
		}
	\end{subfigure}
	\\
	\begin{subfigure}[t]{.45\textwidth}
		\includegraphics[trim={1.5cm .5cm 2cm 1cm}, clip, width=\textwidth]{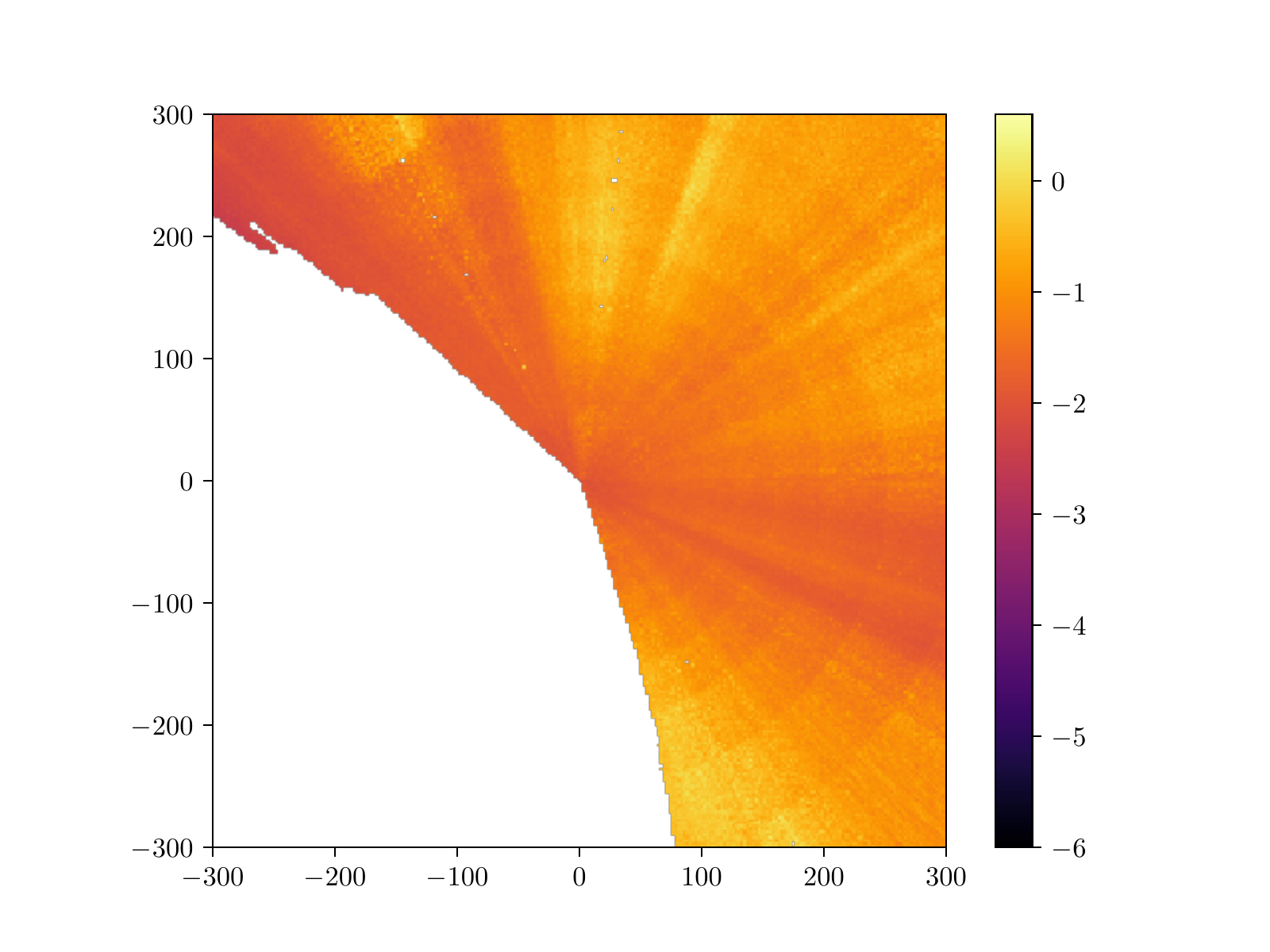}
	\caption{
	\label{fig:log-plane-projection-finkbeiner}
	}
	\end{subfigure}
	~
	\begin{subfigure}[t]{.45\textwidth}
	\includegraphics[trim= {1.1cm 1cm 0.6cm 3cm} , clip, width=\textwidth]{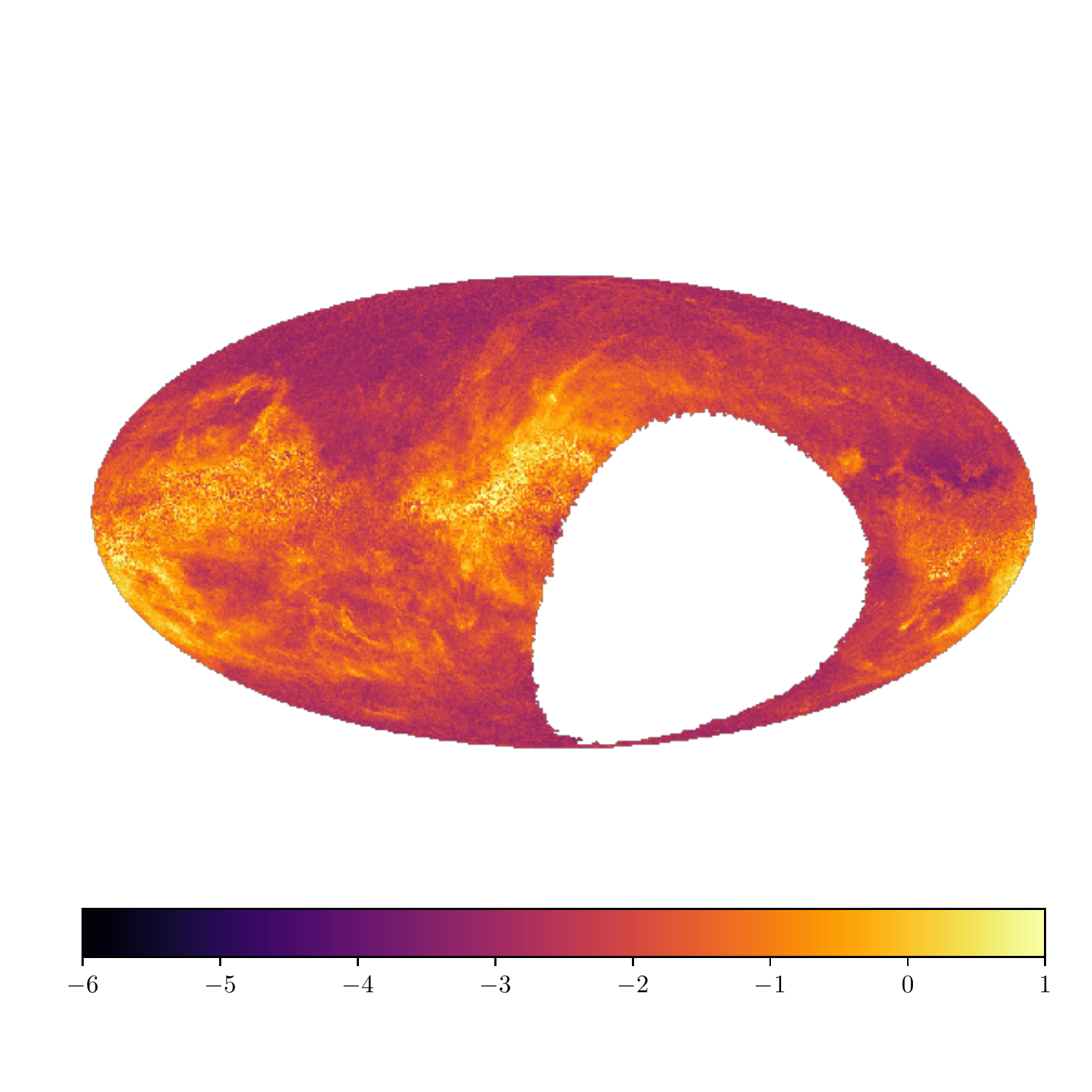}
	\caption{
	\label{fig:log-hp-projection-finkbeiner}
		}
	\end{subfigure}
	\caption{
		A natural logarithmic version of Fig.\,\ref{fig:comparison-plot}.
		\label{fig:log-comparison-plot}
	}
\end{figure*}

In Fig.\,\ref{fig:hp-projection} one can see a projection of the reconstructed dust onto the sky in galactic coordinates. 
Fig.\,\ref{fig:log-hp-projection} shows the corresponding expected logarithmic dust density.

\begin{figure*}[p]
	\centering
	\begin{subfigure}[t]{.46\textwidth}
		\includegraphics[trim={1.5cm .5cm 2cm 1cm}, clip, width=.95\textwidth]{our_mean_plane_projection.pdf}
	\caption{
	\label{fig:our-mean-plane-projection}
	}
	\end{subfigure}
	~
	\begin{subfigure}[t]{.46\textwidth}
		\includegraphics[trim={1.5cm .5cm 2cm 1cm}, clip, width=.95\textwidth]{our_mean_log_plane_projection.pdf}
		\caption{
		\label{fig:our-mean-log-plane-projection}}
	\end{subfigure}
	\\
	\begin{subfigure}[t]{.46\textwidth}
		\includegraphics[trim={1.5cm .5cm 2cm 1cm}, clip, width=.95\textwidth]{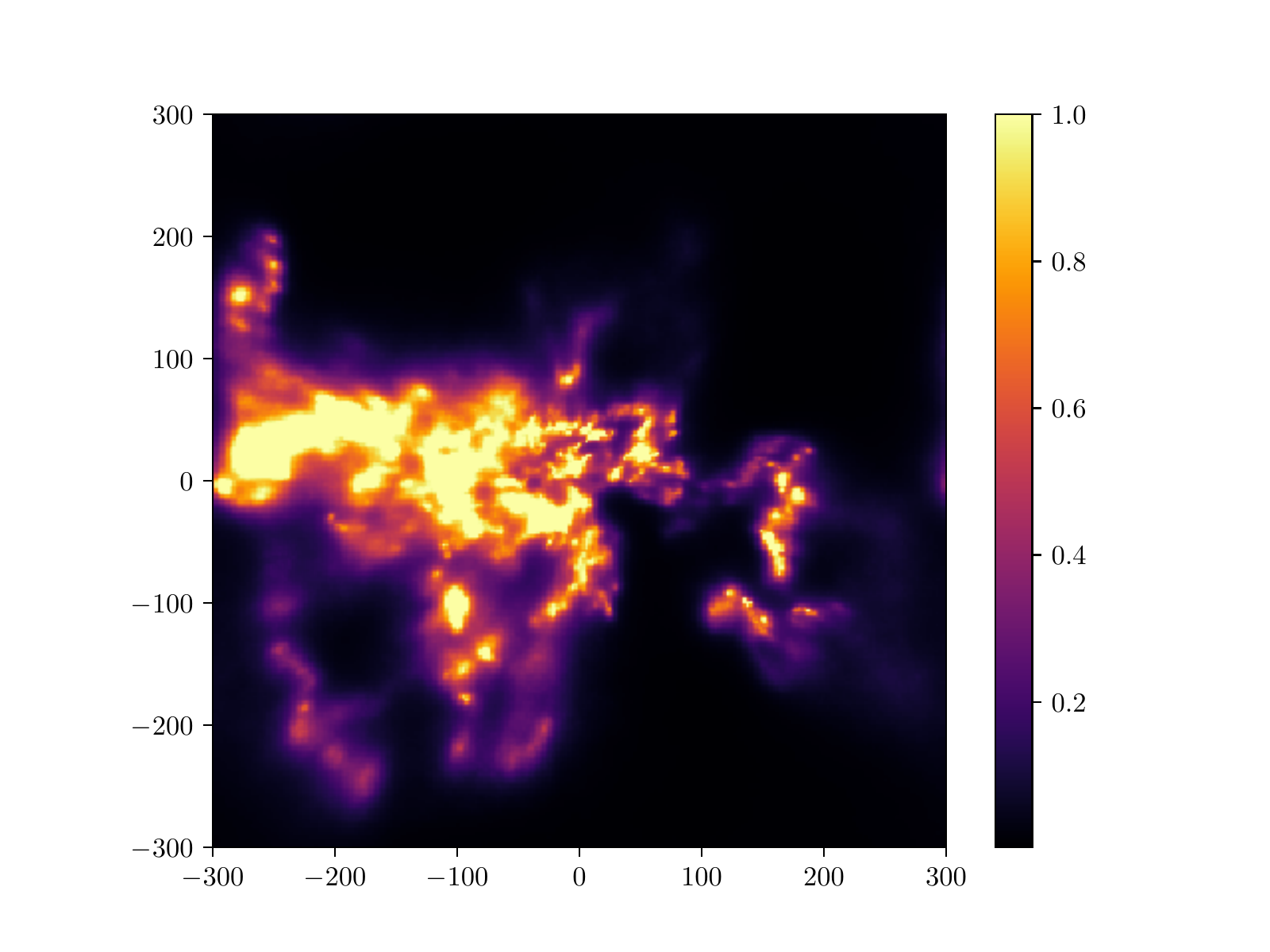}
	\caption{
	\label{fig:our-mean-x-projection}
	}
	\end{subfigure}
	~
	\begin{subfigure}[t]{.46\textwidth}
		\includegraphics[trim={1.5cm .5cm 2cm 1cm}, clip, width=.95\textwidth]{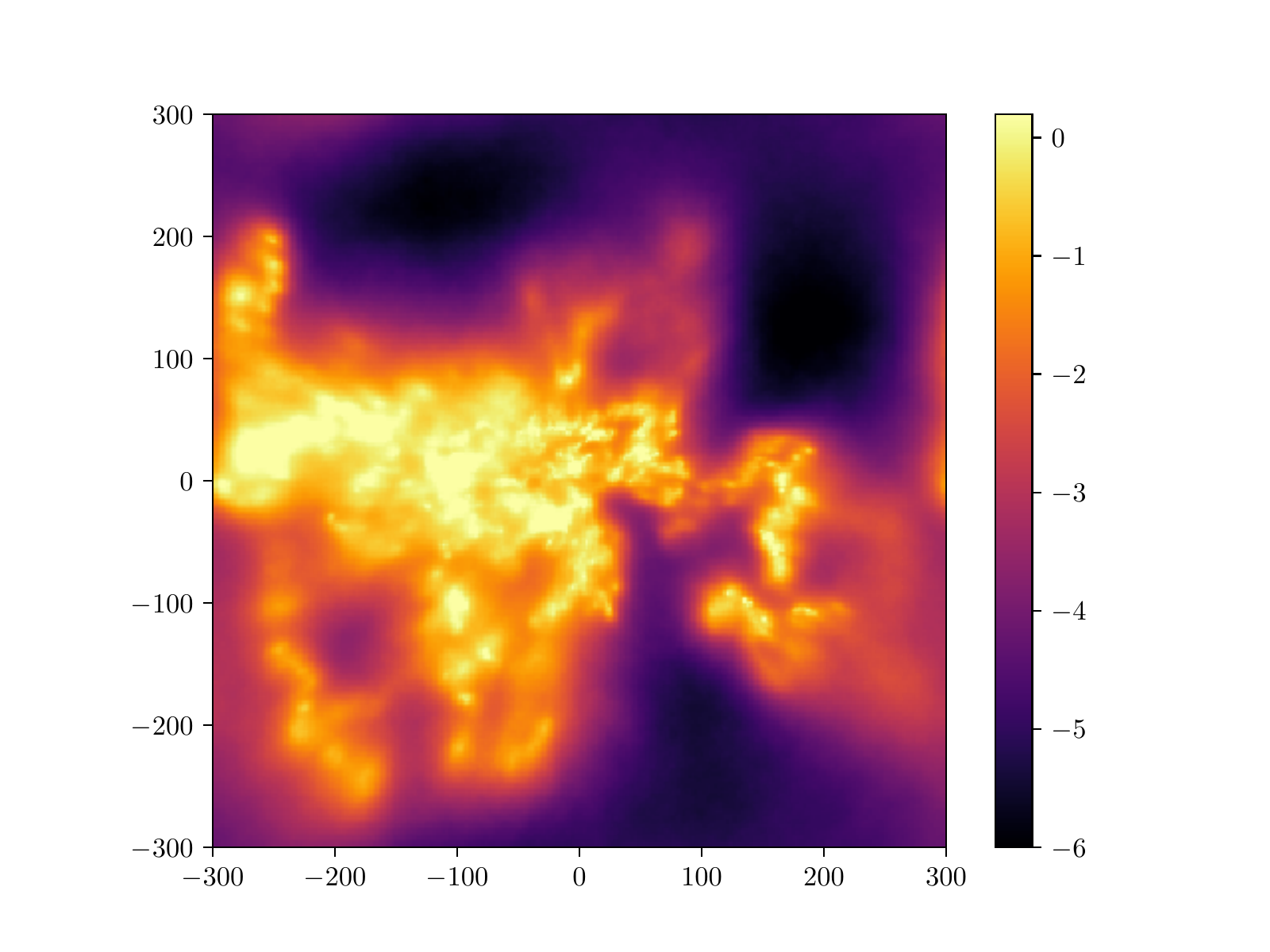}
	\caption{
	\label{fig:our-mean-log-x-projection}
		}
	\end{subfigure}
	\\
	\begin{subfigure}[t]{.46\textwidth}
		\includegraphics[trim={1.5cm .5cm 2cm 1cm}, clip, width=.95\textwidth]{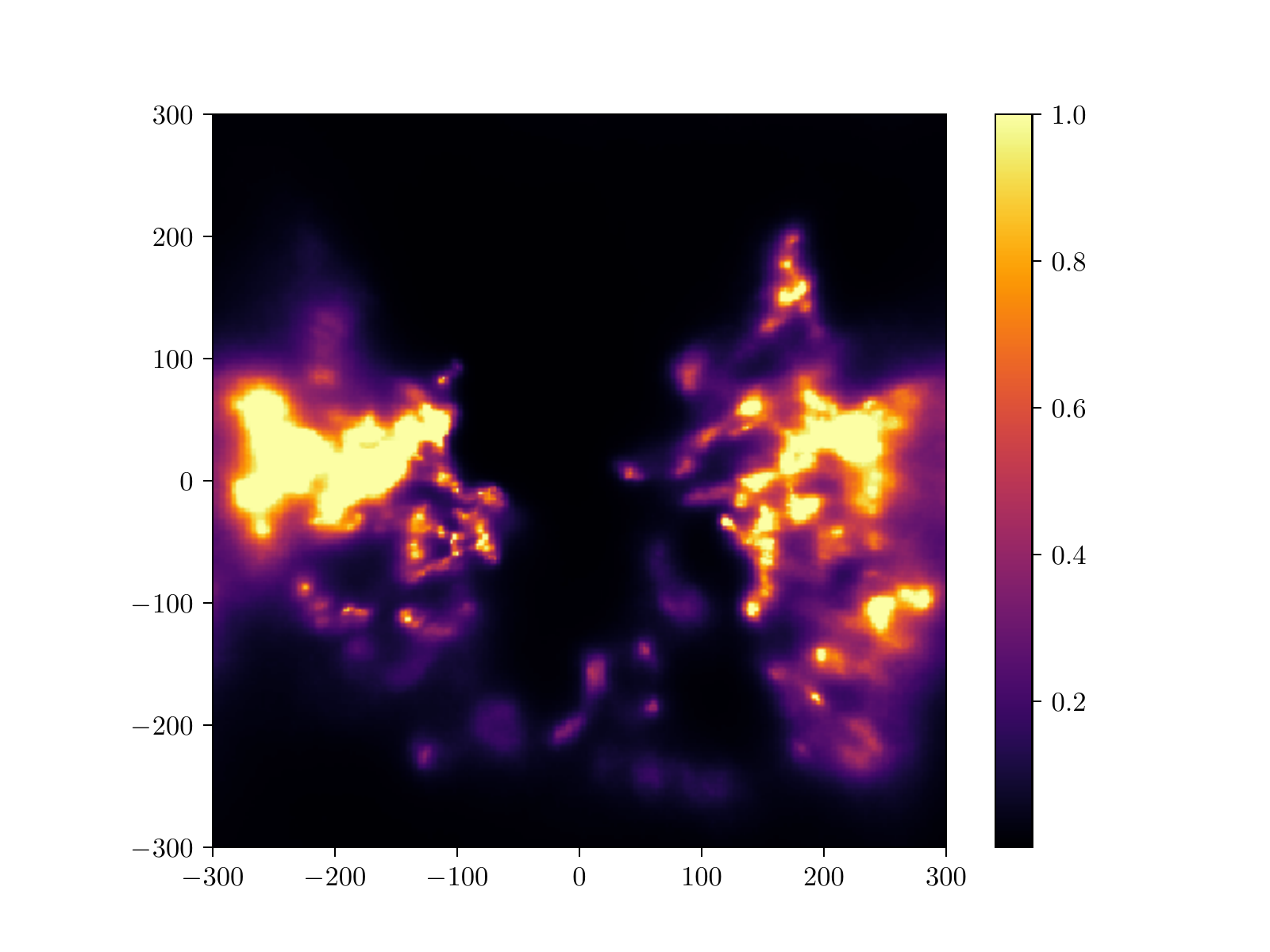}
	\caption{
	\label{fig:our-mean-y-projection}
	}
	\end{subfigure}
	~
	\begin{subfigure}[t]{.46\textwidth}
		\includegraphics[trim= {1.5cm .5cm 2cm 1cm} , clip, width=.95\textwidth]{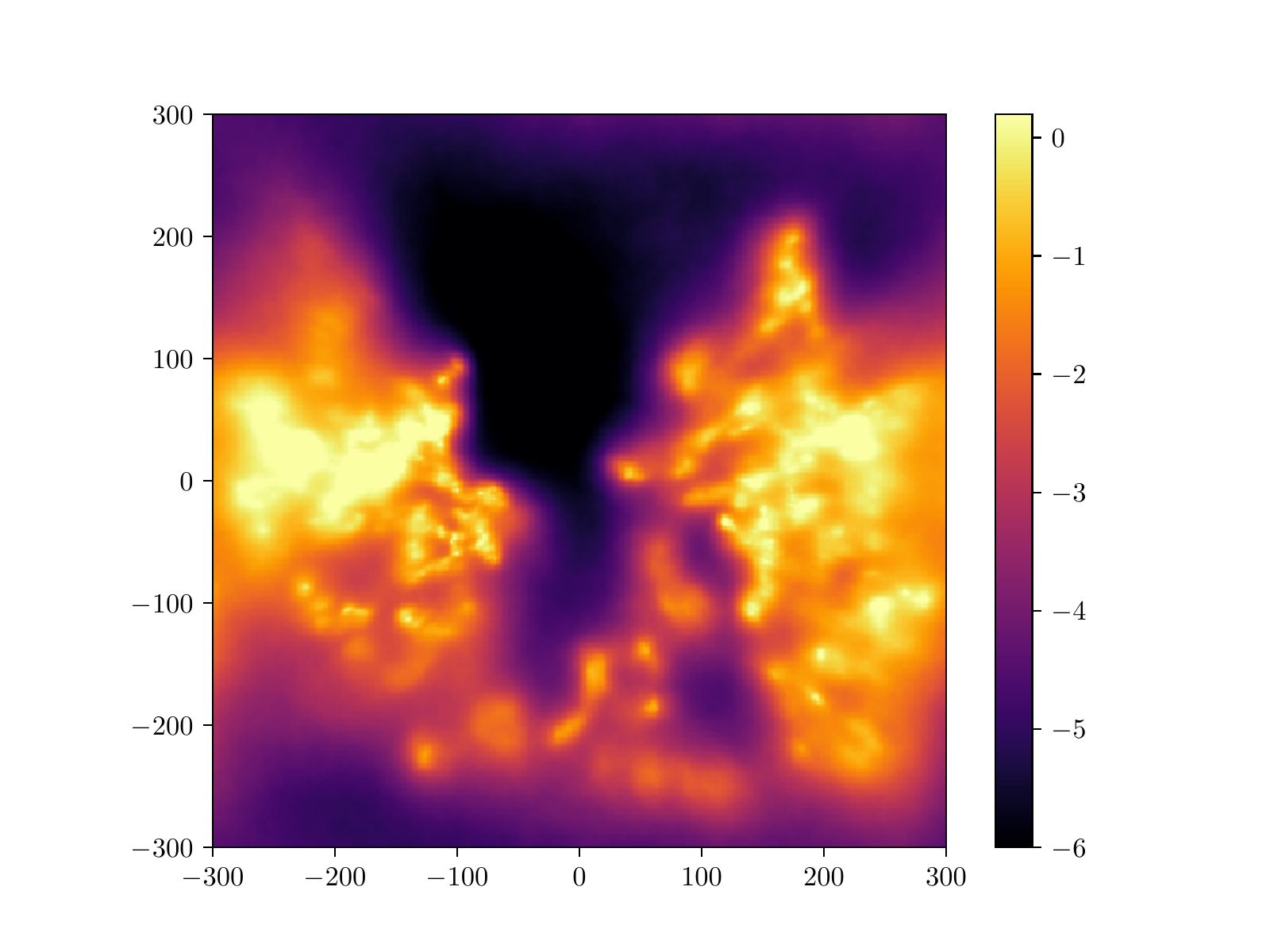}
	\caption{
	\label{fig:our-mean-log-y-projection}
	}
	\end{subfigure}
	\caption{
		\label{fig:different-angles}
		The reconstructed dust density in different projections.
	The rows show integrated dust extinction for sightlines parallel to the $z$- $x$- and $y$- axis respectively.
	In the first rows, the galactic center is located towards the bottom of the plot, in the other two rows the galactic North is located towards the top of the plot.
	The cube is in galactic coordinates, thus the $x$-axis is oriented towards the galactic center and the $z$-axis is perpendicular to the galactic plane.
	The first column shows the integrated G-band extinction in $e$-folds of extinction, the second column is a logarithmic version of the first column.
	}
\end{figure*}

In Fig.\,\ref{fig:plane-projection} the projection of the dust reconstruction on the galactic plane is shown. 
This view is especially interesting to study the dust morphology as this projection introduces no perspective-based distortion.
It is especially suited to spot underdense regions such as the local bubble in high resolution.
A logarithmic plot of the projection on the galactic plane can be seen in Fig.\,\ref{fig:log-plane-projection}.
We show integrated dust density for sightlines parallel to the $x$-, $y$-, and $z$-axis in Fig.\,\ref{fig:different-angles}.

\begin{figure*}[hp]
	\centering
	\begin{subfigure}[t]{.45\textwidth}
		\includegraphics[trim={1.1cm 1cm 0.6cm 3.5cm}, clip, width=\textwidth]{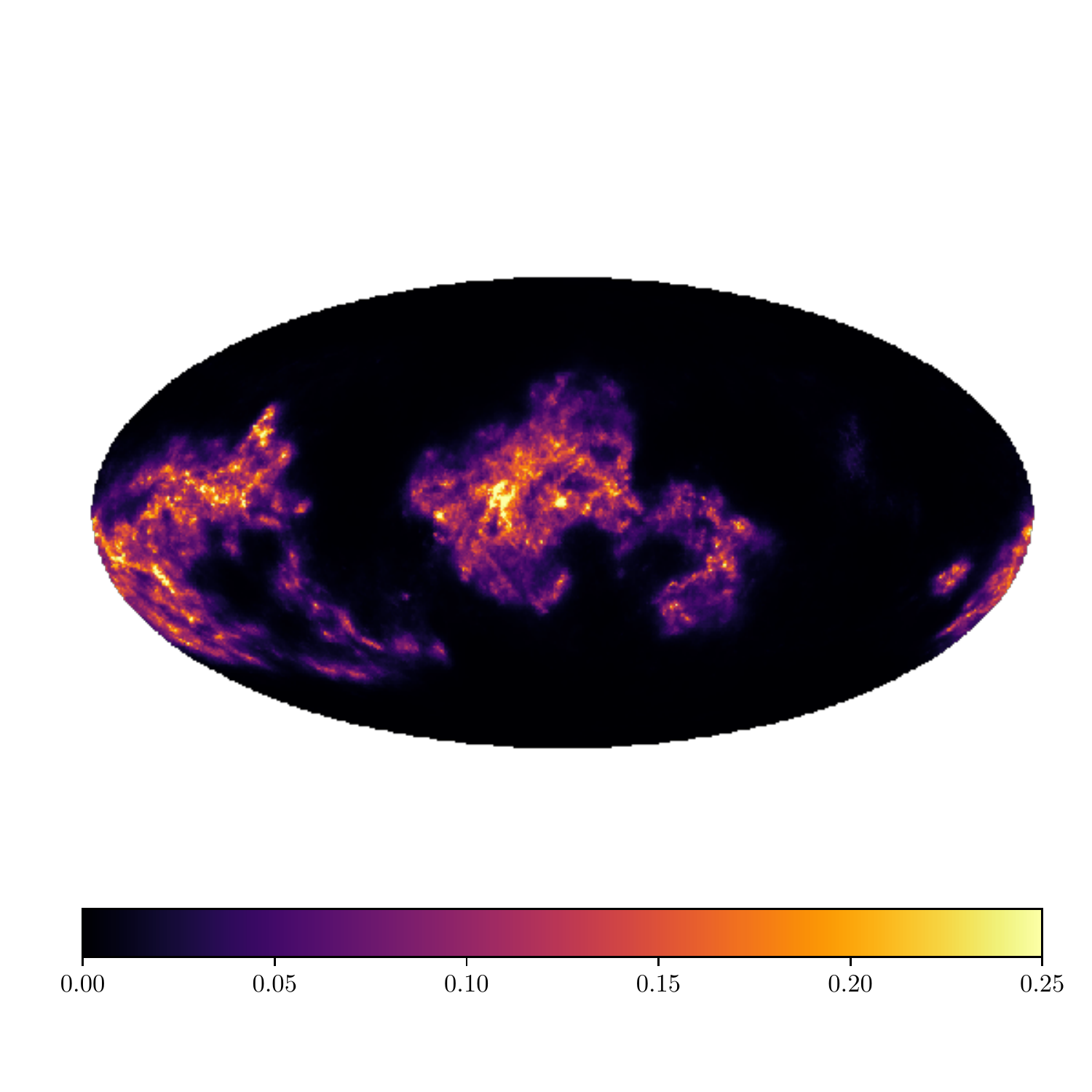}
		\caption{
			\label{fig:our-err-hp}
			}
	\end{subfigure}
	~
	\begin{subfigure}[t]{.45\textwidth}
		\includegraphics[trim={1.1cm 1cm 0.6cm 3.5cm}, clip, width=\textwidth]{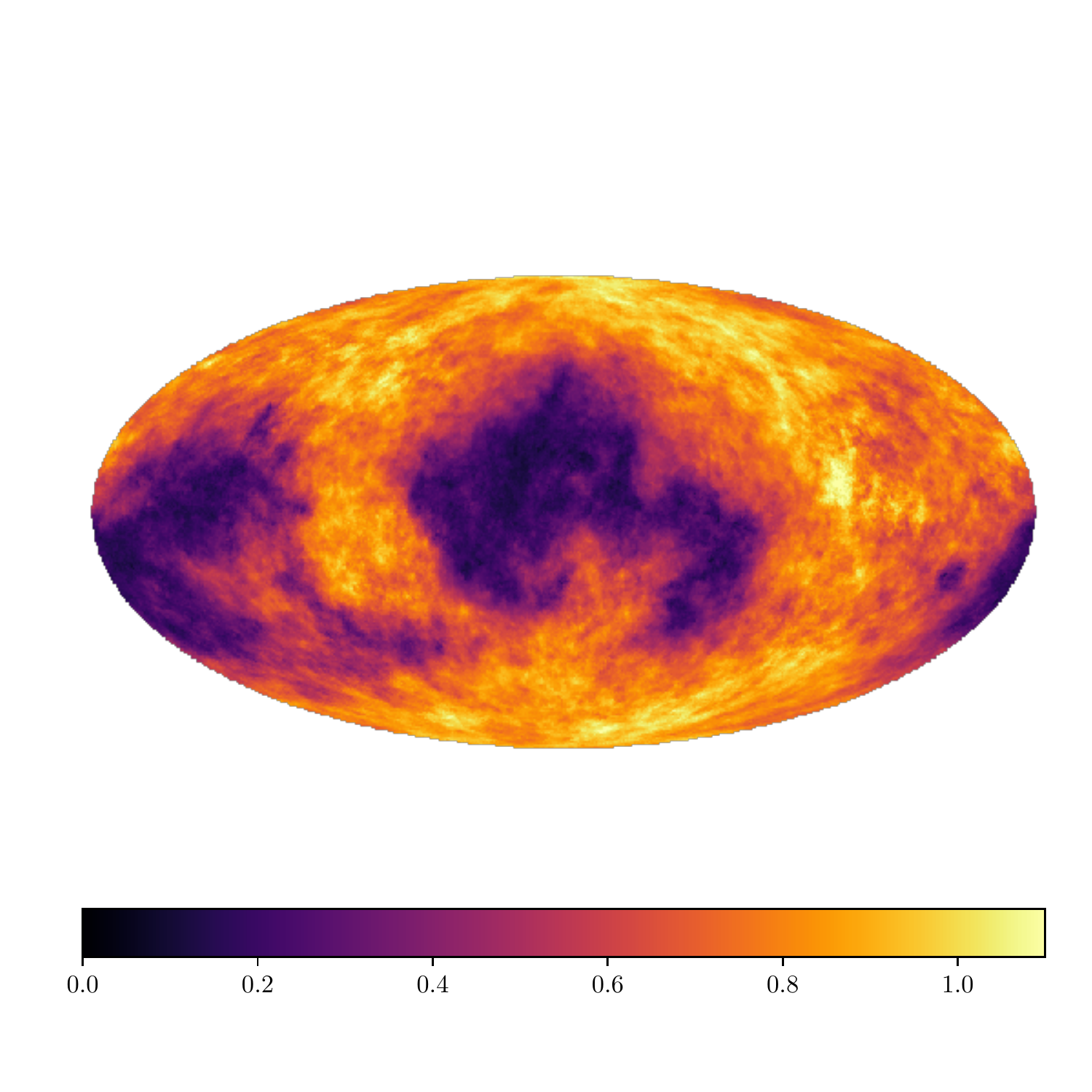}
		\caption{
			\label{fig:our-log-err-hp}
			}
	\end{subfigure}
	\\
	\begin{subfigure}[t]{.45\textwidth}
		\includegraphics[trim={1.1cm 1cm 0.6cm 3.5cm}, clip, width=\textwidth]{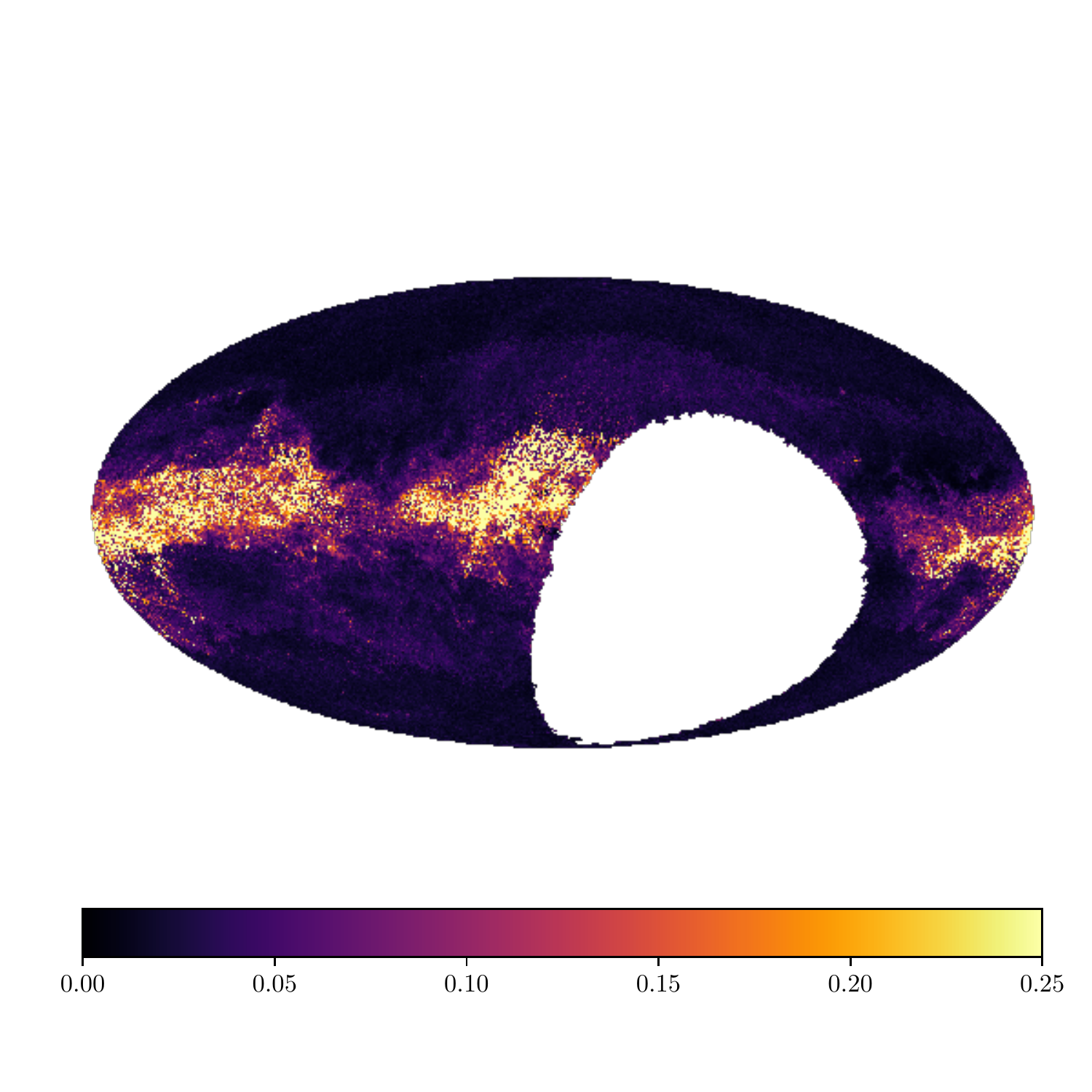}
		\caption{
			\label{fig:finkbeiner-err-hp}
			}
	\end{subfigure}
	~
	\begin{subfigure}[t]{.45\textwidth}
		\includegraphics[trim={1.1cm 1cm 0.6cm 3.5cm}, clip, width=\textwidth]{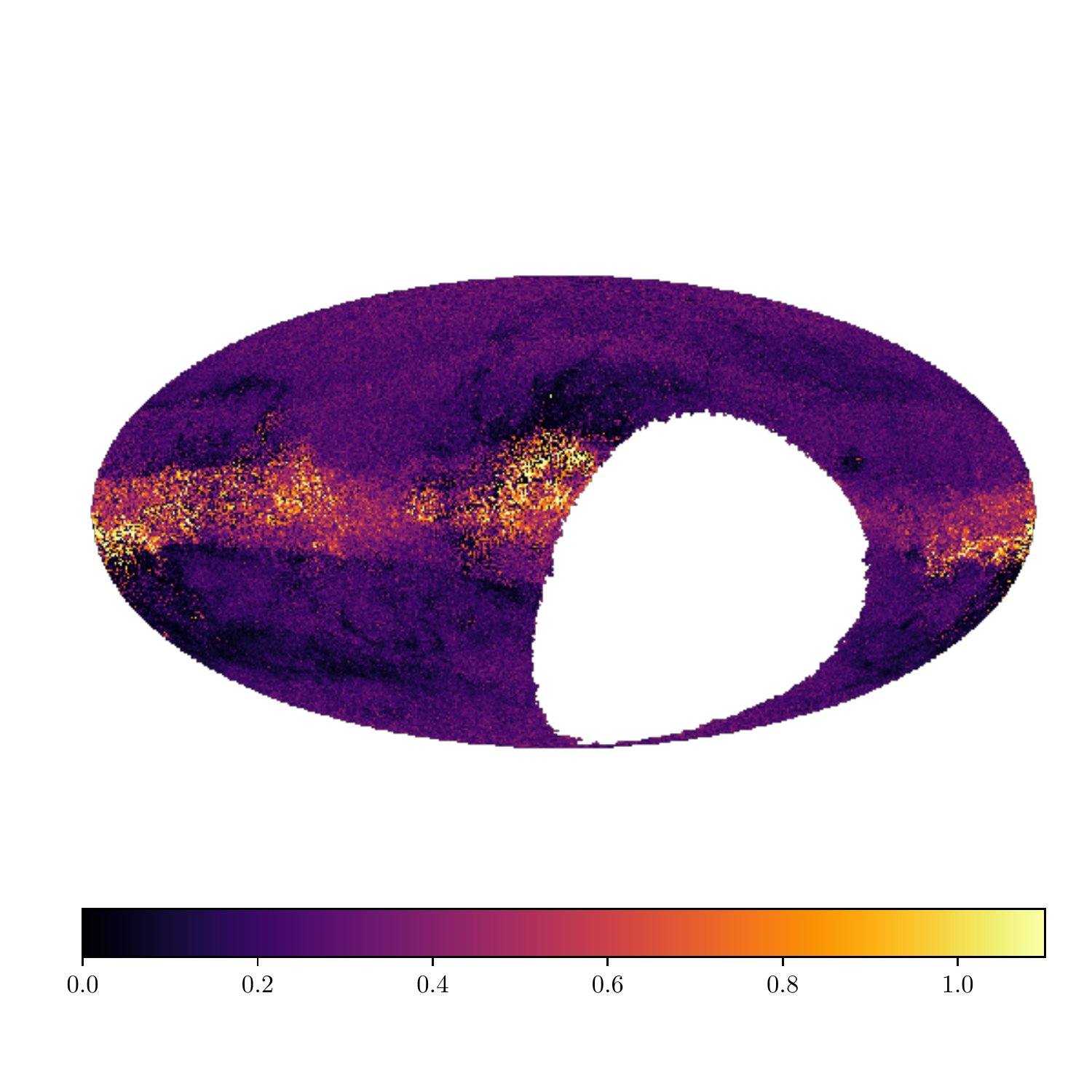}
		\caption{
			\label{fig:finkbeiner-log-err-hp}
			}
	\end{subfigure}
	\caption{
		\label{fig:err-hp-projections}
		Uncertainty of the reconstruction of this paper derived from posterior samples (first row) and of the reconstruction of \cite{green2018galactic} (second row), both in the sky projection.
		The uncertainties are in the same unit as the corresponding maps in Fig.\,\ref{fig:comparison-plot}, or dimensionless for logarithmic uncertainties.
		The first column shows the variance for the dust extinction and the second column shows the variance of the logarithmic projected dust density on natural log-scale, which can be interpreted as a relative error.
		}
\end{figure*}

\begin{figure*}[hp]
	\begin{subfigure}[t]{.45\textwidth}
		\includegraphics[trim={1.5cm .5cm 2cm 1cm}, clip, width=\textwidth]{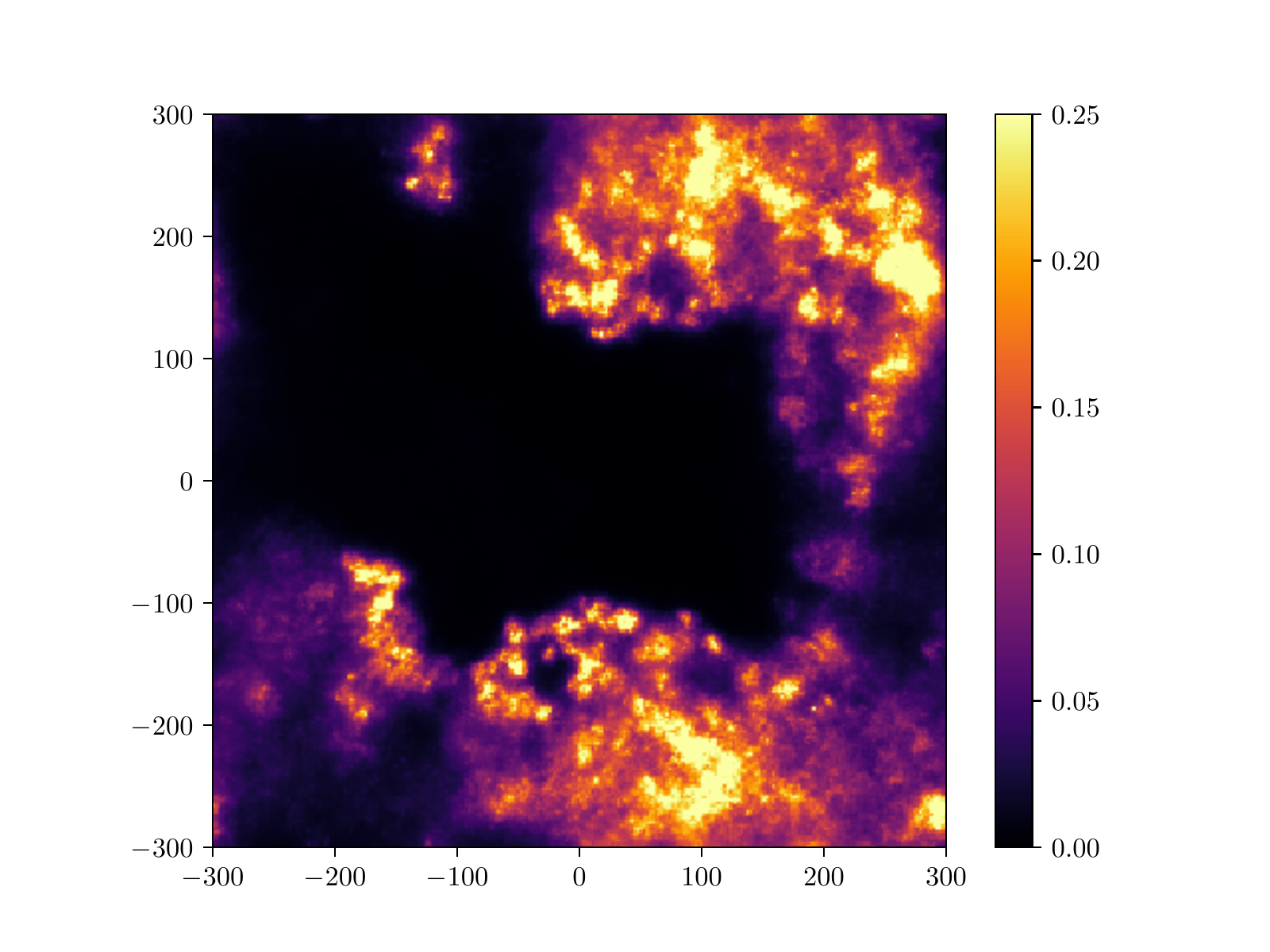}
		\caption{
			\label{fig:our-err-plane}
			}
	\end{subfigure}
	~
	\begin{subfigure}[t]{.45\textwidth}
		\includegraphics[trim={1.5cm .5cm 2cm 1cm}, clip, width=\textwidth]{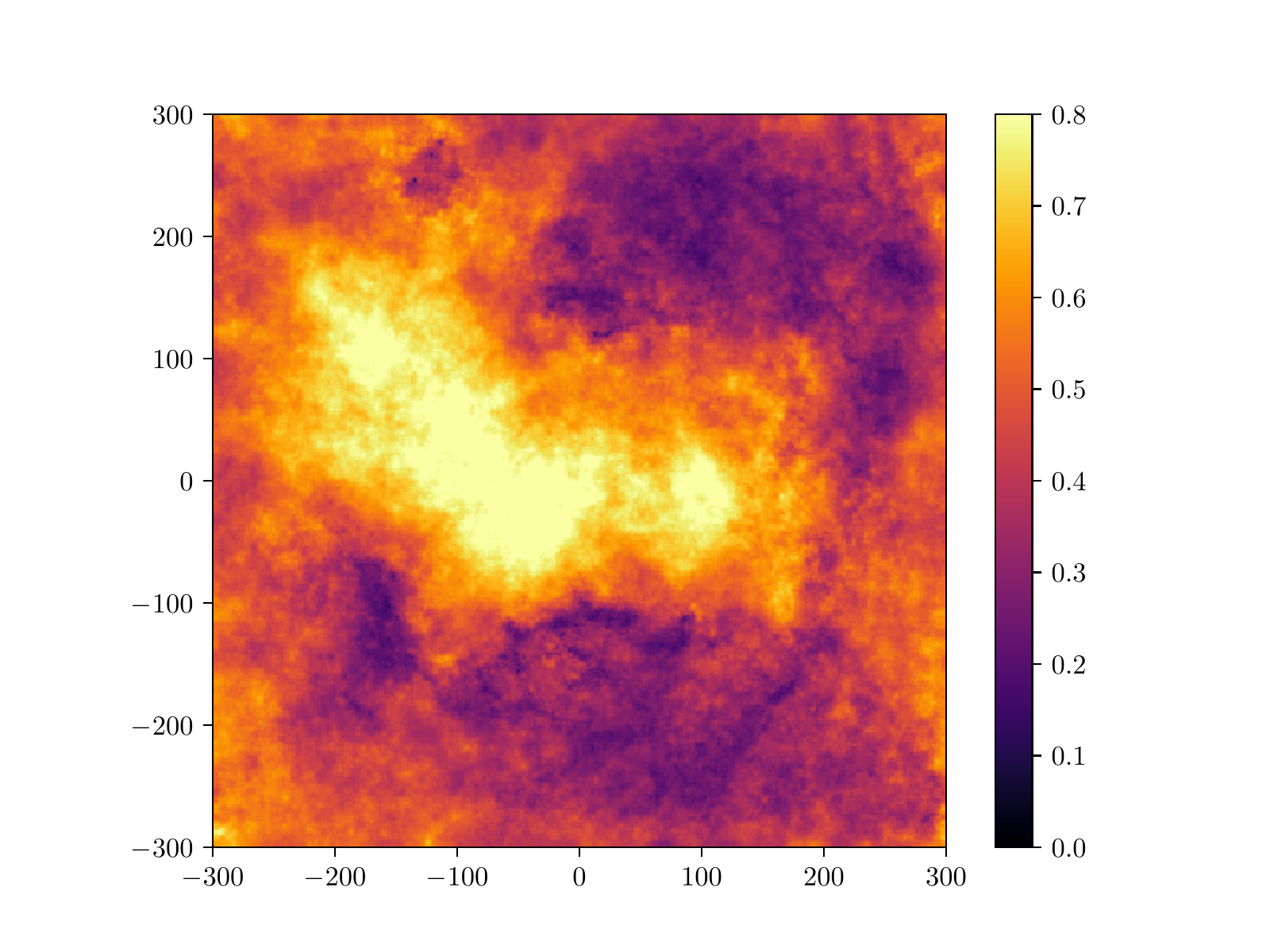}
		\caption{
			\label{fig:our-log-err-plane}
			}
	\end{subfigure}
	\\
	\begin{subfigure}[t]{.45\textwidth}
		\includegraphics[trim={1.5cm .5cm 2cm 1cm}, clip, width=\textwidth]{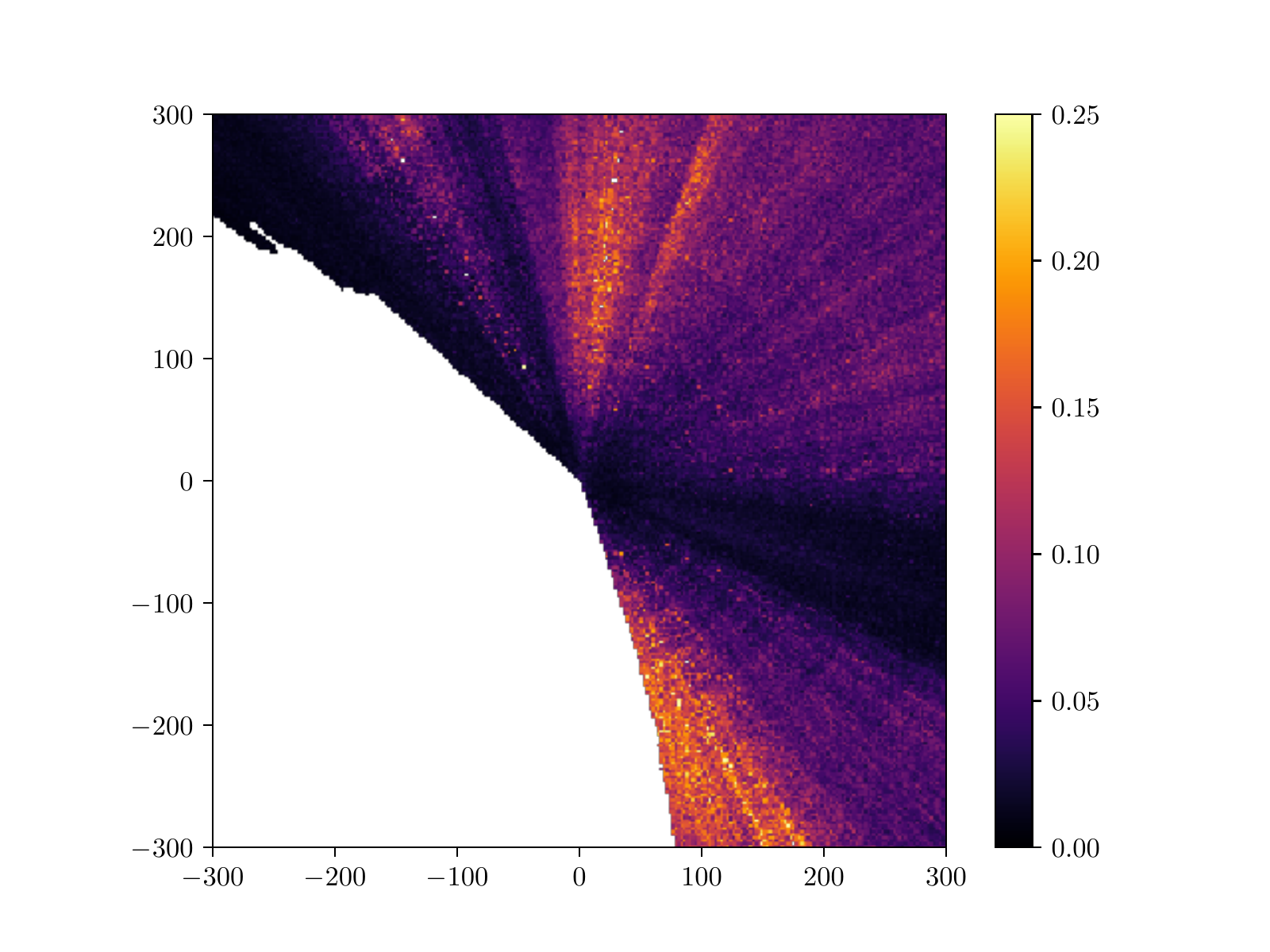}
		\caption{
			\label{fig:finkbeiner-err-plane}
			}
	\end{subfigure}
	~
	\begin{subfigure}[t]{.45\textwidth}
		\includegraphics[trim={1.5cm .5cm 2cm 1cm}, clip, width=\textwidth]{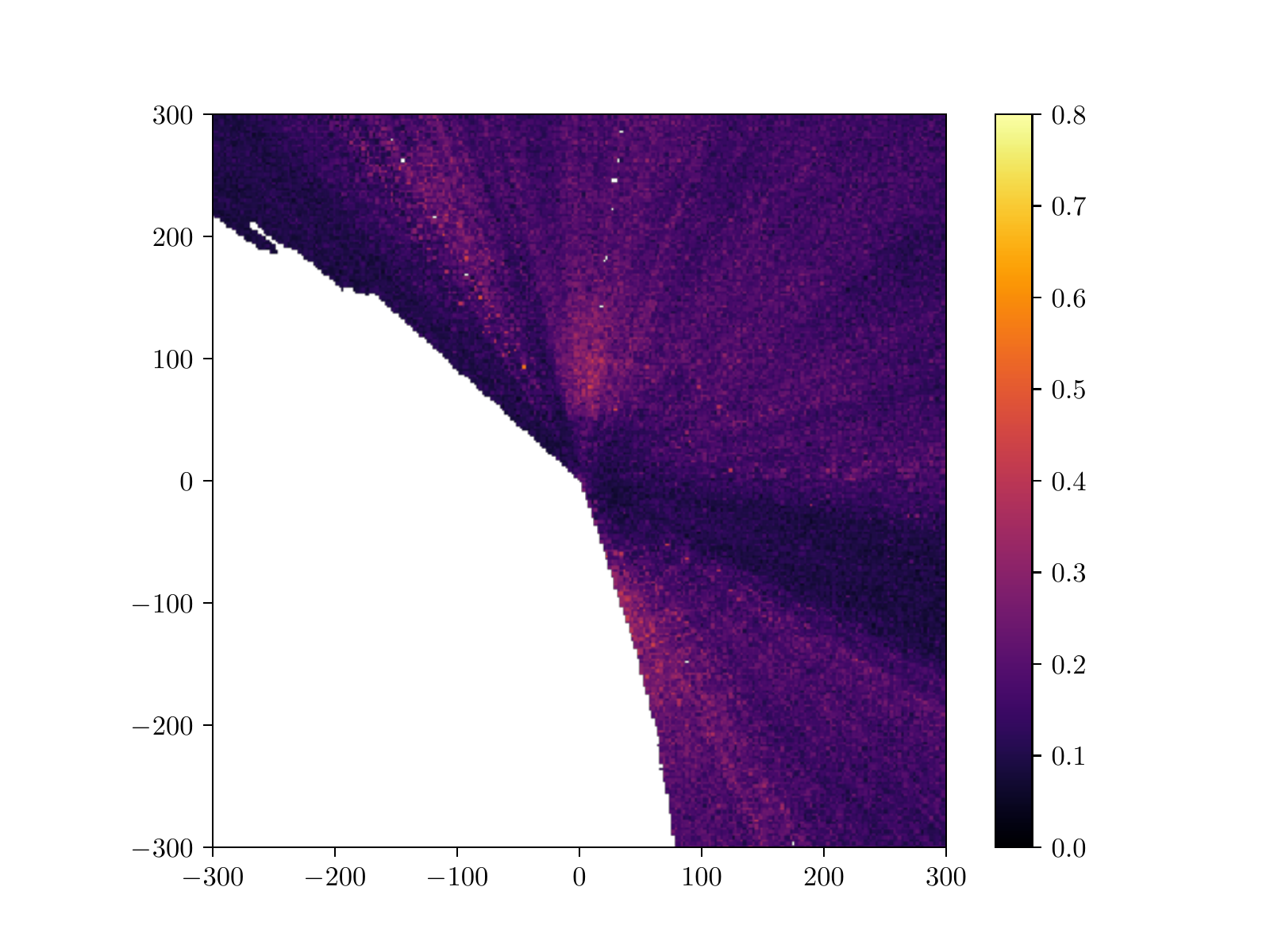}
		\caption{
			\label{fig:finkbeiner-log-err-log-plane}
			}
	\end{subfigure}
	\caption{
		\label{fig:err-plane-projections}
		Posterior uncertainty of the reconstruction of this paper derived from samples (first row) and of the reconstruction of \cite{green2018galactic} (second row) in the plane projection.
		The uncertainties are in the same unit as the corresponding maps in Fig.\,\ref{fig:comparison-plot}, or dimensionless for logarithmic uncertainties.
		The first column shows the variance for the dust extinction and the second column shows the variance of the logarithmic projected dust density on natural log-scale which can be interpreted as a relative error.
		}
\end{figure*}

We provide posterior uncertainty estimation via samples.
One should note that these uncertainties might be underestimated due to
the variational approach taken in this paper.
One can see a map of the expected posterior variance of the sky projection in Fig.\,\ref{fig:our-err-hp} and in the plane projection in Fig.\,\ref{fig:our-err-plane}.


\subsection{Using the reconstruction}

The results of the reconstruction are provided online on \url{https://wwwmpa.mpa-garching.mpg.de/~ensslin/research/data/dust.html}, or by its doi:10.5281/zenodo.2577337, and can be used under the terms of the ODC-By 1.0 license.
Proper attribution should be given to this paper as well as to the Gaia collaboration \citep{collaboration2016description}.

We give an overview of known systematic effects and advice on how to use the provided dust map.
\begin{itemize}
	\item We do not recommend to use the outer 15pc of the reconstruction. Periodic boundary conditions were assumed for algorithmic reasons, which leads to correlations leaking around the border of the cube. The inferred prior correlation kernel (Fig.\,\ref{fig:kernels}) suggests that correlations are vanishing after 30pc.
	\item We provide posterior samples. When doing further analysis of our reconstruction we recommend doing so for every sample in order to propagate errors.
	\item It was observed in Sec.\ref{sec:mock-results} that there can be a small number of outliers, that is differential dust extinction values that are much larger than the reconstructed value, by amounts that cannot be explained by the reconstructed uncertainty.
	\item We anticipate a perception threshold that leads to the absence of extremely low density dust clouds. 
		The two main reasons for this are that the truncated Gaussian likelihood provides less evidence in the regime where the extinction is close to zero and that variational Bayesian schemes are known to underestimate errors.
		Studying a larger volume will shed further light on this subject, as sightlines for more distant stars still provide information about nearby dust clouds.
		One should note that the overall Gaia extinction data provides 20 times more sightlines than were used in this reconstruction.
\end{itemize}

\section{Discussion}
\label{sec:discussion}

Here we discuss qualitative, quantitative, and methodological differences to other dust mapping efforts.
In table \ref{table:method-comparison} a detailed break down of methodological differences to other papers are shown.
There are three notable differences of our method to other methods that we would like to stress.
\begin{enumerate}
	\item{The here used dataset is one of largest one used so far.}
	\item{We use a high amount of data while still taking 3D correlations into account.}
	\item{We reconstruct the spatial correlation power spectrum. The motivation and impact of this already briefly discussed in Sec.\,\ref{sec:prior}}
\end{enumerate}
    \clearpage%
    \thispagestyle{empty}%
    \begin{landscape}%
        \centering %
\begin{tabular}{c|ccccc}
	& this paper & \cite{sale2018large} & \cite{kh2018detection} & \cite{lallement20183d} & \cite{green2018galactic} \\
	\hline
	parallax uncertainty & smoothing only & marginalization by sampling & neglected & neglected & proper uncertainty handling\\
	max distance & $300\sqrt{3}\,\text{pc}$ & $5$\,kpc & $6$\,kpc & $\approx 2\sqrt{2}\,\text{kpc}$ & $3\,\text{kpc}$\\
	max voxel resolution & 2.3\,pc & not applicable & about 200\,pc & 5\,pc & 16.4\,pc/0.063\,pc\\
	number of datapoints & 3.7 million & 6\,349 & 21\,000 & 71\,357 & 806 million\\
	power spectrum inference & yes & no & no & no & no\\
	correlations & 3D & 3D & 2D map only & 3D & 1D correlations only\\
	positiveness & yes & only of reddening & no & yes & yes\\
	statistical method & Variational Bayes & Expectation Propagation & analytic & maximum posterior & Hamiltonian Monte Carlo \\
	data sets & Gaia DR2 & synthetic Gaia data & APOGEE & Gaia DR1 + APOGEE + 2MASS & Pan-STARRS + 2MASS
\end{tabular}
	    \captionof{table}{\label{table:method-comparison}
	    A table comparing different dust inference methods with the one performed in this paper.
	    The first row indicates how the parallax uncertainty of the stars was treated.
	    Hereby smoothing refers to weighting a voxel in the line of sight by the survival function of the star radial distance, as is described in Eq.\,(\ref{eq:los-response}).
	    The distance of the furthest point in the reconstruction is given in the second row.
	    The dimensions of the smallest voxel are given in the third row.
	    For the reconstruction of \cite{sale2018large} the concept of voxel resolution is not readily applicable; \cite{sale2018large} use 140 inducing points spanning a region for which one could evaluate the posterior mean at any point.
	    The resolution for \cite{green2018galactic} contains two values because the resolution is different in radial/angular direction.
	    The fourth row provides the number of used data points.
	    The fifth row indicates whether the power spectrum is inferred.
	    The sixth row states which kind of correlations are assumed for the reconstruction.
	    Whether positivity of dust density is enforced can be read in the seventh row.
	    The second to last row states the method, with which the posterior summary statistics was calculated from the unnormalized log posterior.
	    In the last row the data sets used for the reconstruction are listed.
	    }
    \end{landscape}
We compare our dust map to other maps.
Comparisons to 2D dust maps are only possible on a qualitative level, since it is not clear whether structures visible in the 2D maps that are not present in the 3D map are simply further away or are too noisy in the data for the algorithm to pick them up.
On a qualitative level it is possible to see several morphological similarities of our reconstruction in Fig.\,\ref{fig:hp-projection} to the Planck dust map \citep{akrami2018planck} in Fig.\,\ref{fig:Planck-and-data}.
These figures also show that many dust structures that are not inside the galactic plane are local features.

The two 3D dust maps mentioned in Sec.\,\ref{sec:introduction} permit a more thorough analysis.
Fig.\,\ref{fig:comparison-plot} shows a compilation of projected dust densities for our reconstruction as well as the reconstruction of \cite{lallement20183d} and \cite{green2018galactic}\footnote{It should be noted that there is a new version \citep{green20193d} that appeared during the revision of this paper.}, restricted to the same volume as the reconstruction discussed in this paper.
A logarithmic version of this figure is provided by Fig.\,\ref{fig:log-comparison-plot}.

While our map seems to agree on large scales with the other maps, there seems to be a pronounced tension in the predictions of the position of some dust clouds compared to the reconstruction of \cite{lallement20183d}. 
Compared to the map of \cite{green2018galactic} we recover the small scales significantly better and suffer far less from radial smearing. It should be noted that \cite{green2018galactic} mapped a significantly larger part of our galaxy, and that the region that overlaps with our map was declared to be not that reliable by the authors themselves.
The differences are probably due to the different nature of the used datasets. 
The Gaia DR2 data used in our reconstruction has a vastly higher amount of data points than those used for the other reconstructions.
These data points, taken from Gaia DR2, have a very small parallax error.
Additionally our reconstruction takes the full 3D correlation structure into account.

Our reconstruction as well as the reconstruction of \cite{green2018galactic} permit quantifying uncertainties using samples.
A plot of uncertainties of the dust density reconstructions projected into the galactic plane can be seen in Fig.\,\ref{fig:err-plane-projections}.
Uncertainties of the dust density reconstructions in the sky projection can be seen in Fig.\,\ref{fig:err-hp-projections}.

\begin{figure*}[hp]
	\centering
	\begin{subfigure}[t]{.45\textwidth}
		\includegraphics[width=\textwidth]{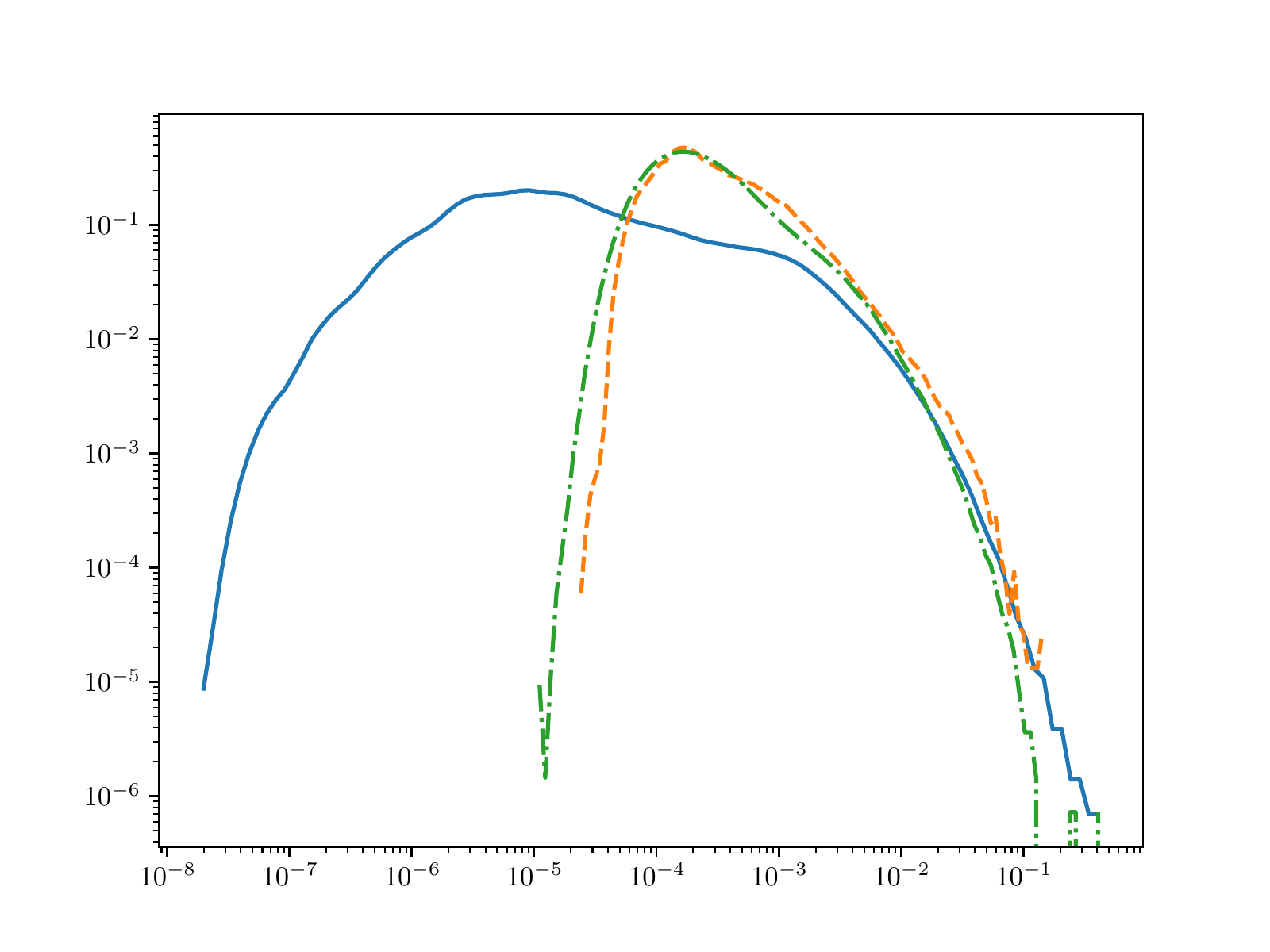}
		\caption{
			\label{fig:histograms}
			}
	\end{subfigure}
	~
	\begin{subfigure}[t]{.45\textwidth}
		\includegraphics[width=\textwidth]{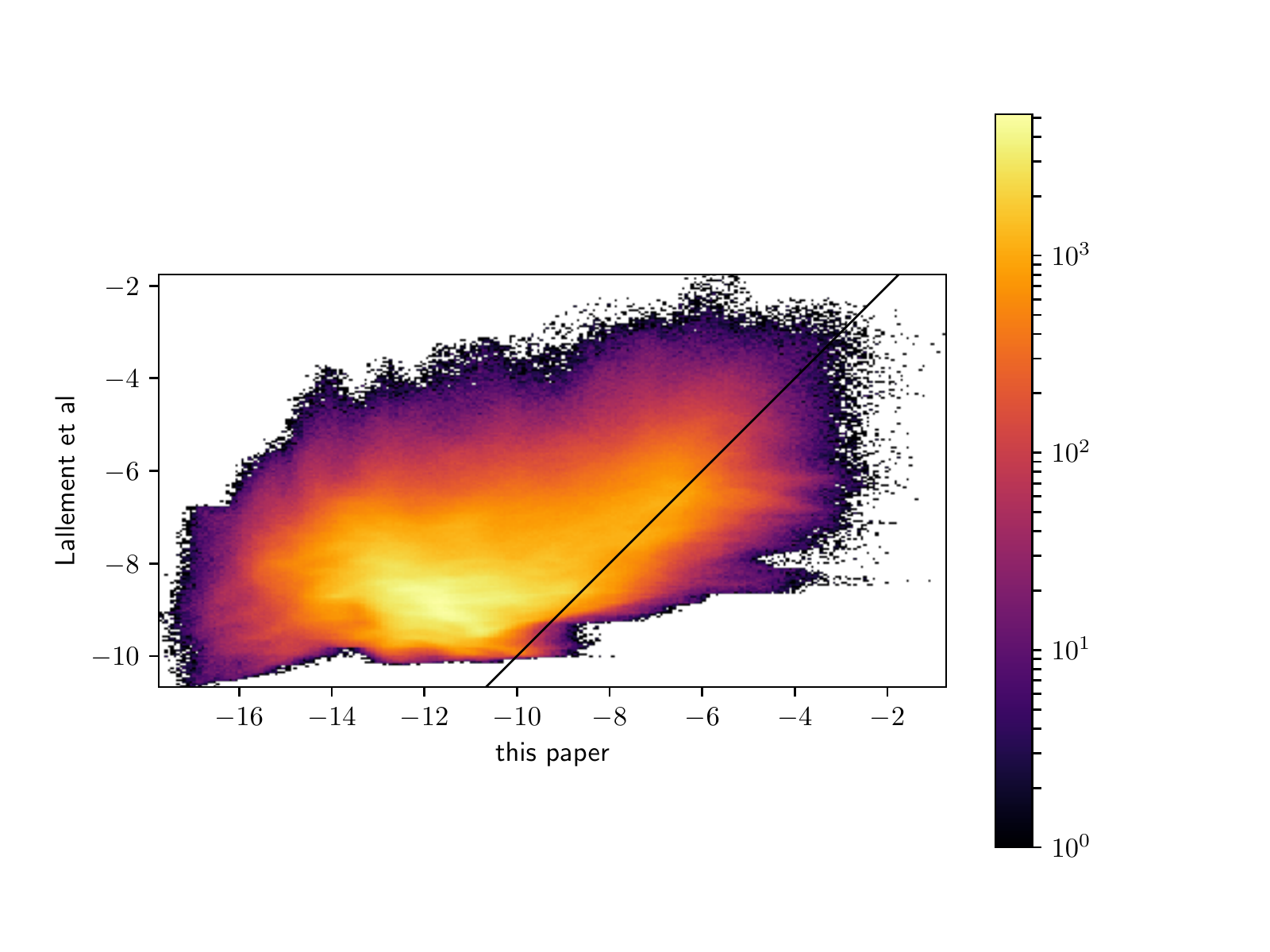}
		\caption{
			\label{fig:scatter-our-lallement}
			}
	\end{subfigure}
	\\
	\begin{subfigure}[t]{.45\textwidth}
		\includegraphics[width=\textwidth]{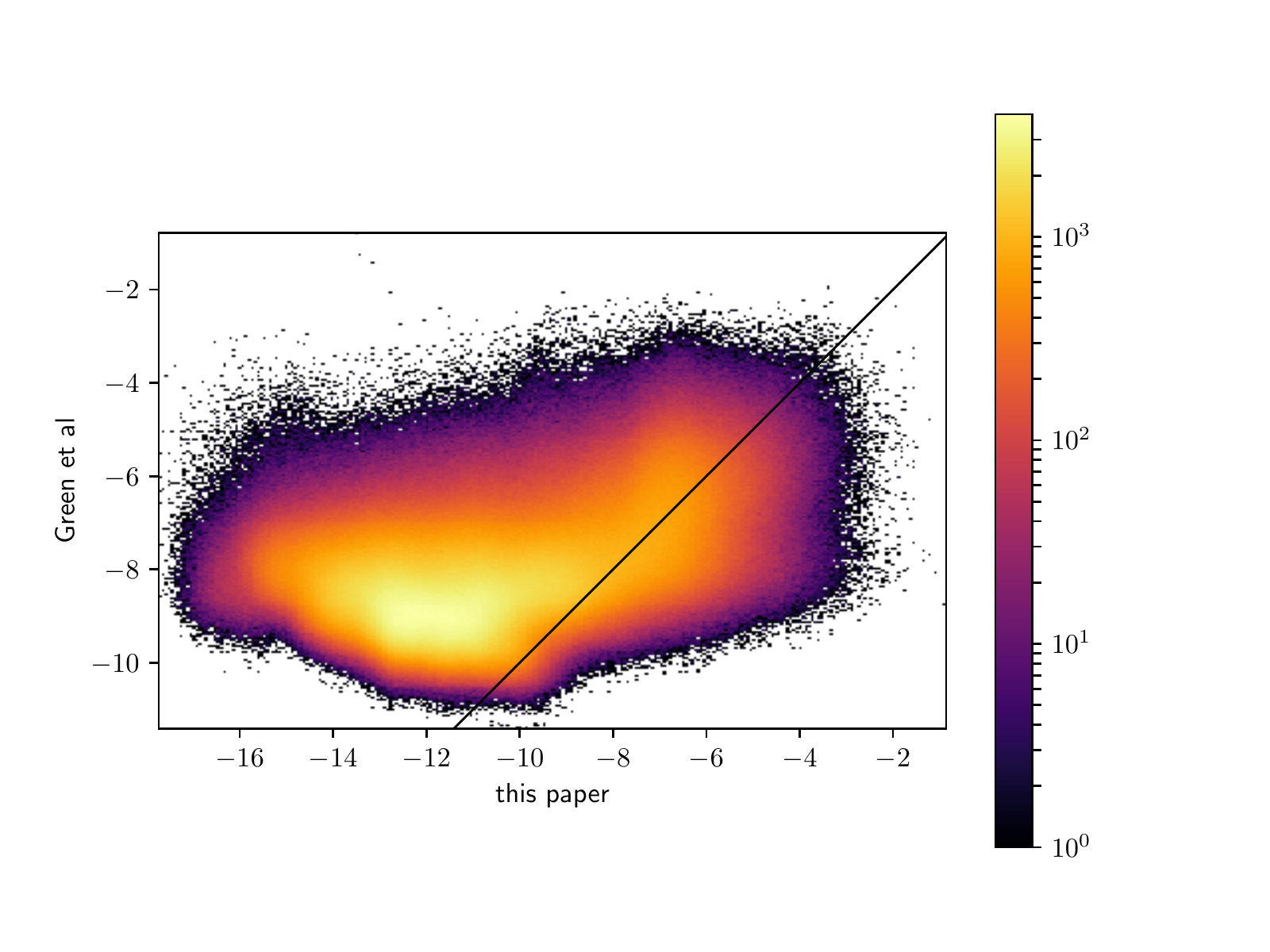}
		\caption{
			\label{fig:scatter-our-finkbeiner}
			}
	\end{subfigure}
	~
	\begin{subfigure}[t]{.45\textwidth}
		\includegraphics[width=\textwidth]{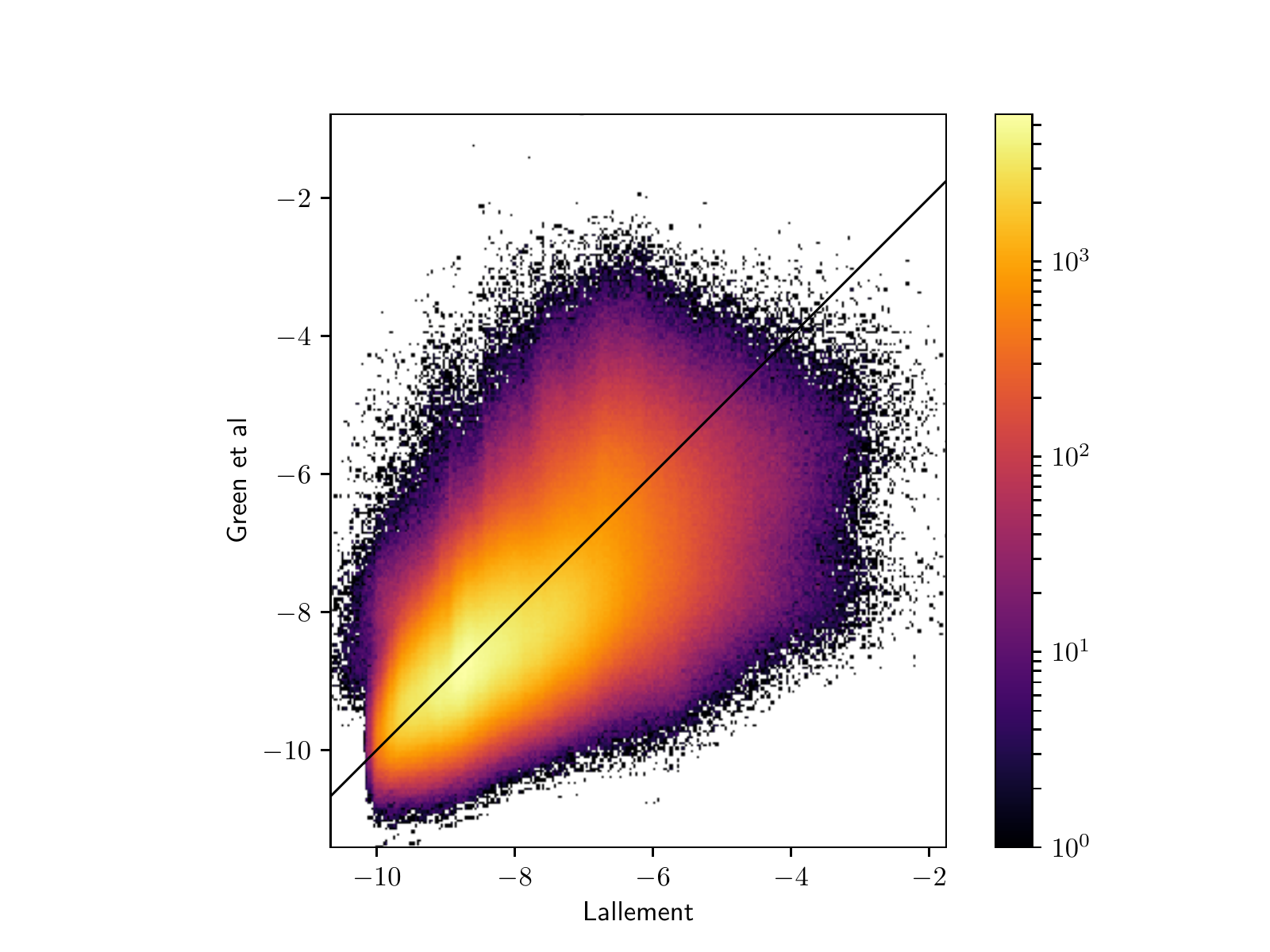}
		\caption{
			\label{fig:scatter-lallement-finkbeiner}
			}
	\end{subfigure}
	\caption{\label{fig:heatmaps}
		Panel a shows normalized histograms of dust densities. The solid line corresponds to our reconstruction, the dashed line is the reconstruction of \cite{lallement20183d} and the dash-dotted line is the reconstruction of \cite{green2018galactic}.
		The other three plots are heatmaps of voxel-wise correlations between reconstructed logarithmic dust densities, where the color shows bin counts.
		The black line in the heatmaps is the identity function, corresponding to perfect correlation.
		}
\end{figure*}

To quantify the dynamic range of the reconstruction and as a prediction on the variability of the logarithmic dust density we calculated histograms of dust density which show how many voxels have which dust density.
These histograms can be seen in Fig.\,\ref{fig:histograms}.
One can see that the histogram of our reconstruction extends slightly more towards high dust densities and substantially towards low dust densities.
This is possibly because our reconstruction is more sharply resolved, thus regions of high dust density get captured better and bleed less into the regions where dust is absent.

We characterize how much pairs of those reconstructions agree by the heatmaps of their voxel-wise value pairs.
These heatmaps can be seen in Fig.\,\ref{fig:heatmaps}.
For two perfectly agreeing reconstructions the heatmap would show a line with slope 1. 
Again it can be seen that the dust density in our reconstruction varies significantly more than in the two other reconstruction. 
While all maps agree more or less for high dust densities,
our dust map exhibits vastly more volume with low dust density.

\begin{figure*}[p]
	\centering
	\begin{subfigure}[t]{.46\textwidth}
		\includegraphics[trim={1.5cm .5cm 2cm 1cm}, clip, width=.95\textwidth]{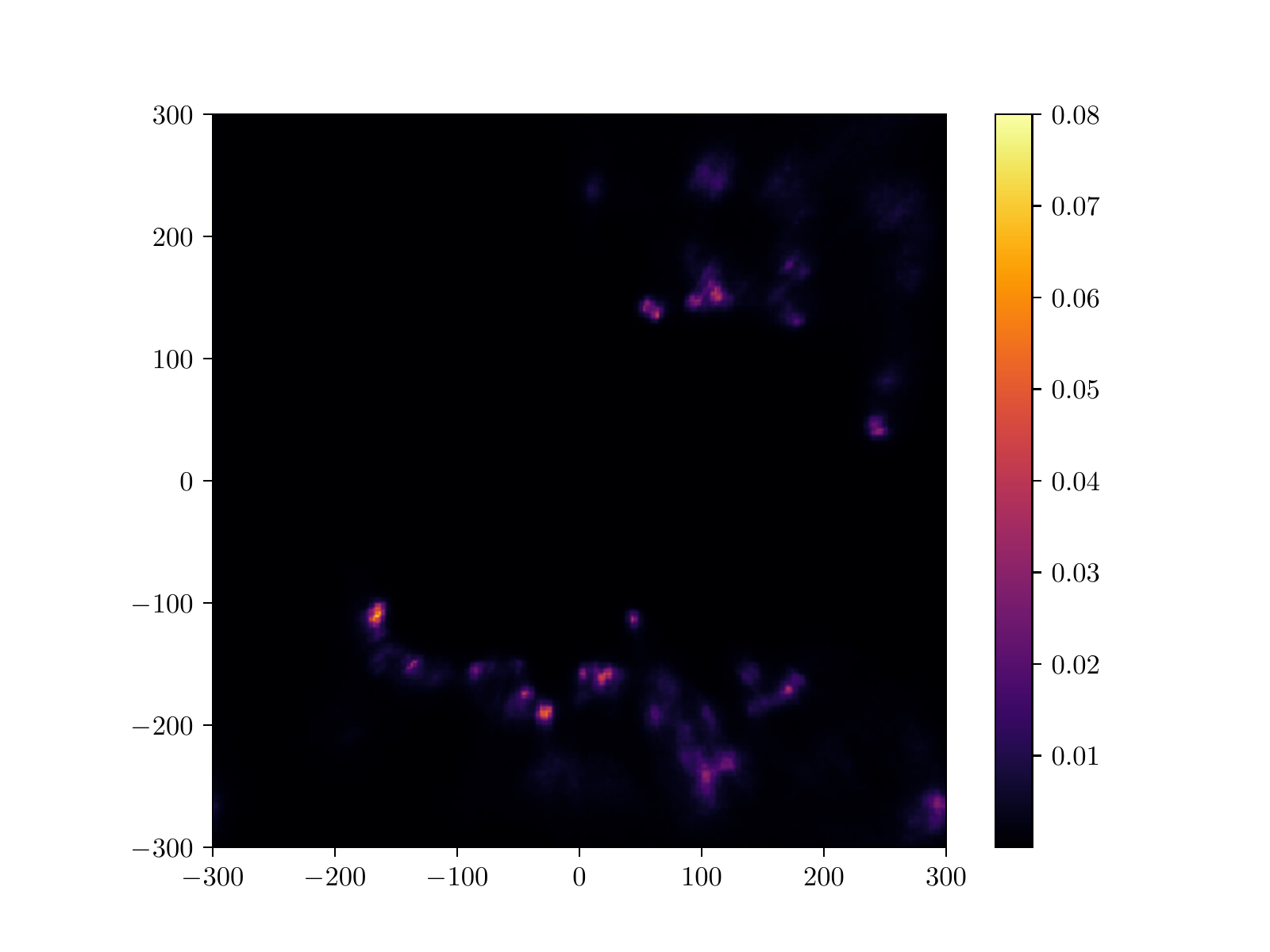}
	\caption{
	\label{fig:our-mean-plane-slice}
	}
	\end{subfigure}
	~
	\begin{subfigure}[t]{.46\textwidth}
		\includegraphics[trim={1.5cm .5cm 2cm 1cm}, clip, width=.95\textwidth]{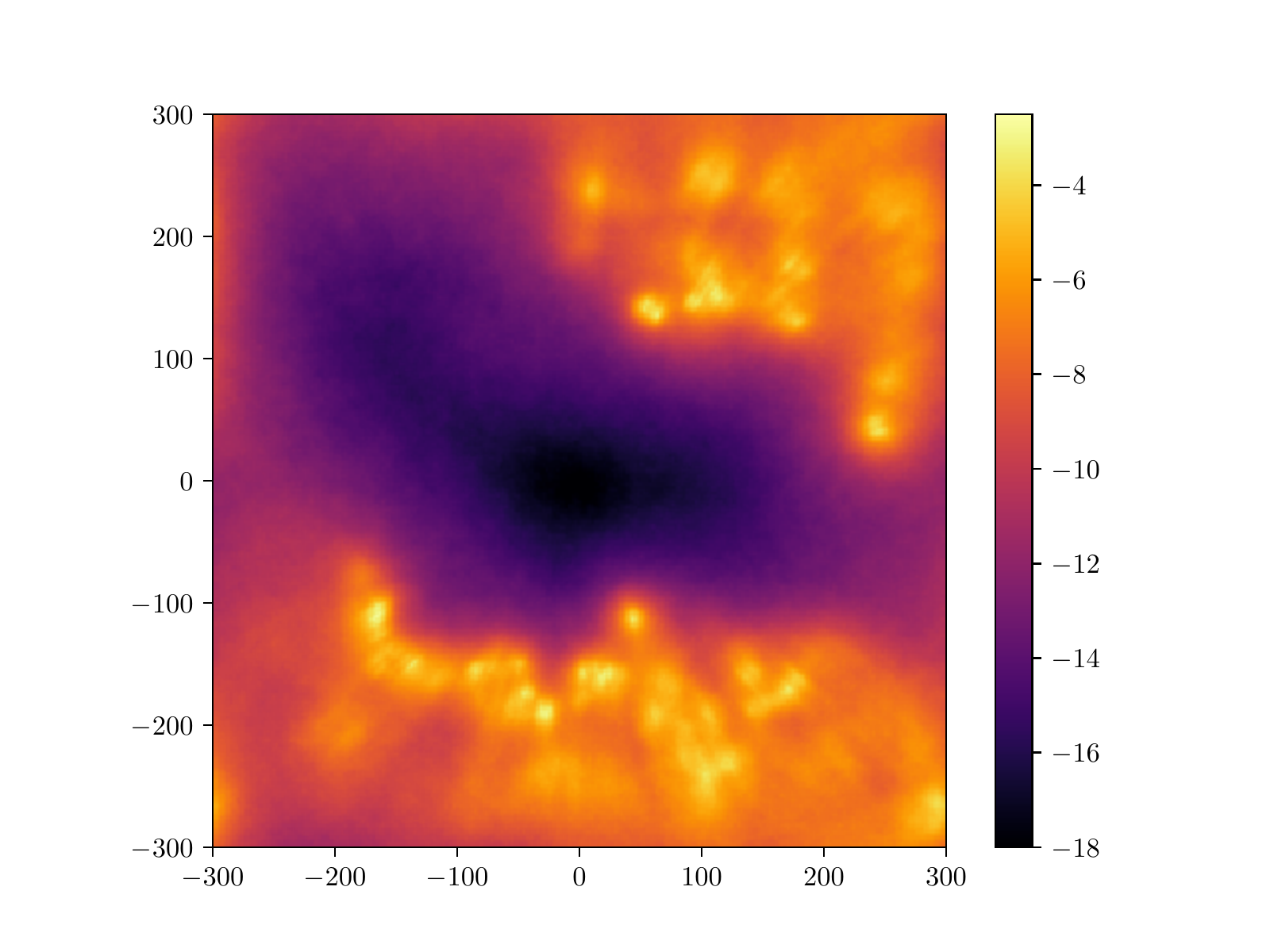}
		\caption{
		\label{fig:our-mean-log-plane-slice}}
	\end{subfigure}
	\\
	\begin{subfigure}[t]{.46\textwidth}
		\includegraphics[trim={1.5cm .5cm 2cm 1cm}, clip, width=.95\textwidth]{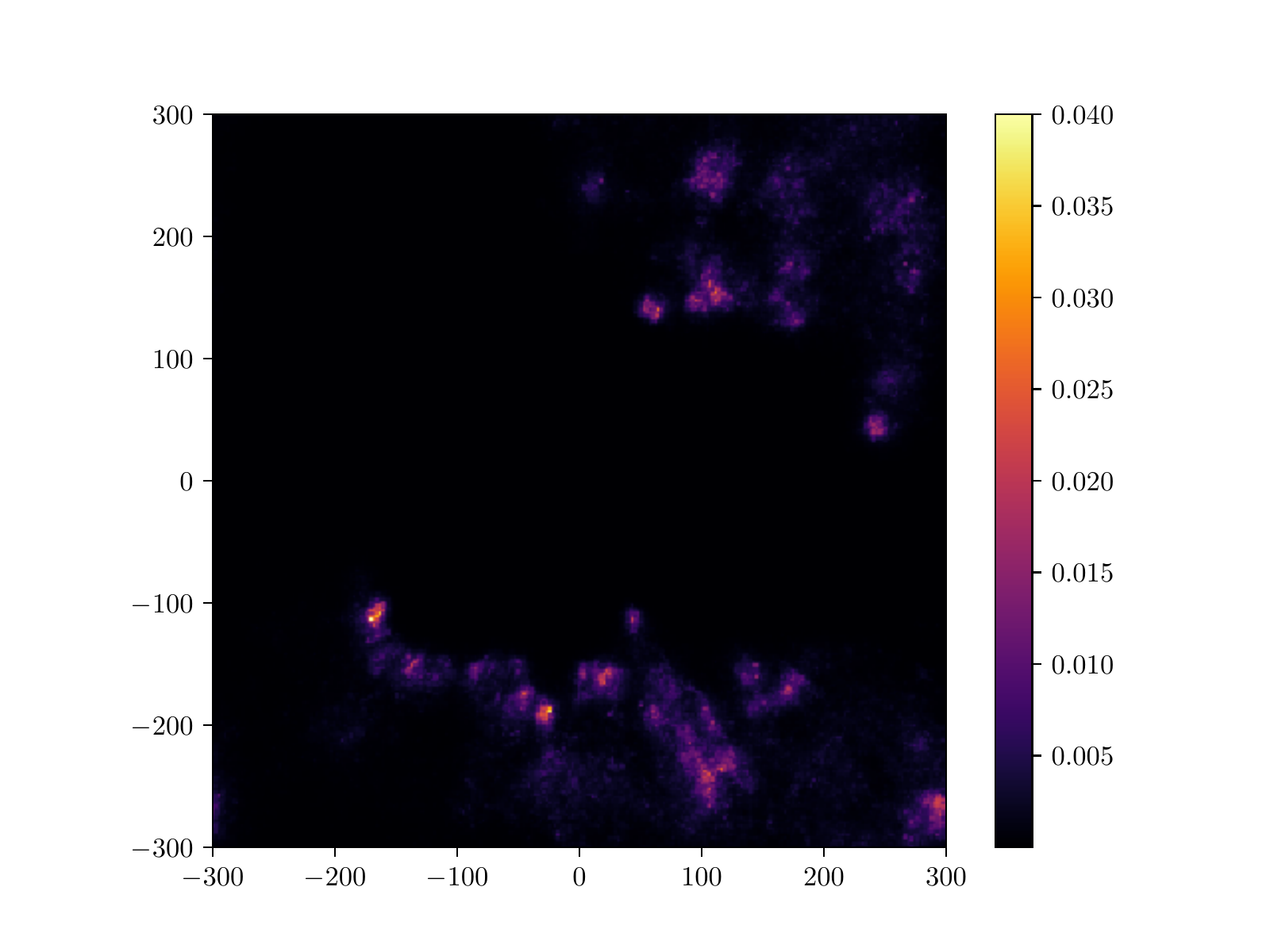}
	\caption{
	\label{fig:our-err-plane-slice}
	}
	\end{subfigure}
	~
	\begin{subfigure}[t]{.46\textwidth}
		\includegraphics[trim={1.5cm .5cm 2cm 1cm}, clip, width=.95\textwidth]{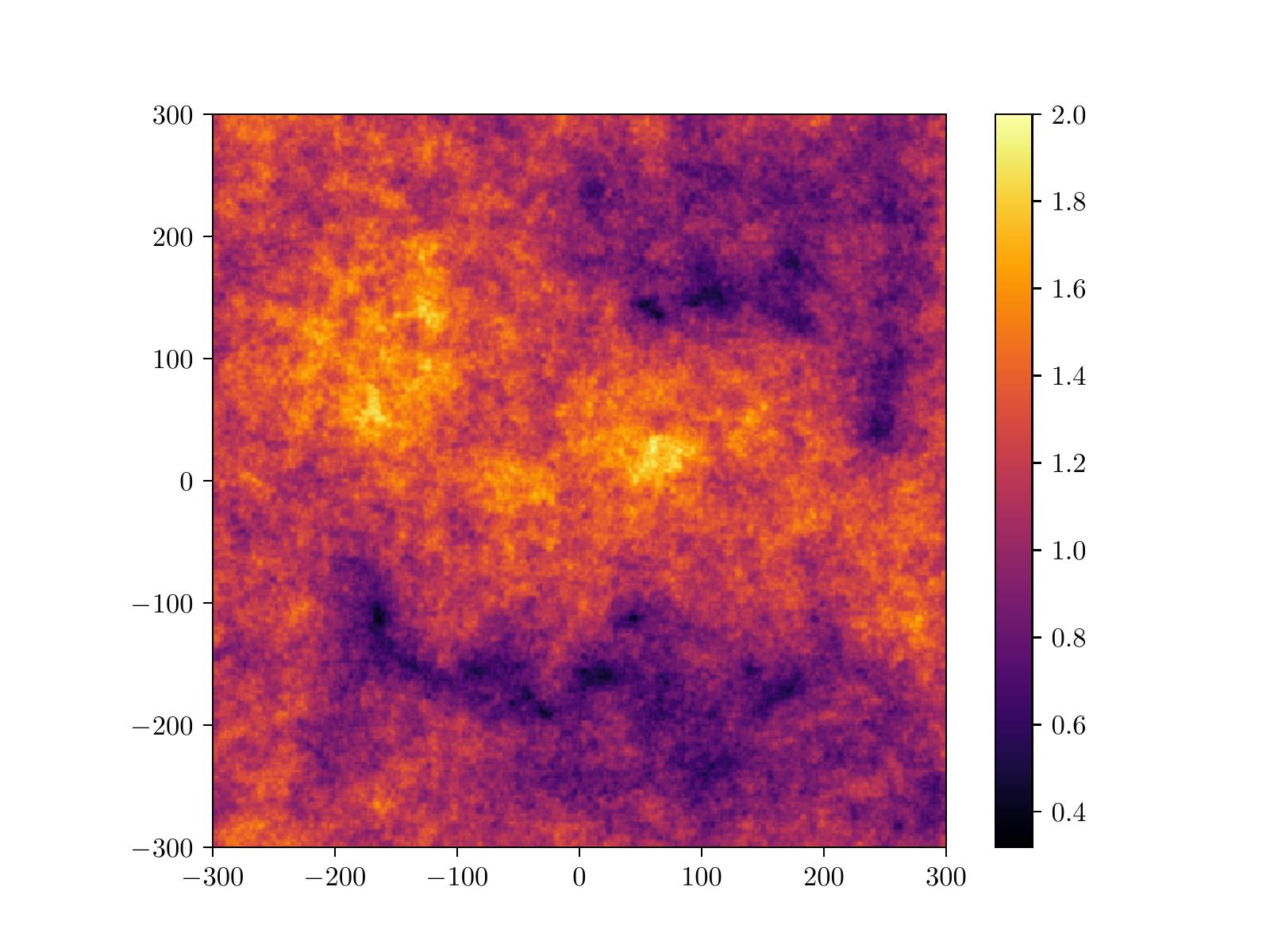}
	\caption{
	\label{fig:our-err-log-plane-slice}
		}
	\end{subfigure}
	\caption{
		\label{fig:plane-slices}
		The reconstructed dust pseudo-density in a slice of the galactic plane.
	The first plot shows differential dust extinction in the plane containing the Sun. The second plot of the first row is a logarithmic version of the first plot.
	The second row shows the corresponding uncertainty maps.
	The unit of the dust is G-band extinction in $e$-folds per parsec.
	}
\end{figure*}

The reconstruction of \cite{lallement20183d} is performed using Gaussian process regression, as is ours.
Thus one can compute the prior Gaussian process correlation power spectrum used in their reconstruction.
Fig.\,\ref{fig:prior-spectra} shows both our inferred power spectrum as well as their assumed power spectrum.
These two power spectra agree more or less for the larger modes (low $k$), where the data is very constraining.

\begin{figure}[ht]
	\includegraphics[width=0.5\textwidth]{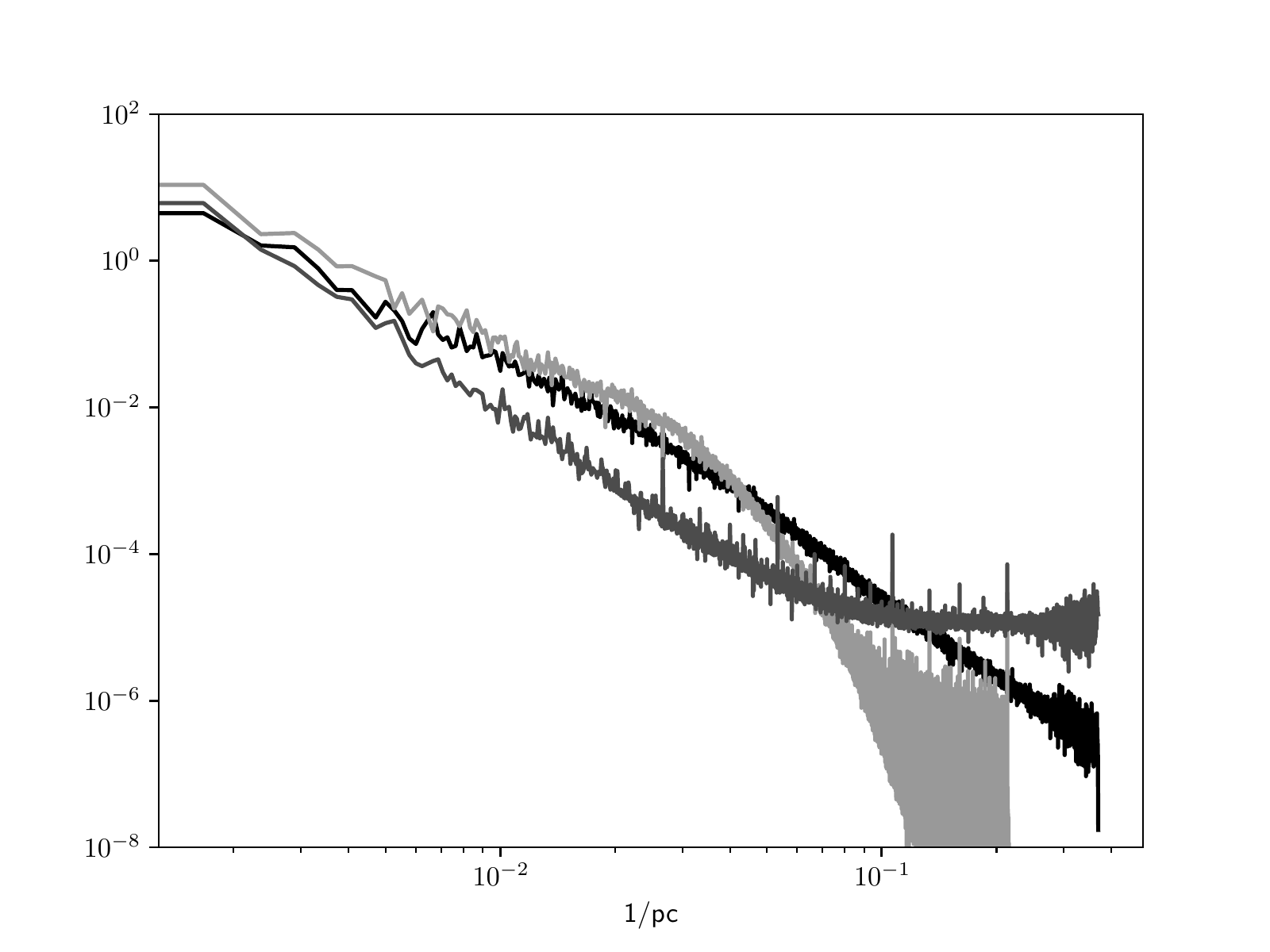}
	\caption{
		Empirical spatial correlation power spectra of the reconstructed mean dust density in units of $\text{pc}$.
		The black line was computed from our reconstruction, the dark-grey line is computed from the reconstruction of \cite{lallement20183d} and the light-grey line is computed from the reconstruction of \cite{green2018galactic}.
		For the reconstruction of \cite{green2018galactic} unspecified voxel values on sightlines that lacked data were replaced with 0.
	\label{fig:powers}
	}
\end{figure}

One can empirically compute power spectra of the dust density using a Fourier transformation. 
A comparison plot with all the three mentioned reconstruction can be found in Fig.\,\ref{fig:powers}. 
This shows a white noise floor in the reconstruction of \cite{green2018galactic}, which can visually also be seen as small scale structures in the plane projections shown in Figs.\,\ref{fig:comparison-plot} and \ref{fig:log-comparison-plot}.

\section{Conclusions}
\label{sec:conclusions}
   \begin{enumerate}
      \item We provide a highly resolved map of the local dust density using only Gaia data.
	      This map agrees on large scales with previously published maps of \cite{lallement20183d} and \cite{green2018galactic}, but also shows significant differences on small scales.
	      These differences might to a large degree stem from the different data used.
	 Our map shows many structures visible in the Planck dust map \citep{akrami2018planck}.
      \item In comparison to previous maps, we were able to improve on 3D resolution while still being mostly consistent on the large scales.
	      A comparison to 2D maps like the Planck dust map seems to confirm the features present in our map.
      \item We find that the logarithmic density of dust exhibits a power-law power spectrum with a 3D spectral index of $3.1$, corresponding to a 1D index of $1.1$. This is a significantly harder spectrum as that expected for a passive tracer in Kolmogorov turbulence, which would be a 1D index of $5/3$.
	      The harder spectrum is probably caused by the sharp edges of the local bubble and other ionization or dust evaporation fronts.
      \item In contrast to other dust reconstructions, we predict very low dust densities inside the local bubble.
	      This discrepancy is possibly an artifact of our reconstruction as there are known dust clouds in our vicinity, for example the northern high latitude shells \citep{puspitarini2012distance} and the local Leo cold cloud \citep{peek2011local}. 
	The Leo cold cloud is however considerably smaller than a voxel of our simulation.
	The possibility that Gaia extinction estimates are biased for small distances can also not be excluded.
      \item We hope that by providing accurate reconstructions of the nearby dust clouds, further studies of dust morphology will be possible as well as the construction of more accurate extinction models for photon observations in a large range of frequency bands.
   \end{enumerate}

\begin{acknowledgements}
	We acknowledge helpful commentaries and suggestion from our anonymous referee as well as fruitful discussions with S. Hutschenreuter, J. Knollm\"uller, P. Arras and others from the information field theory group at the MPI for astrophysics, Garching.

This work has made use of data from the European Space Agency (ESA) mission
{\it Gaia} (\url{https://www.cosmos.esa.int/gaia}), processed by the {\it Gaia}
Data Processing and Analysis Consortium (DPAC,
\url{https://www.cosmos.esa.int/web/gaia/dpac/consortium}). Funding for the DPAC
has been provided by national institutions, in particular the institutions
participating in the {\it Gaia} Multilateral Agreement.
\end{acknowledgements}


\bibliography{ift}
\bibliographystyle{aa}
\end{document}